\documentclass[useAMS, usenatbib]{mn2e}
\usepackage{color}
\usepackage{epsfig}
\usepackage{epstopdf}
\usepackage{amsmath, amssymb}
\usepackage{hyperref}
\usepackage{wasysym} 
\usepackage{subfigure}
\usepackage{multirow}
\usepackage{enumerate}

\usepackage{setspace}  

\newlength{\plotwidth}
\newlength{\fullwidth}
\renewcommand{\topmargin}{-1cm} 
 
\title[The BOSS-WiggleZ overlap region I]{The BOSS-WiggleZ overlap region I: Baryon Acoustic Oscillations}
\author[Florian Beutler et al.]
{\parbox{\textwidth}{Florian Beutler$^{1}$\thanks{E-mail: \texttt{fbeutler@lbl.gov}},
Chris Blake$^2$, Jun Koda$^{2,3}$, Felipe A.\ Mar\'in$^2$, Hee-Jong Seo$^{4}$, Antonio J. Cuesta$^5$, Donald P. Schneider$^{6,7}$}\vspace{0.4cm}\\
\parbox{\textwidth}{
$^{1}$Lawrence Berkeley National Laboratory, 1 Cyclotron Road, Berkeley, CA 94720, USA\\
$^{2}$Centre for Astrophysics \& Supercomputing, Swinburne University of Technology, P.O. Box 218, Hawthorn, VIC 3122, Australia\\
$^{3}$INAF, Osservatorio Astronomico di Brera, via E. Bianchi 46, 23807\\
$^{4}$Department of Physics and Astronomy, Ohio University, 251B Clippinger Labs, Athens, OH 45701\\
$^{5}$ICC, University of Barcelona, IEEC-UB, Mart{\'\i} i Franqu{\`e}s 1, E-08028, Barcelona, Spain\\
$^{6}$Department of Astronomy and Astrophysics, The Pennsylvania State University, University Park, PA 16802\\
$^{7}$ Institute for Gravitation and the Cosmos, The Pennsylvania State University, University Park, PA 16802}}

\begin{document}

\label{firstpage}

\maketitle

\begin{abstract}
We study the large-scale clustering of galaxies in the overlap region of the Baryon Oscillation Spectroscopic Survey (BOSS) CMASS sample and the WiggleZ Dark Energy Survey.  We calculate the auto-correlation and cross-correlation functions in the overlap region of the two datasets and detect a Baryon Acoustic Oscillation (BAO) signal in each of them.  The BAO measurement from the cross-correlation function represents the first such detection between two different galaxy surveys.  After applying density-field reconstruction we report distance-scale measurements $D_V r_s^{\rm fid} / r_s = (1970 \pm 45, 2132 \pm 65, 2100 \pm 200)$ Mpc from CMASS, the cross-correlation and WiggleZ, respectively. The distance scales derived from the two datasets are consistent, and are also robust against switching the displacement fields used for reconstruction between the two surveys. We use correlated mock realizations to calculate the covariance between the three BAO constraints. This approach can be used to construct a correlation matrix, permitting for the first time a rigorous combination of WiggleZ and CMASS BAO measurements.  Using a volume-scaling technique, our result can also be used to combine WiggleZ and future CMASS DR12 results. Finally, we show that the relative velocity effect, a possible source of systematic uncertainty for the BAO technique, is consistent with zero for our samples.
\end{abstract}

\begin{keywords}
surveys, cosmology: observations, dark energy, distance scale, large scale structure of Universe
\end{keywords}

\section{Introduction}

The Baryon Acoustic Oscillation (BAO) signal is a relict of the early Universe, where photon pressure caused sound-waves to move out of over-densities~\citep{Peebles:1970ag,Sunyaev:1970eu,Bond:1987ub}. These sound-waves became imprinted in the distribution of Cosmic Microwave Background (CMB) photons as well as in the matter density field. Over time, the density field evolved through gravitational collapse and cosmic expansion. While gravitational interaction can smear out the BAO signal, a complete destruction would require interactions over very large scales (today $\approx 150\,$Mpc), making the BAO feature a very robust observable. 

The BAO signal in the density field at different redshifts can be related to the BAO signal measured in the CMB and therefore allows employment of the so called standard ruler technique~\citep{Blake:2003rh, Seo:2003pu}. We can compare the apparent size of the BAO signal measured in galaxy surveys with the absolute size of this signal measured in the CMB and use this to map out the expansion history of the Universe. Simulations have shown that the BAO signal is unaffected by systematic uncertainties down to the sub-percent level~\citep{Eisenstein:2006nj,Guzik:2006bu,Smith:2006ne,Smith:2007gi,Angulo:2007fw,Padmanabhan:2009yr,Mehta:2011xf} and hence represents one of the most reliable tools available for precision cosmology.

The most precise BAO measurement has recently been reported by the Baryon Oscillation Spectroscopic Survey (BOSS) collaboration ($1\%$,~\citealt{Anderson:2013zyy}) at a redshift of $z=0.57$. BOSS also achieved a $2\%$ BAO distance constraint with the LOWZ sample at $z = 0.32$~\citep{Tojeiro:2014eea}. The WiggleZ galaxy survey~\citep{Drinkwater:2009sd} produced a $4\%$ constraint at redshift $z=0.6$~\citep{Blake:2011en,Kazin:2014qga} and the 6-degree Field Galaxy Survey (6dFGS)~\citep{Jones:2009yz} yielded a $5\%$ constraint at $z=0.1$~\citep{Beutler:2011hx}. Using the SDSS-DR7~\citep{Abazajian:2008wr} main sample,~\citet{Ross:2014qpa} reported a $4\%$ BAO constraint. Future galaxy surveys point to Baryon Acoustic Oscillations as their main science driver (e.g. DESI,~\citealt{Schlegel:2011zz}) and will push well below percent-level precision over a large redshift range. 

In this paper we study the BAO signal in the overlap region of the two largest galaxy surveys available to date, the CMASS sample of BOSS DR11~\citep{Ahn:2013gms} and the WiggleZ galaxy survey~\citep{Drinkwater:2009sd}. We use the BOSS DR11 sample, since the overlap between BOSS and WiggleZ does not increase with DR12. The two surveys have been produced independently, being subject to different seeing conditions, instrumental noise, reduction pipelines and observation strategies. While BOSS targeted mainly Luminous Red Galaxies (LRGs) and has a galaxy bias around $b\approx2$, WiggleZ selected blue Emission Line Galaxies (ELGs) with a bias around $b\approx 1$. Since we restrict our analysis to the overlap region between the two surveys, our datasets have the same sample variance. This allows a test of possible systematic uncertainties in galaxy clustering measurements. In this study we focus on BAO measurements while a companion paper~\citep{Marin:2015} investigates redshift-space distortions in the overlap region.

The typical candidates for possible systematic errors in galaxy clustering studies are redshift space distortions as well as non-linear clustering,  accompanied by a non-linear galaxy bias. Non-linear evolution is normally connected to the density peaks of the matter density field and hence we expect that galaxies which sit in high density regions, such as LRGs, should have stronger non-linear effects compared to field galaxies, such as ELGs. On the other hand, field galaxies have a larger redshift space distortion signal, which is a source of possible systematics if not modeled correctly. In addition to the properties of the density field itself, there are also survey specific aspects such as incompleteness that can modify the measured clustering statistic.

Another possible systematic bias for BAO measurements is the relative velocity effect. The different velocities of baryons and dark matter after decoupling can affect early galaxy formation processes. At places where the relative velocity is large, baryons can escape gravitational potentials. The first stars in the Universe will form wherever the relative velocity is low, since this is where the baryons will condense into the gravitational potentials of the dark matter. The relative velocity decays with redshift and is negligible at the redshift of the BOSS and WiggleZ galaxies. The question is whether the modulation of structure formation due to the relative velocity effect at high redshift is carried in the tracer galaxies we observe today to measure BAO. \citet{Yoo:2013qla} argue that galaxies which consist predominantly of old stars (such as LRGs) could carry this effect, while young galaxies might not. This would result in a modulation of the BAO signal in BOSS, but not in WiggleZ. In the overlap region between BOSS and WiggleZ we can measure the BAO position in the same volume and compare whether the two surveys yield the same result, allowing the placement of constraints on the possible impact of the relative velocity effect.

In the next section we will introduce the two datasets used in our analysis, BOSS-CMASS and WiggleZ. Section~\ref{sec:cor} describes our correlation function estimate followed by a discussion of our mock catalogues in Section~\ref{sec:mocks}. In Section~\ref{sec:recon} we present our technique of density field reconstruction followed by a discussion of our model for the correlation functions in Section~\ref{sec:model}. We then compare the obtained displacement fields and perform the correlation function fits in Section~\ref{sec:sys}. In Section~\ref{sec:covariance} we determine the correlation between BOSS-CMASS and WiggleZ. In Section~\ref{sec:relvel} we introduce the relative velocity effect and perform fits to the data to constrain the relative velocity bias. We conclude in Section~\ref{sec:conclusion}.

For clarity we will use the name CMASS-BW, WiggleZ-BW and cc-BW for the CMASS, WiggleZ and cross-correlation results limited to the overlap region between the two surveys. We adopt a flat $\Lambda$CDM cosmology with $\Omega_m = 0.27$. The same model is used to construct templates for the BAO fits and hence our measurements should be used in conjunction with $r^{\rm fid}_s(z_d) = 150.18\,$Mpc\footnote{Sound horizon scale calculated with CAMB~\citep{Lewis:1999bs}.}.

\section{Datasets}
\label{sec:datasets}

\subsection{The BOSS survey}
\label{sec:BOSS}

\begin{figure*}
\begin{center}
\epsfig{file=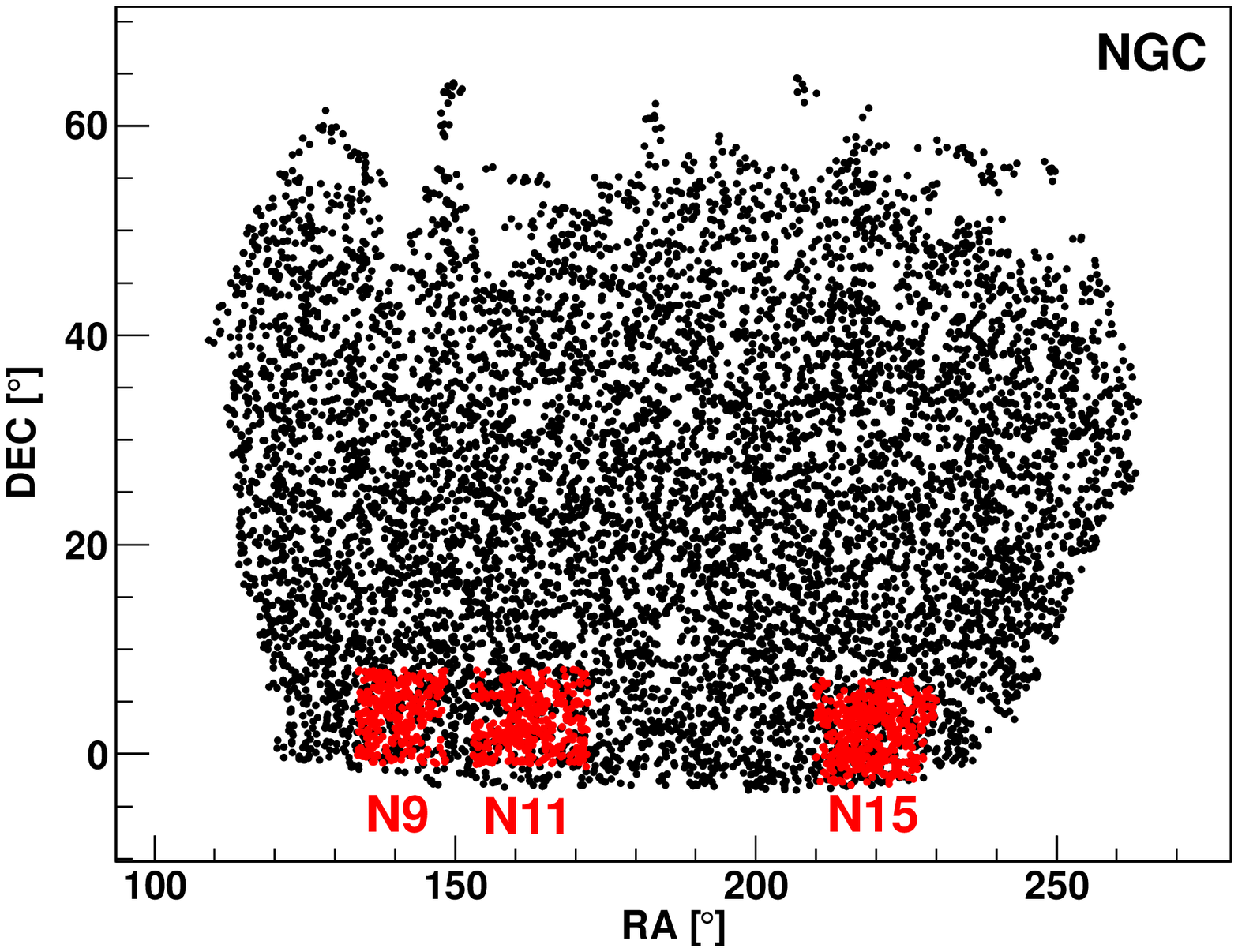,width=8cm}
\epsfig{file=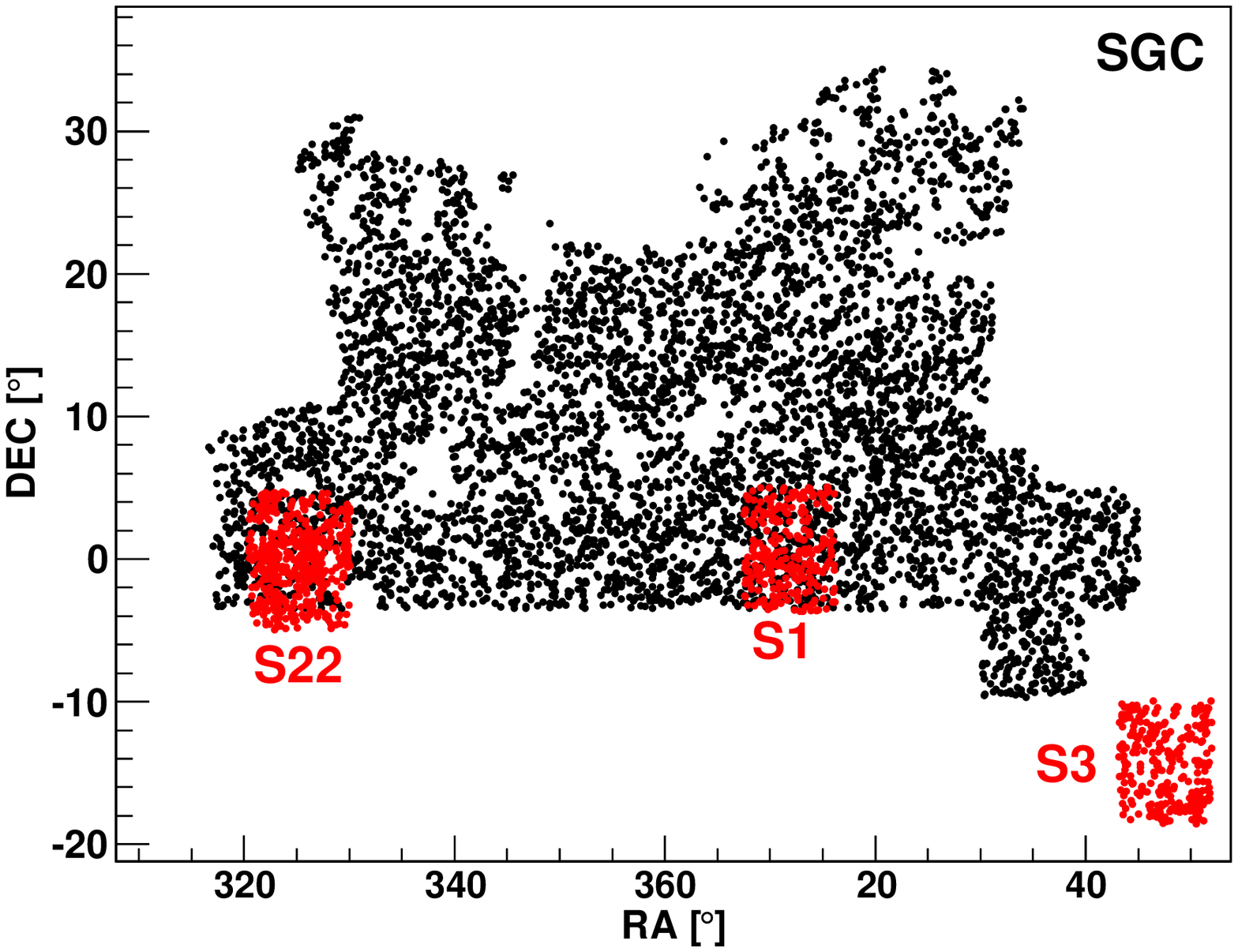,width=8cm}
\caption{Sky coverage of BOSS-CMASS DR11 (black) and WiggleZ (red). The left plot shows the north galactic cap (NGC), while the right plot shows the south galactic cap (SGC). Five of the six WiggleZ regions are covered by CMASS, with region S22 being only partly covered. We only plot a random fraction of $3\%$ of all galaxies.}
\label{fig:sky}
\end{center}
\end{figure*}

BOSS, as part of SDSS-III~\citep{Eisenstein:2011sa,Dawson:2012va} measured spectroscopic redshifts of $\approx 1.5$ million galaxies (and $150\,000$ quasars) making use of the SDSS multi-fibre spectrographs~\citep{Bolton:2012hz,Smee:2012wd}. The galaxies are selected from multi-color SDSS imaging~\citep{Fukugita:1996qt,Gunn:1998vh,Smith:2002pca,Gunn:2006tw,Doi:2010rf} and cover a redshift range of $z = 0.15$ - $0.7$, where the survey is split into two samples called LOWZ ($z=0.15$ - $0.43$) and CMASS ($z=0.43$ - $0.7$). In this analysis we only use the CMASS sample. The survey is optimized for the measurement of the BAO scale and hence covers a large cosmic volume ($V_{\rm eff} = 2.31\times 10^9[$Mpc$/h]^3$) with a density of $\overline{n} \approx 3\times10^{-4}[h/$Mpc$]^{3}$, high enough to ensure that shot noise is not the dominant error contribution at the BAO scale~\citep{White:2010ed}. Most CMASS galaxies are red with a prominent $4000\,$\AA\; break in their spectral energy distribution. Halo Occupation studies have shown that galaxies selected like the CMASS galaxies are mainly central galaxies residing in dark matter halos of $10^{13}M_{\odot}/h$, with a $5$ - $10\%$ satellite fraction~\citep{White:2010ed}. CMASS galaxies are highly biased ($b\sim 2$), which boosts the clustering signal including BAO with respect to the shot noise level.

We include three different incompleteness weights to  account for shortcomings of the CMASS dataset (see~\citealt{Ross:2012qm} and~\citealt{Anderson:2013zyy} for details): A redshift failure weight, $w_{\rm rf}$, a fibre collision weight, $w_{\rm fc}$, and a systematics weight, $w_{\rm sys}$, which is a combination of a stellar density weight and a seeing condition weight. Each galaxy is thus counted as 
\begin{equation}
w_c = (w_{\rm rf} + w_{\rm fc} - 1)w_{\rm sys}.
\label{eq:cweight}
\end{equation}
Figure~\ref{fig:sky} shows the sky coverage of BOSS-CMASS with the north galactic cap (NGC) on the left and the south galactic cap (SGC) on the right.

\subsection{The WiggleZ survey}
\label{sec:WiggleZ}

The WiggleZ Dark Energy Survey~\citep{Drinkwater:2009sd} is a large-scale galaxy redshift survey of bright emission-line galaxies (ELGs), which was carried out at the Anglo-Australian Telescope between August 2006 and January 2011 using the AAOmega spectrograph~\citep{Sharp:2006gy}. Targets were selected via joint ultraviolet and optical magnitude and color cuts using input imaging from the Galaxy Evolution Explorer (GALEX) satellite~\citep{Martin:2004yr}. The survey is now complete, comprising $240\,000$ redshifts and covering $816\,\deg^2$ in six separate sky areas. The redshift range is roughly $0.1 < z < 1.0$ with a mean redshift at $z=0.6$. Figure~\ref{fig:sky} shows the sky coverage of WiggleZ (red), where we name the different patches S1, S3, S22, N9, N11, and N15.

The strategy for completeness correction in WiggleZ is different to the method used in CMASS. Instead of weighting the data galaxies, the incompleteness has been introduced into the random catalogues~\citep{Blake:2010xz} and hence no completeness weighting is needed for this dataset.

\subsection{Definition of the overlap region}
\label{sec:overlap}

\begin{figure*}
\epsfig{file=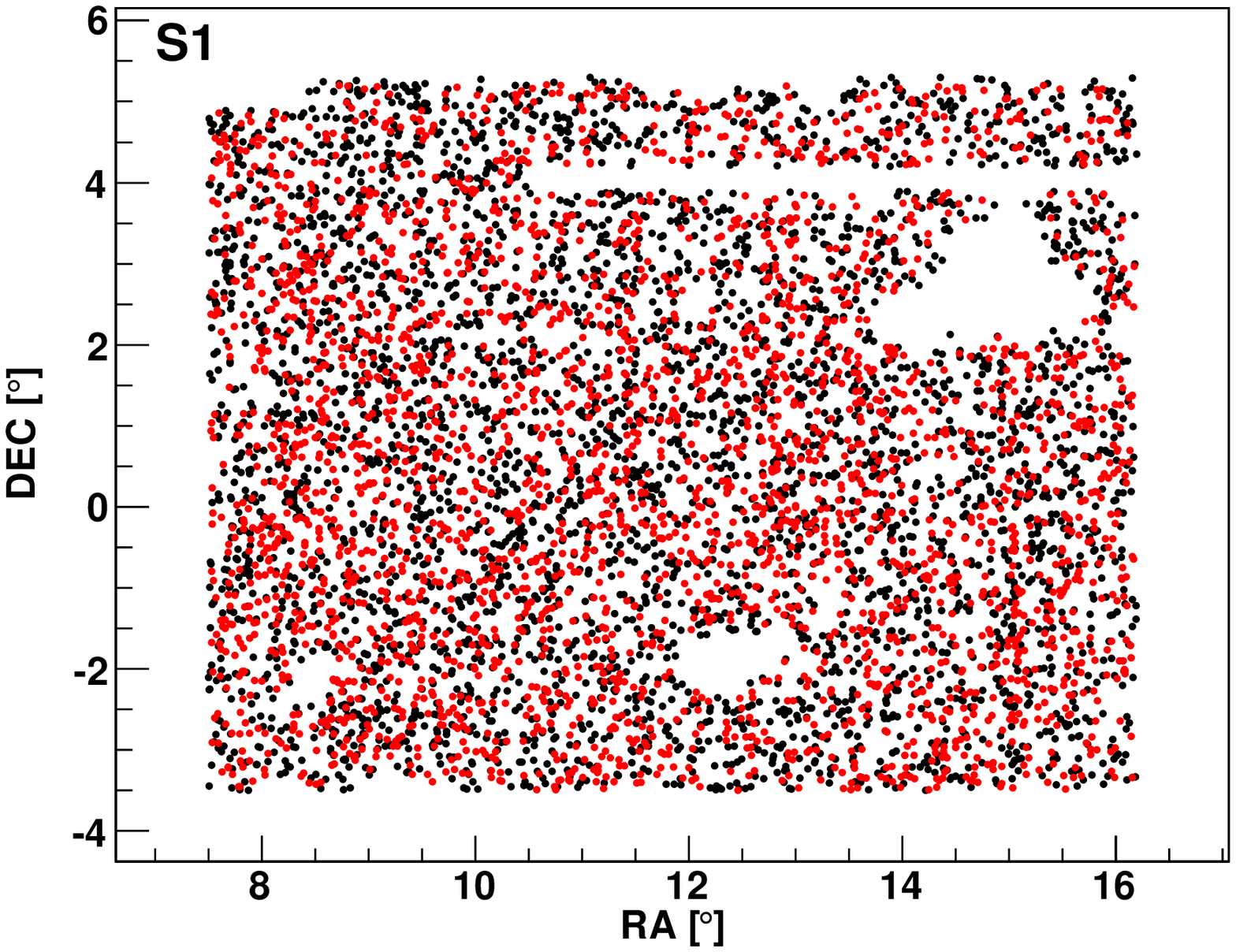,width=5.8cm}
\epsfig{file=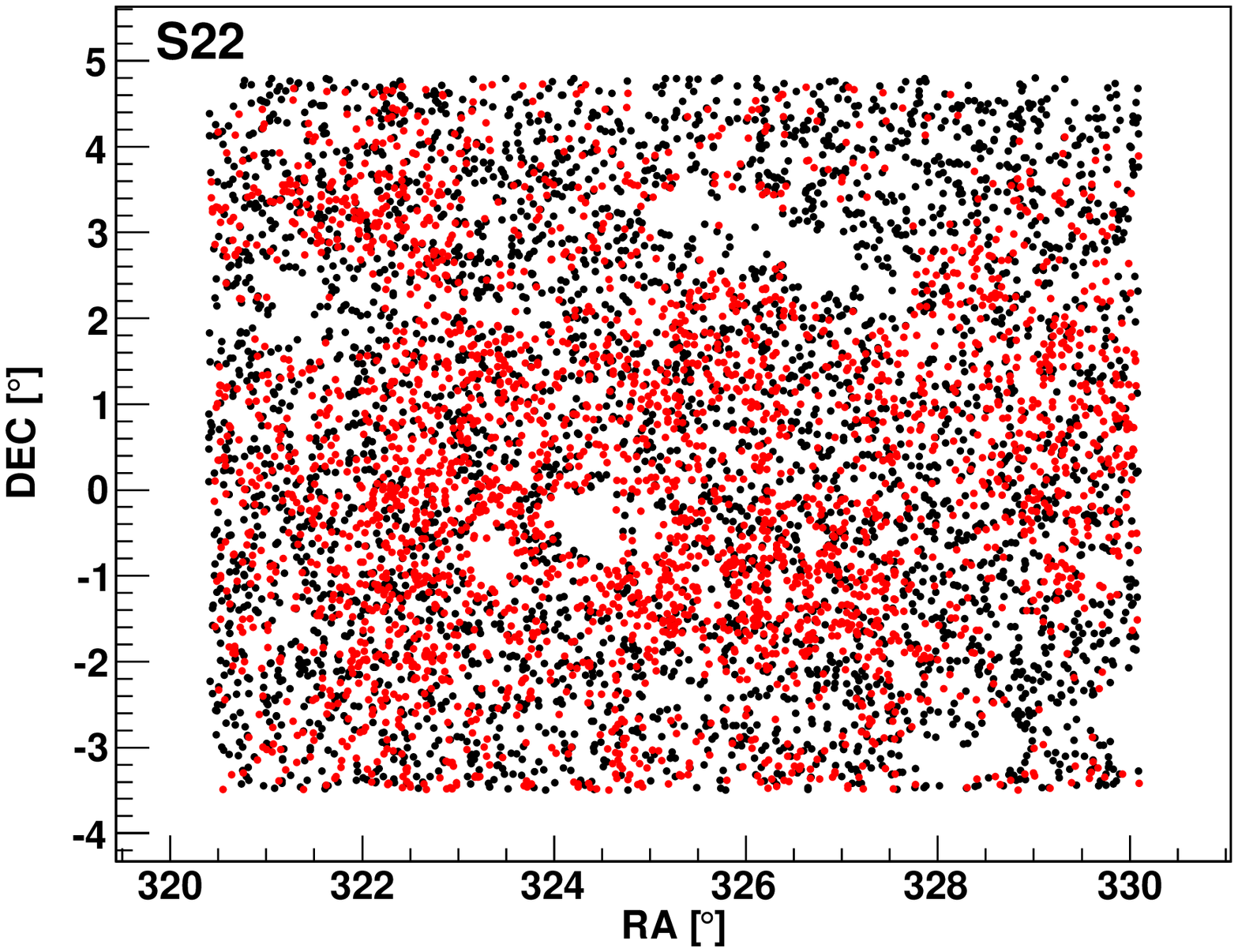,width=5.8cm}\\
\epsfig{file=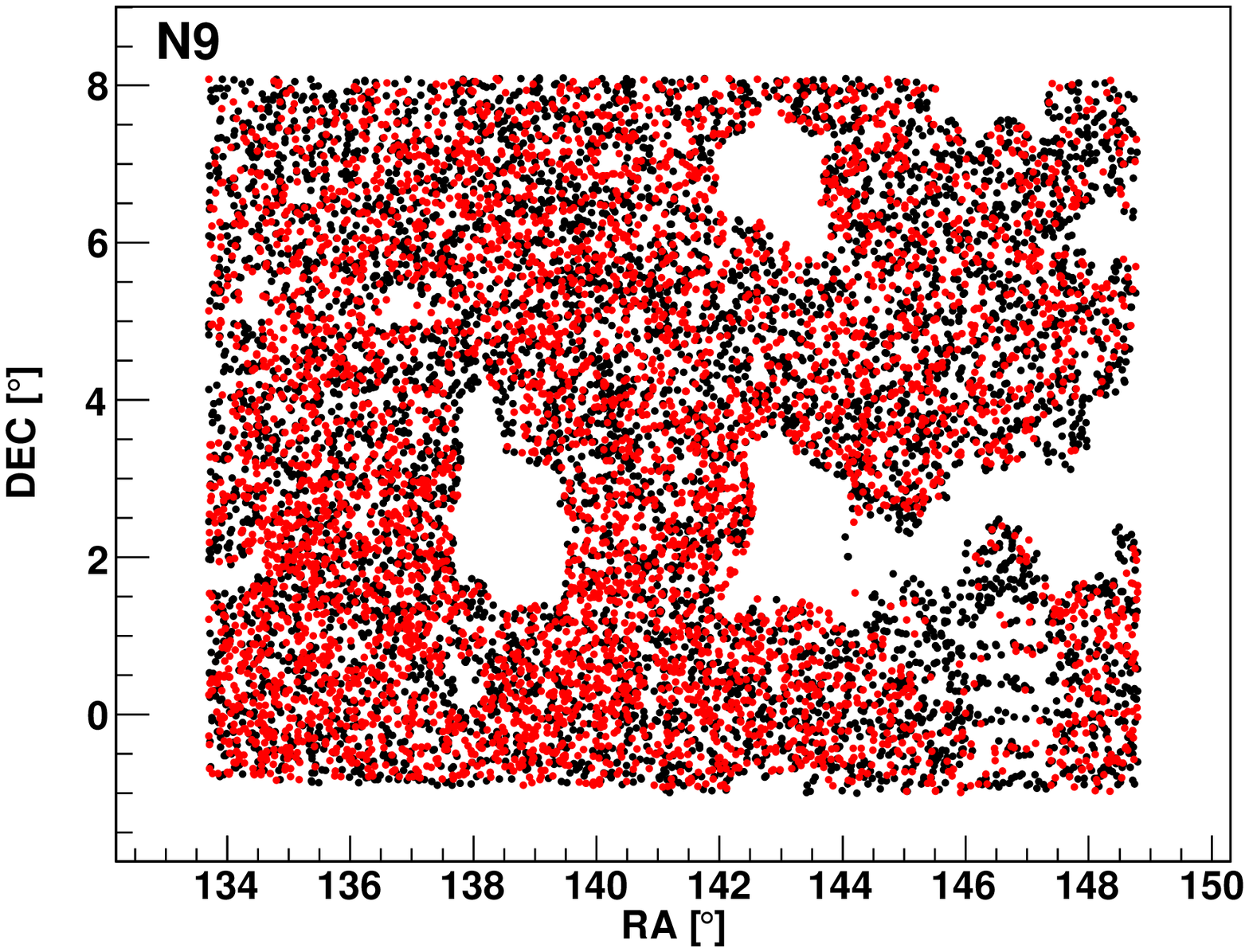,width=5.8cm}
\epsfig{file=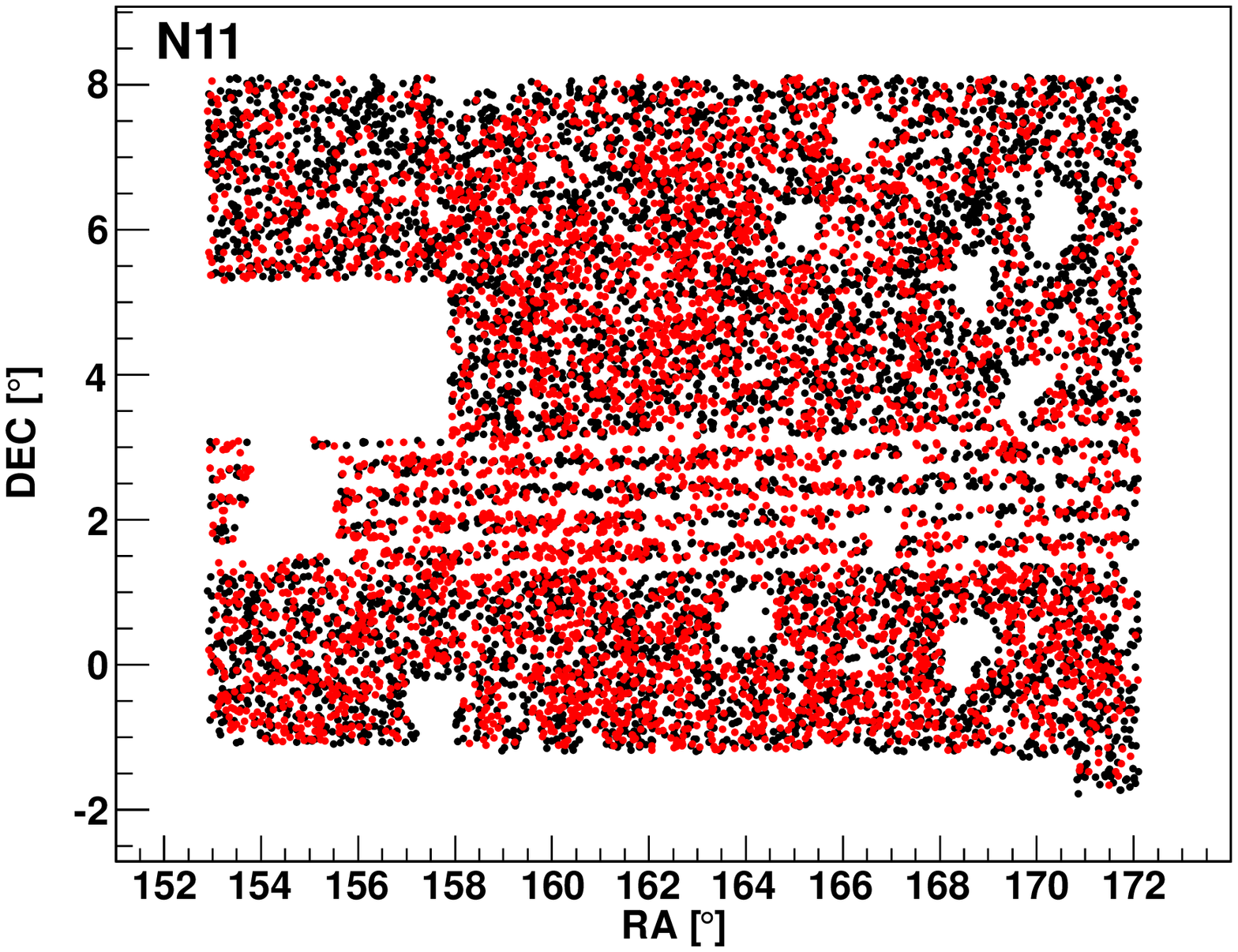,width=5.8cm}
\epsfig{file=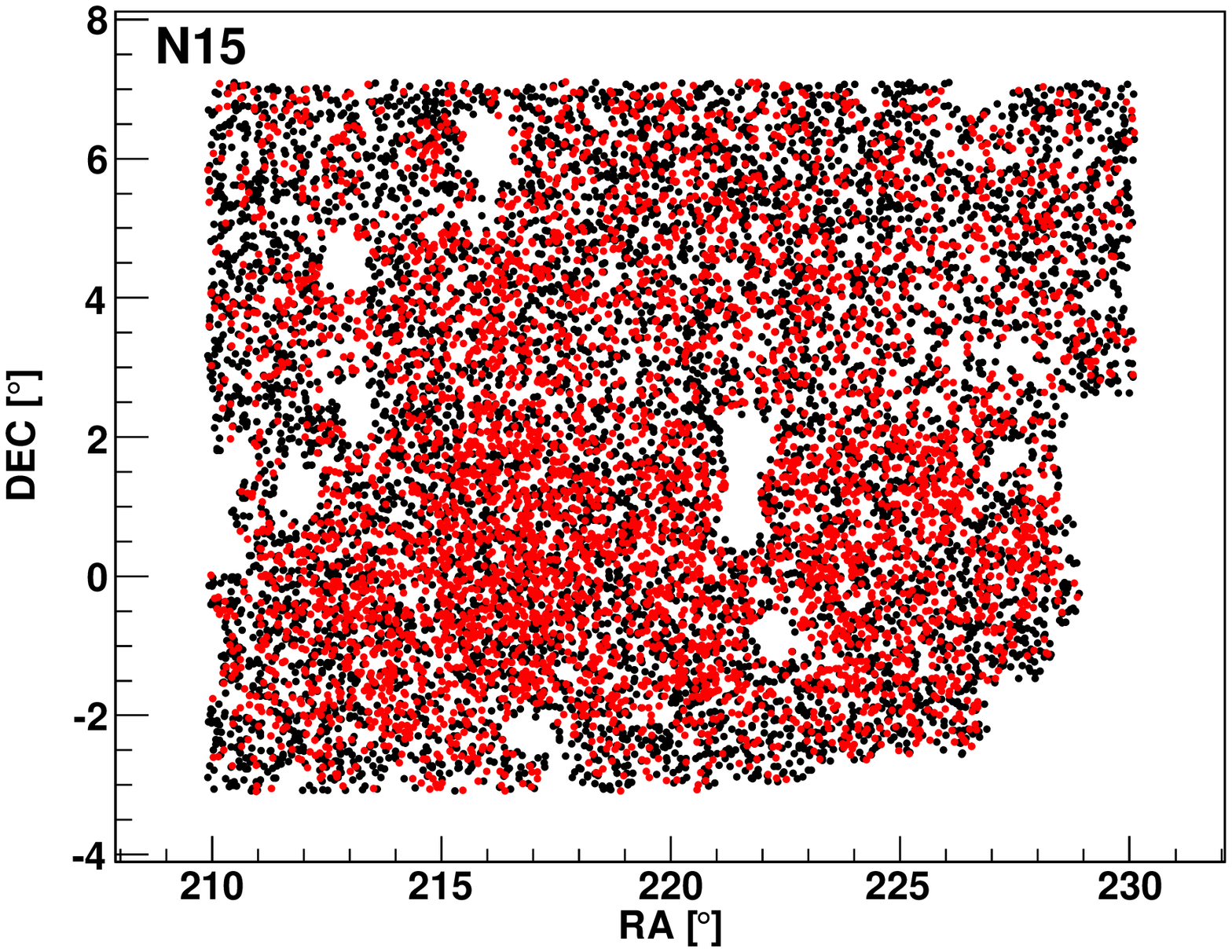,width=5.8cm}
\caption{The overlap region between BOSS-CMASS (black) and WiggleZ (red). Most of the angular incompleteness is caused by WiggleZ, while the empty stripes in region N11 are caused by incomplete photometric data in CMASS. To generate these regions, we divided the sky into $0.1\,\deg^2$ bins and included all bins which contain CMASS as well as WiggleZ random galaxies. We only plot a random fraction of $10\%$ of all galaxies.}
\label{fig:skypro}
\end{figure*}

We define the overlap region between CMASS and WiggleZ by splitting the sky into $0.1\,\deg^2$ bins and selecting all bins which contain CMASS as well as WiggleZ random galaxies. The redshift range is defined by CMASS and is given by $0.43 < z < 0.7$.
Figure~\ref{fig:sky} shows the six WiggleZ regions (red), of which five are covered by the BOSS-CMASS sample (black), with region S22 being only partly covered.

\begin{table}
\begin{center}
\caption{Effective volume and number of galaxies of the five distinct CMASS-WiggleZ overlap regions (see Figure~\ref{fig:skypro}) as well as the total volume of the two surveys. The effective volume is calculated using Eq.~\ref{eq:veff} and $P_0 = 20\,000h^{-3}\,$Mpc$^{3}$ for CMASS and $P_0 = 5000h^{-3}\,$Mpc$^{3}$ for WiggleZ. The names CMASS-BW and WiggleZ-BW stand for the CMASS and WiggleZ samples restricted to the overlap region between the two.}
	\begin{tabular}{llcl}
		\hline
		\multicolumn{2}{c}{Region} & $V_{\rm eff}$ & $N_{\rm gal}$\\
		 & & [$10^7h^{-3}$Mpc$^3$] & \\
		\hline
		\hline	
		\multirow{2}{*}{S1} & CMASS-BW & $1.8$ & $5742$\\
     		 & WiggleZ-BW & $0.8$ & $6621$\\		
		 \hline
		\multirow{2}{*}{S22} & CMASS-BW & $1.9$ & $6070$\\
     		 & WiggleZ-BW & $1.1$ & $10\,339$\\
		\hline
		\multirow{2}{*}{N9} & CMASS-BW & $3.1$ & $9356$\\
     		 & WiggleZ-BW & $1.7$ & $13\,960$\\
		 \hline
		\multirow{2}{*}{N11} & CMASS-BW & $3.7$ & $10\,280$\\
     		 & WiggleZ-BW & 2.0 & 15\,324\\		
		 \hline
	   	\multirow{2}{*}{N15} & CMASS-BW & $4.6$ & $14\,673$\\
     		 & WiggleZ-BW & $2.6$ & $22\,736$\\	
		 \hline
		 \hline
		\multirow{2}{*}{combined} & CMASS-BW & $15.1$ & $48\,570$\\
		& WiggleZ-BW & $8.2$ & $71\,407$\\
		\hline
		\hline
		 & CMASS-DR12 & $232.2$ & $786\,324$\\
		 & CMASS-DR11 & $204.0$ & $690\,827$\\
		 & CMASS-DR10 & $150.6$ & $544\,133$\\
		 & CMASS-DR9 & $76.9$ & $264\,281$\\
		 & WiggleZ & $21.3$ & $191\,732$\\ 
		\hline
	  \end{tabular}
\label{tab:volume}
\end{center}
\end{table}

The five overlap regions are shown separately in Figure~\ref{fig:skypro}. Most of the incompleteness in these plots is caused by the WiggleZ survey, while the empty stripes in region N11 are a result of missing photometry and hence missing galaxies in CMASS. 

\begin{figure}
\begin{center}
\epsfig{file=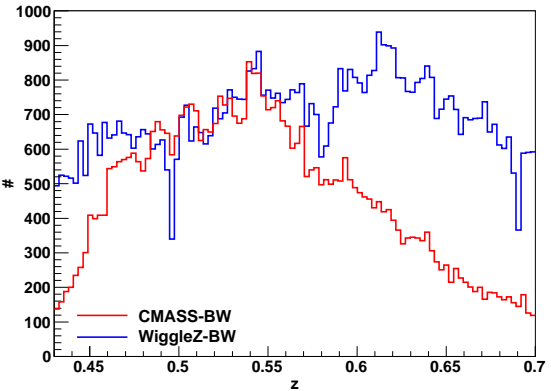,width=8cm}
\caption{Redshift distribution of CMASS-BW (red) and WiggleZ-BW (blue) combining the five separate regions.}
\label{fig:red}
\end{center}
\end{figure}

The relative importance of sample variance and shot noise in a galaxy clustering measurement is determined by the quantity $n(z)P(k)$, where $n(z)$ is the galaxy number density and $P(k)$ is the galaxy power spectrum amplitude at the BAO scale. Therefore we can trade a smaller galaxy density with a larger galaxy bias and vice versa. The WiggleZ survey has a higher galaxy number density compared to CMASS, while CMASS galaxies have a larger bias and hence a larger power spectrum amplitude. The CMASS sample has been designed with the target $nP=3$, while WiggleZ has aimed for $nP = 1$. The best quantity to compare the two surveys is the effective volume, where we use the equation suggested by~\citet{Tegmark:1997rp}:
\begin{equation}
V_{\rm eff} = \int d^3x \left[\frac{n(\vec{x})P_0}{1 + n(\vec{x})P_0}\right]^2.
\label{eq:veff}
\end{equation}
Here $P_0$ is fixed to the amplitude of the power spectrum at the wavenumber of the first BAO peak, $k\approx 0.06h/$Mpc, which turns out to be $P_0 = 20\,000h^{-3}\,$Mpc$^{3}$ for CMASS and $P_0=5000h^{-3}\,$Mpc$^{3}$ for WiggleZ. The larger value of $nP$ in the CMASS sample leads to a larger effective volume compared to WiggleZ (by about a factor of $2$). The different volumes for CMASS and WiggleZ in each overlap region, as well as the combined volumes, are summarized in Table~\ref{tab:volume}. The redshift distribution for the two samples limited to the overlap region is plotted in Figure~\ref{fig:red}.

\section{Estimating the correlation function}
\label{sec:cor}

We calculate the correlation function by counting the number of galaxy-galaxy pairs, $DD(s)$, as a function of scale $s$, as well as galaxy-random, $DR(s)$, and random-random, $RR(s)$ pairs. We then use the correlation function estimator suggested by~\citet{Landy:1993yu}:
\begin{equation}
\xi(s) = 1 + \frac{DD(s)}{RR(s)} \left(\frac{n_r}{n_d} \right)^2 - 2\frac{DR(s)}{RR(s)} \left(\frac{n_r}{n_d} \right),
\label{eq:LS2}
\end{equation}
where $n_r=\sum^{N_{\rm r}}w_i(\vec{x})$ and $n_d=\sum^{N_{\rm d}}w_i(\vec{x})$ represent the sums over the weights for all random and data galaxies, respectively. We include the inverse density weighting of~\citet{Feldman:1993ky}:
\begin{equation}
w_i(\vec{x}) = \frac{1}{1 + n(\vec{x})P_0},
\label{eq:weight}
\end{equation}
with $P_0 = 20\,000h^{-3}$\;Mpc$^{3}$ for CMASS and $P_0 = 5\,000h^{-3}$\;Mpc$^{3}$ for WiggleZ. In the case of CMASS we also include the completeness weighting of Eq.~\ref{eq:cweight}.

\begin{figure*}
\begin{center}
\epsfig{file=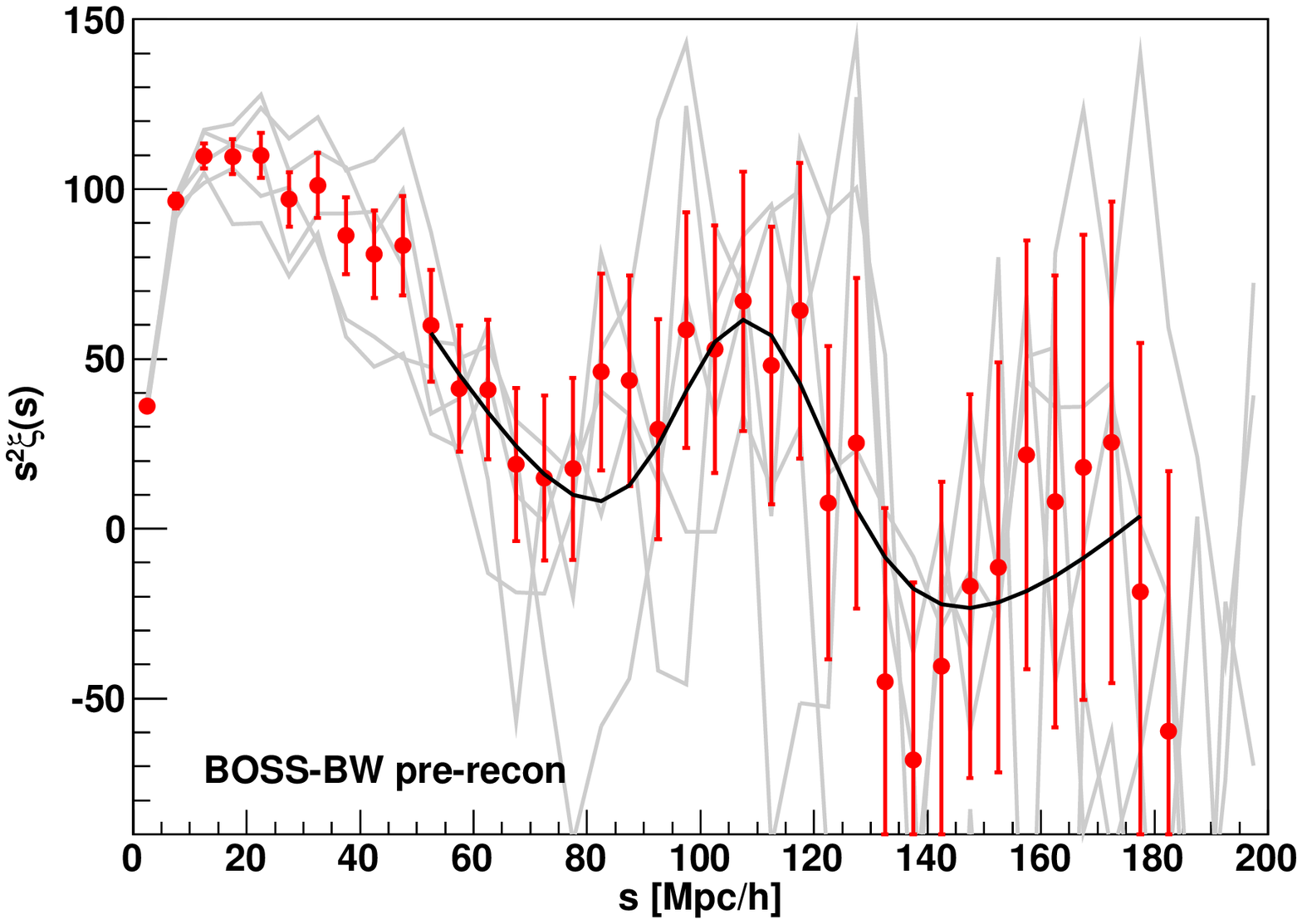,width=7cm}
\epsfig{file=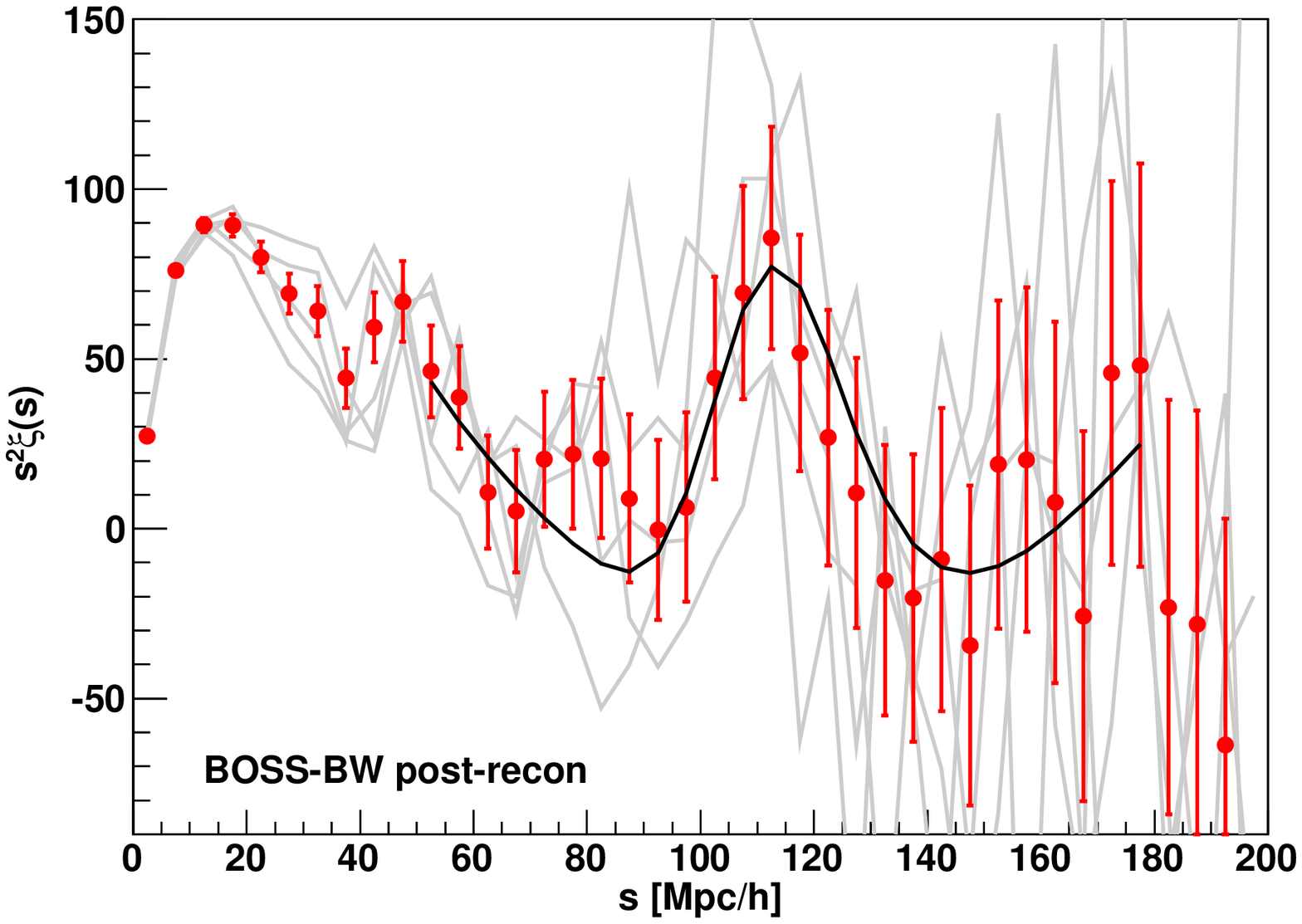,width=7cm}\\
\epsfig{file=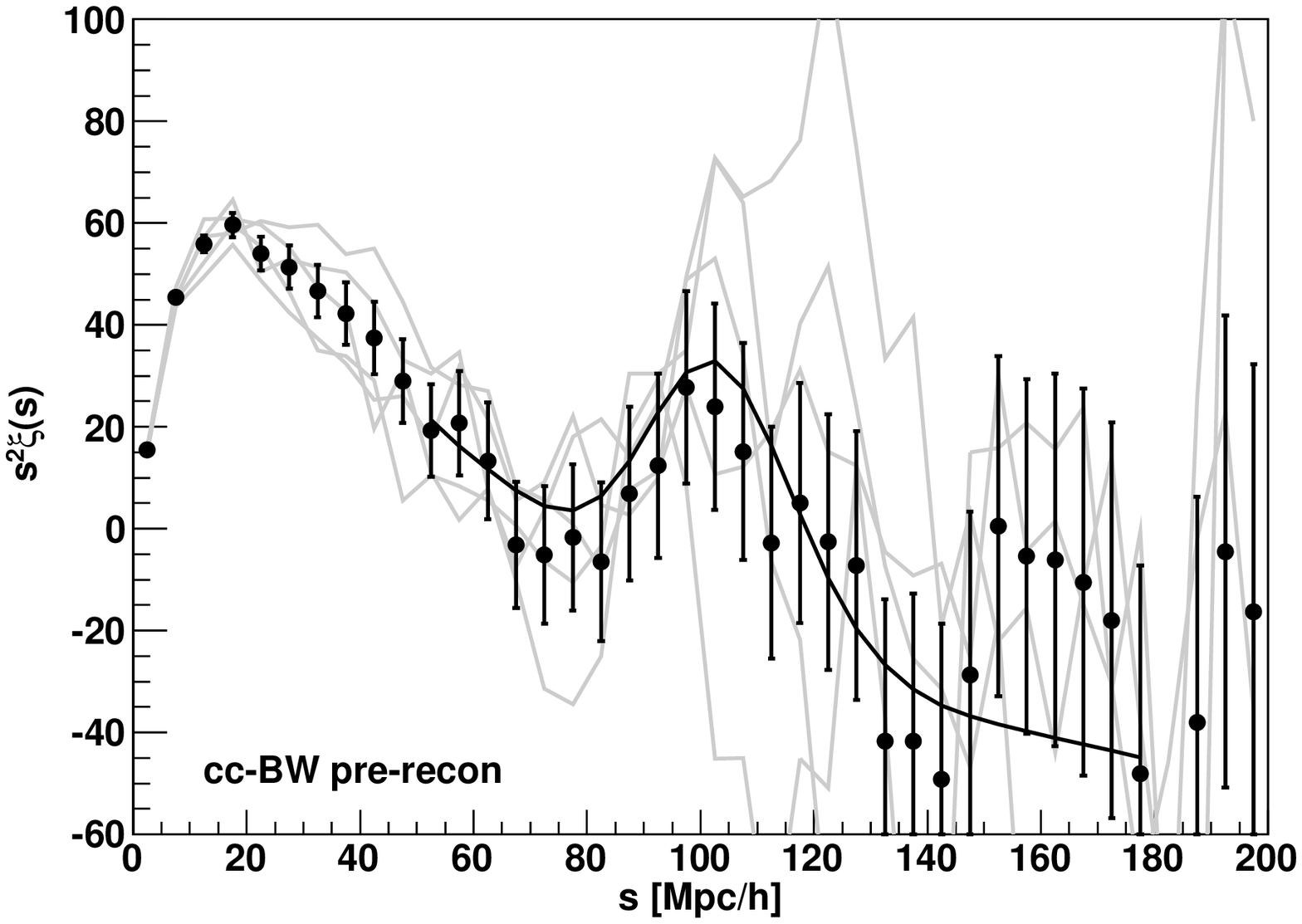,width=7cm}
\epsfig{file=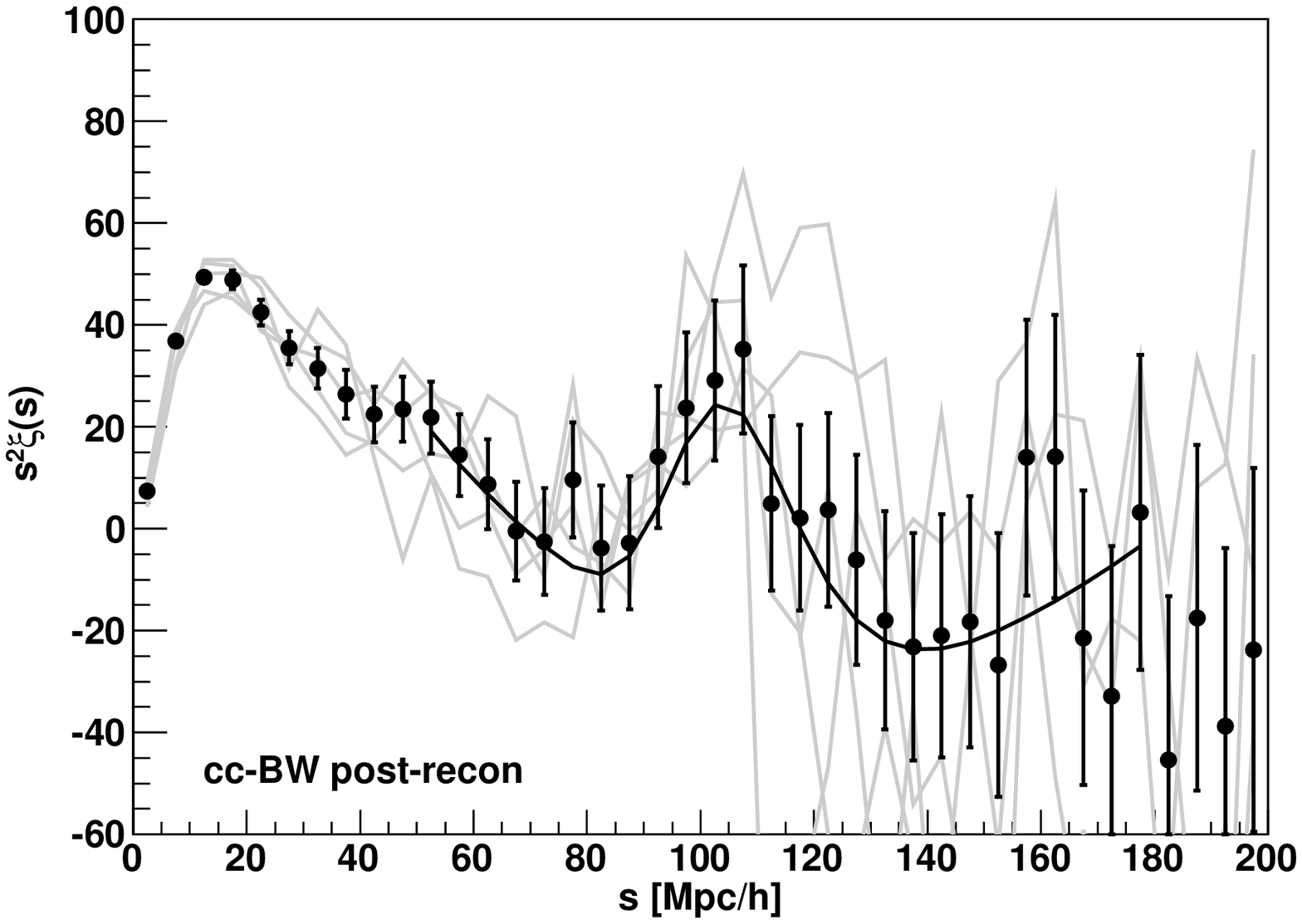,width=7cm}\\
\epsfig{file=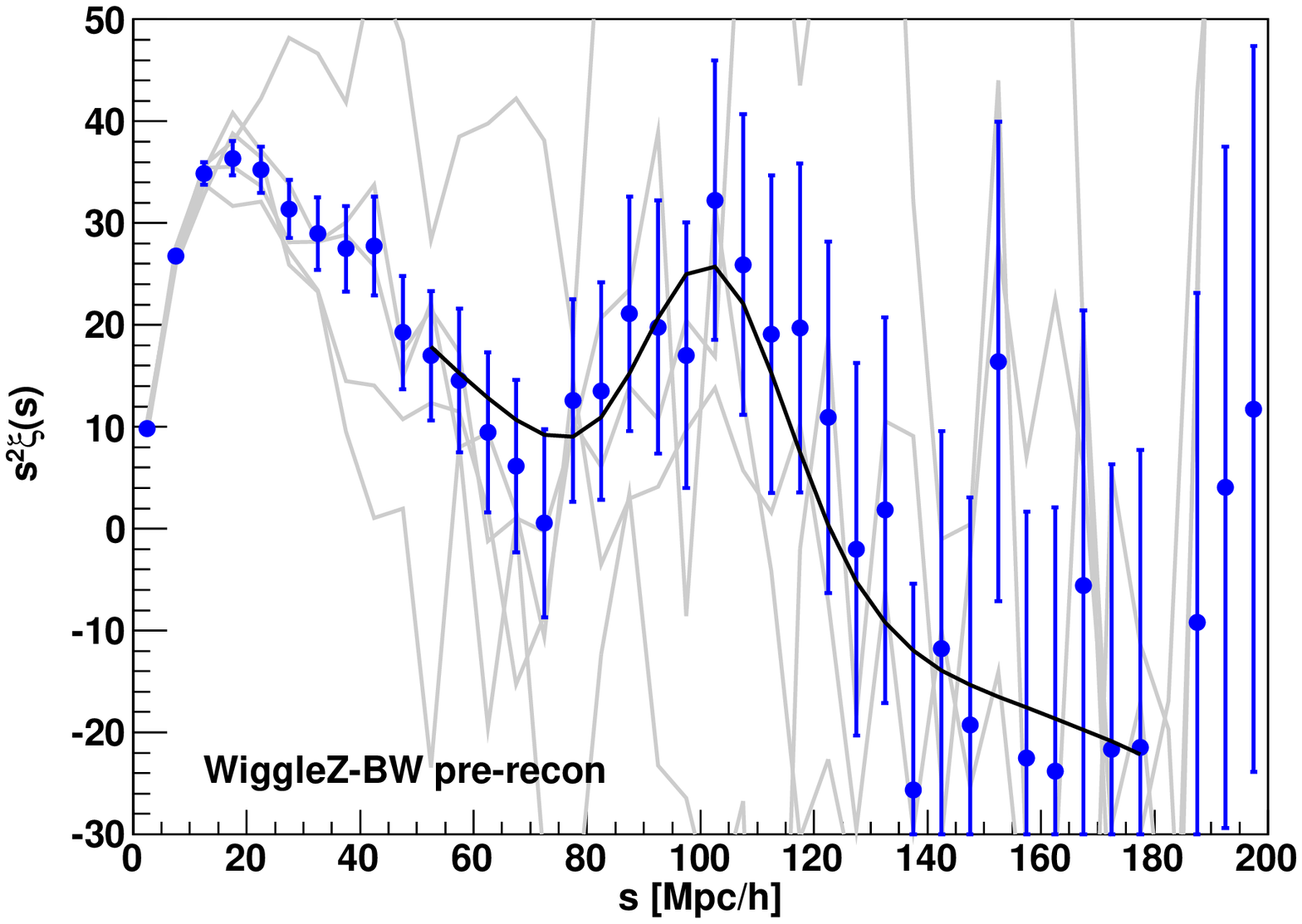,width=7cm}
\epsfig{file=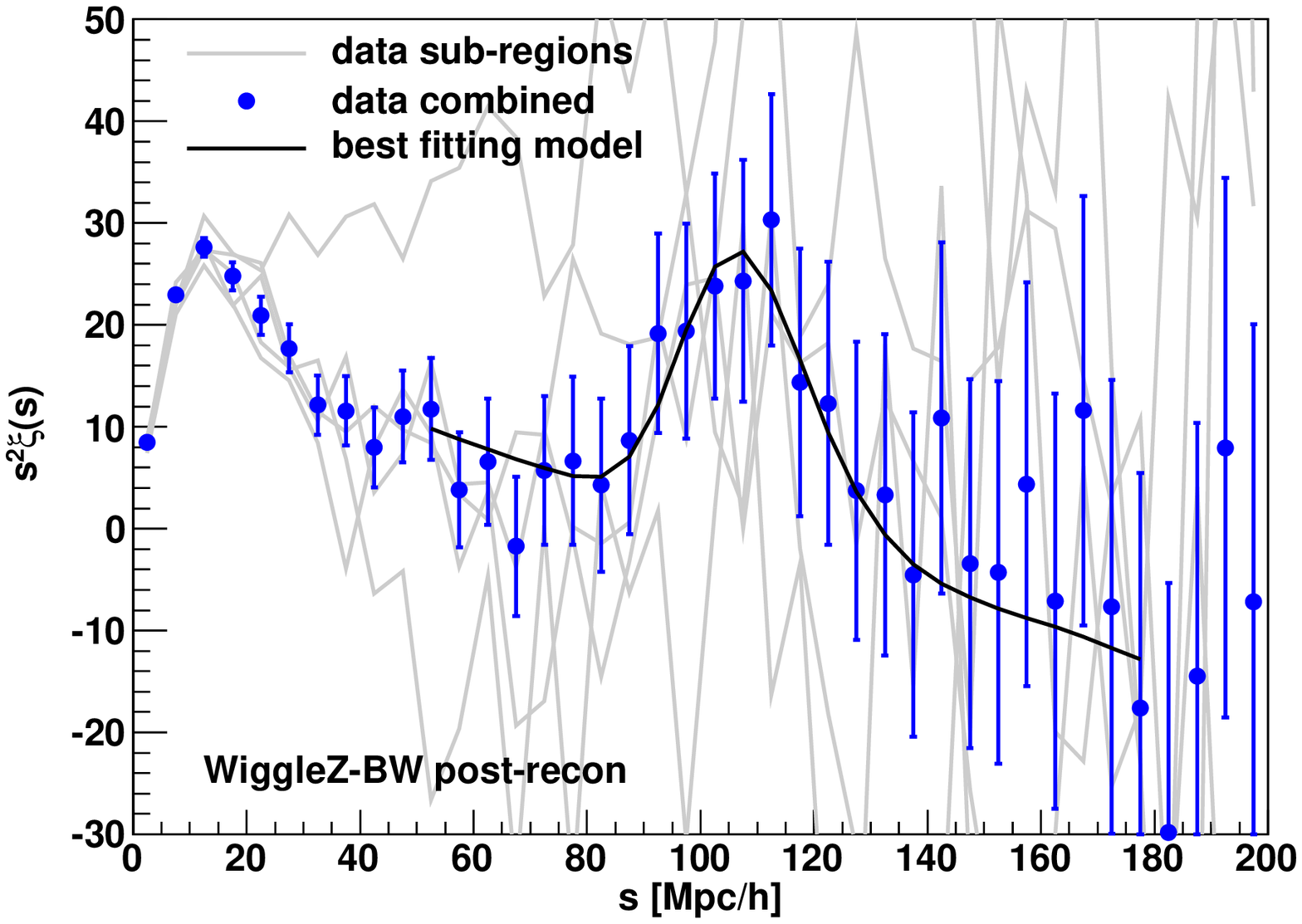,width=7cm}
\caption{The correlation functions of CMASS-BW (top), WiggleZ-BW (bottom)  and the cross-correlation (middle) in the overlap regions between CMASS and WiggleZ. The grey lines show the correlation functions for the five individual sub-regions (see Figure~\ref{fig:sky}), while the colored data points show the combined correlation functions calculated from Eq.~\ref{eq:combine}. The error bars are the diagonal of the combined covariance matrices (see Figure~\ref{fig:covs}). Note, that the scatter in the grey lines does not represent the error in the data points, since each grey line corresponds to a different volume and is weighted accordingly. The black lines show the best fit to the individual correlation functions corresponding to the upper part of Table~\ref{tab:results}.}
\label{fig:cf_cmp1}
\end{center}
\end{figure*}

The cross-correlation function between two tracers A and B can be calculated as
\begin{equation}
\begin{split}
\xi_{\rm AB}(s) &= 1 + \frac{D^{\rm A}D^{\rm B}(s)}{R^{\rm A}R^{\rm B}(s)} \left(\frac{n_r^{\rm A}n_r^{\rm B}}{n_d^{\rm A}n_d^{\rm B}} \right) - \frac{D^{\rm A}R^{\rm B}(s)}{R^{\rm A}R^{\rm B}(s)} \left(\frac{n_r^{\rm A}}{n_d^{\rm A}}\right)\\
&- \frac{R^{\rm A}D^{\rm B}(s)}{R^{\rm A}R^{\rm B}(s)} \left(\frac{n_r^{\rm B}}{n_d^{\rm B}} \right).
\end{split}
\label{eq:LS2}
\end{equation}
Figure~\ref{fig:cf_cmp1} shows the correlation functions for CMASS-BW (top), WiggleZ-BW (bottom) and the cross-correlation (cc-BW, middle). The three figures on the left present the results before density field reconstruction, while the figures on the right display the result after reconstruction (see Section~\ref{sec:recon} for a discussion of our reconstruction technique). The five correlation functions for the individual regions of CMASS-BW and WiggleZ-BW are indicated as grey lines. Using the covariance matrix (see next section) we can combine the correlation functions of the five sub-regions (colored data points). The auto-correlation functions of both surveys, as well as the cross-correlation function before and after reconstruction, show a clear BAO signal at around $100\,$Mpc$/h$. 

\begin{figure}
\begin{center}
\epsfig{file=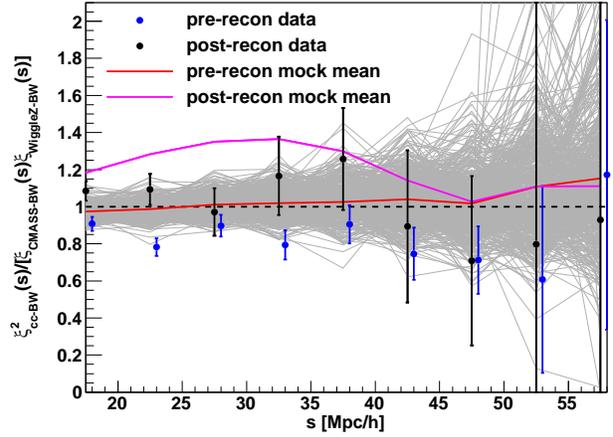,width=9cm}
\caption{The measured correlation coefficient before (blue) and after (black) density field reconstruction. The dashed line shows the expectation of linear theory. The blue data points are shifted by $0.5\,$Mpc$/h$ to the right for clarity. The solid lines indicate the mean correlation coefficient of the mock realizations before (red) and after (magenta) density field reconstruction. The error on the data points is derived from the variations in the $480$ mock catalogues (grey lines). When fitting the correlation coefficient in Section~\ref{sec:relvel} we only use the data before reconstruction, since we do not have a model for the correlation function post reconstruction.}
\label{fig:ratio2}
\end{center}
\end{figure}

We also measured the cross-correlation coefficient defined as 
\begin{equation}
r^2(s) = \frac{\xi^2_{\rm cc-BW}(s)}{\xi_{\rm CMASS-BW}(s)\xi_{\rm WiggleZ-BW}(s)}
\end{equation}
and presented in Figure~\ref{fig:ratio2}. In linear theory we expect this quantity to be 
\begin{equation}
r_{\rm theory}^2(s) = \frac{\left[1 + \frac{1}{3}(\beta_{\rm A} + \beta_{\rm B}) + \frac{1}{5}(\beta_{\rm A}\beta_{\rm B})\right]^2}{(1 + \frac{2}{3}\beta_{\rm A} + \frac{1}{5}\beta^2_{\rm A})(1 + \frac{2}{3}\beta_{\rm B} + \frac{1}{5}\beta^2_{\rm B})}.
\end{equation} 
Assuming $b_{\rm A} = 2$,  $b_{\rm B} = 1$ and $f = b\beta= 0.76$ results in $r^2_{\rm theory} = 0.997$. This expectation is included in Figure~\ref{fig:ratio2} (black dashed line). The mock realizations show a large correlation coefficient after density field reconstruction. We currently do not have a model for the correlation function shape post reconstruction~\citep{White:2015eaa} and therefore we only use the pre-reconstruction result in our fitting in Section~\ref{sec:relvel}.

Figure~\ref{fig:ratio2} also shows a small correlation coefficient before reconstruction (blue data points). We used Gaussian error distribution to obtain the uncertainties on the data points in Figure~\ref{fig:ratio2}, however, the errors on $r$ have a significant non-Gaussian component. In~\citet{Marin:2015} we performed fits to the auto- and cross-correlation functions having $r$ as a free parameter. We find that $r$ is consistent with $1$ for scales above $20\,$Mpc$/h$ (see Figure 5 in~\citealt{Marin:2015}). 

\section{Mock realisations}
\label{sec:mocks}

\begin{figure*}
\begin{center}
\epsfig{file=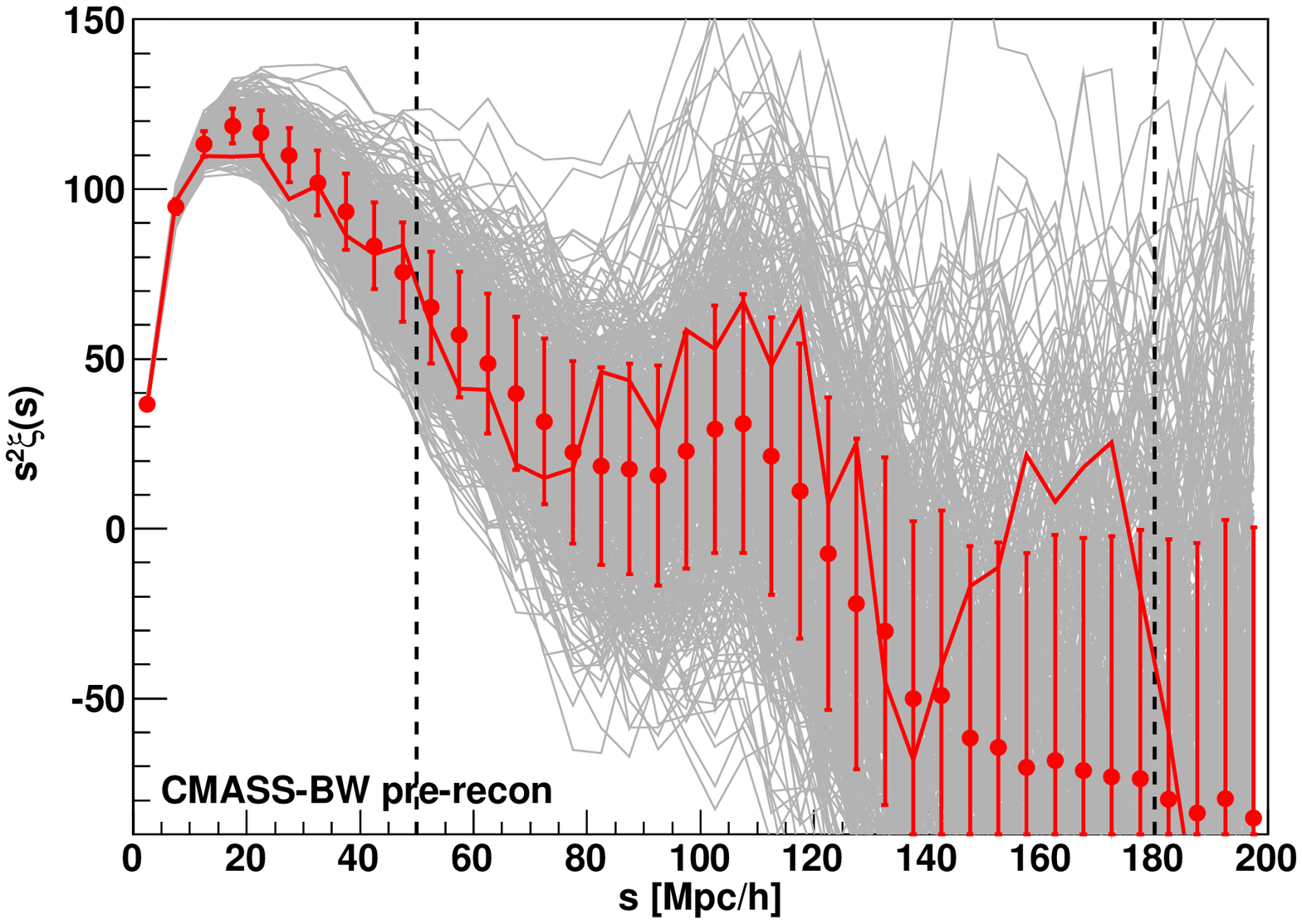,width=8cm}
\epsfig{file=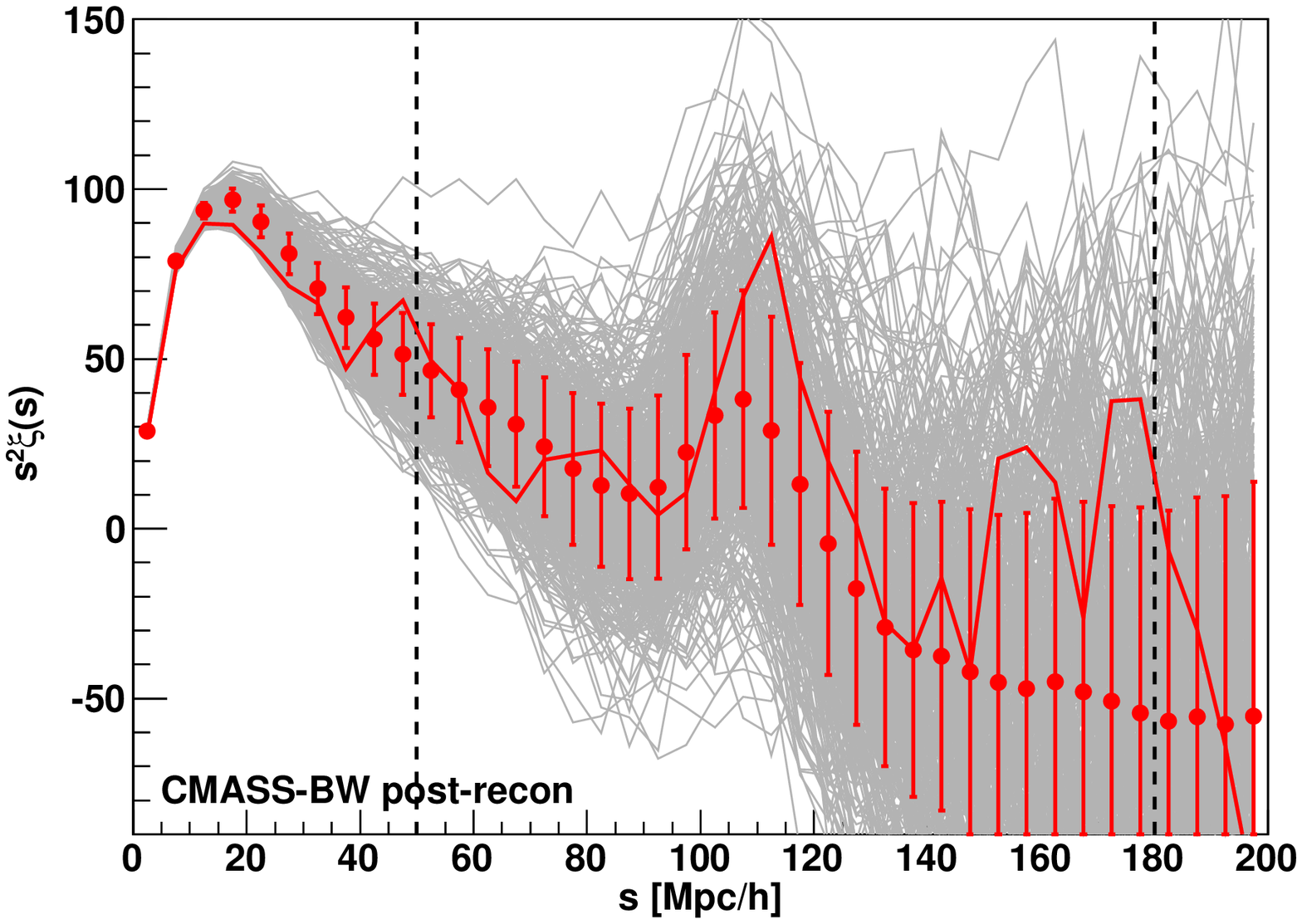,width=8cm}
\epsfig{file=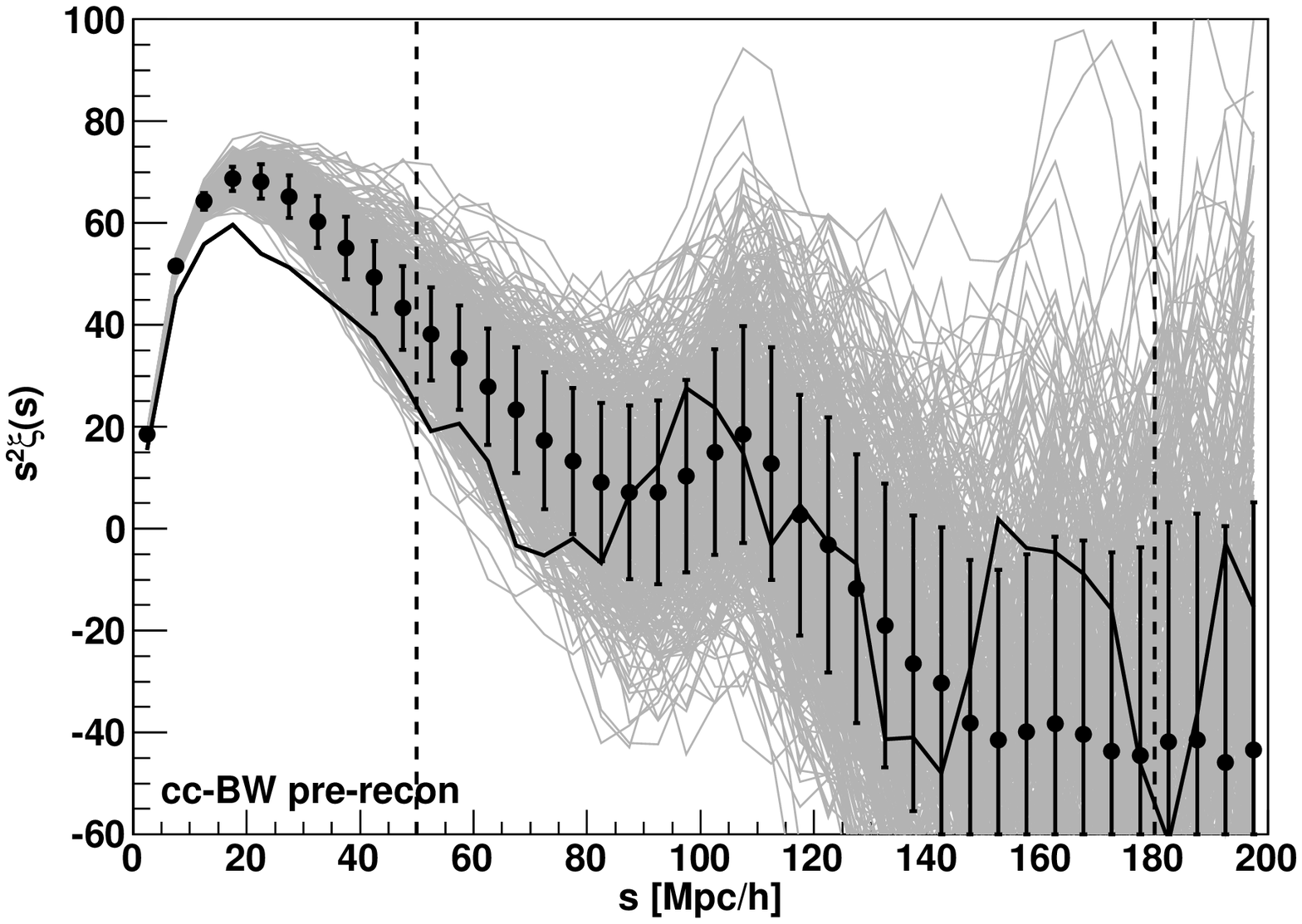,width=8cm}
\epsfig{file=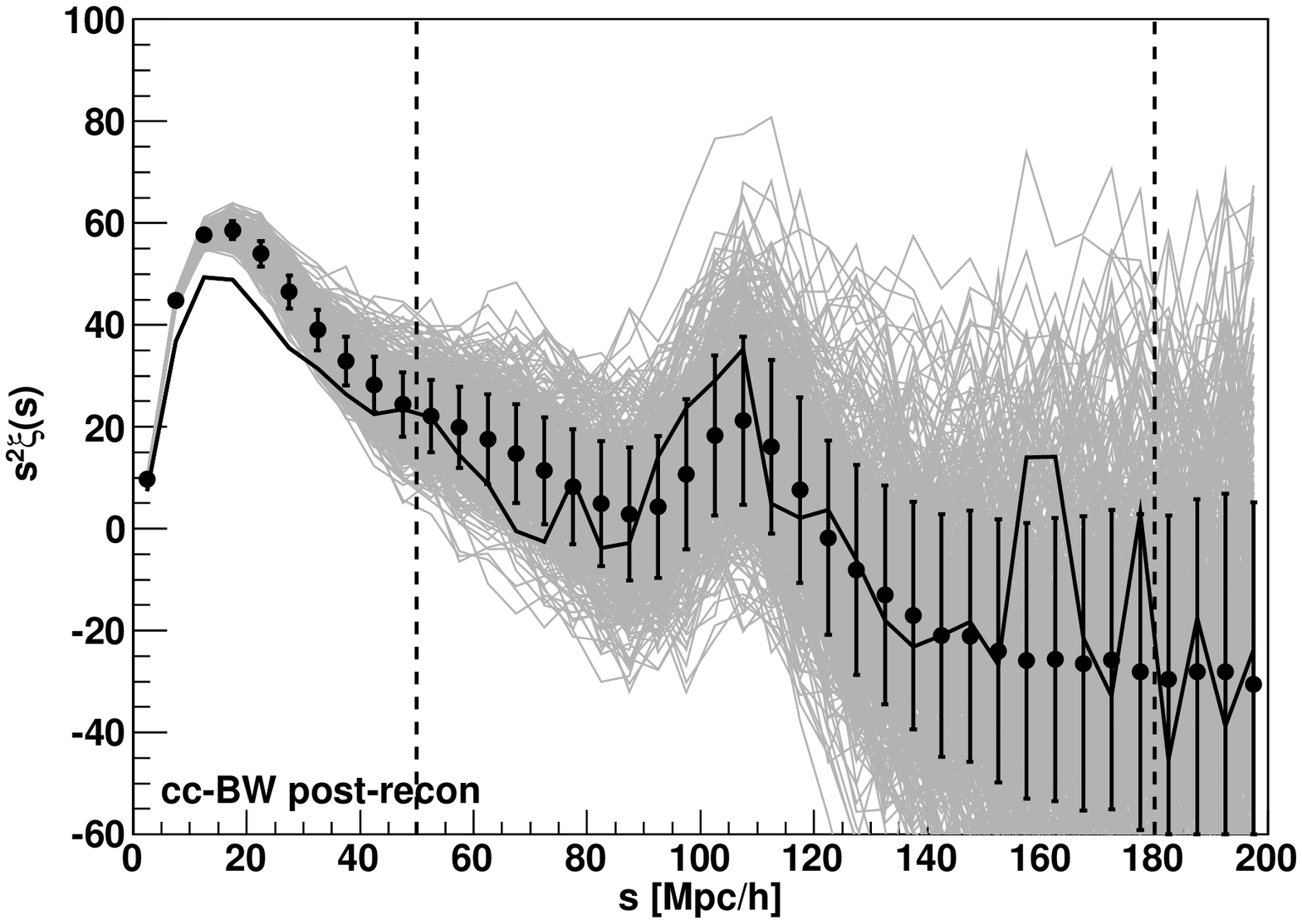,width=8cm}
\epsfig{file=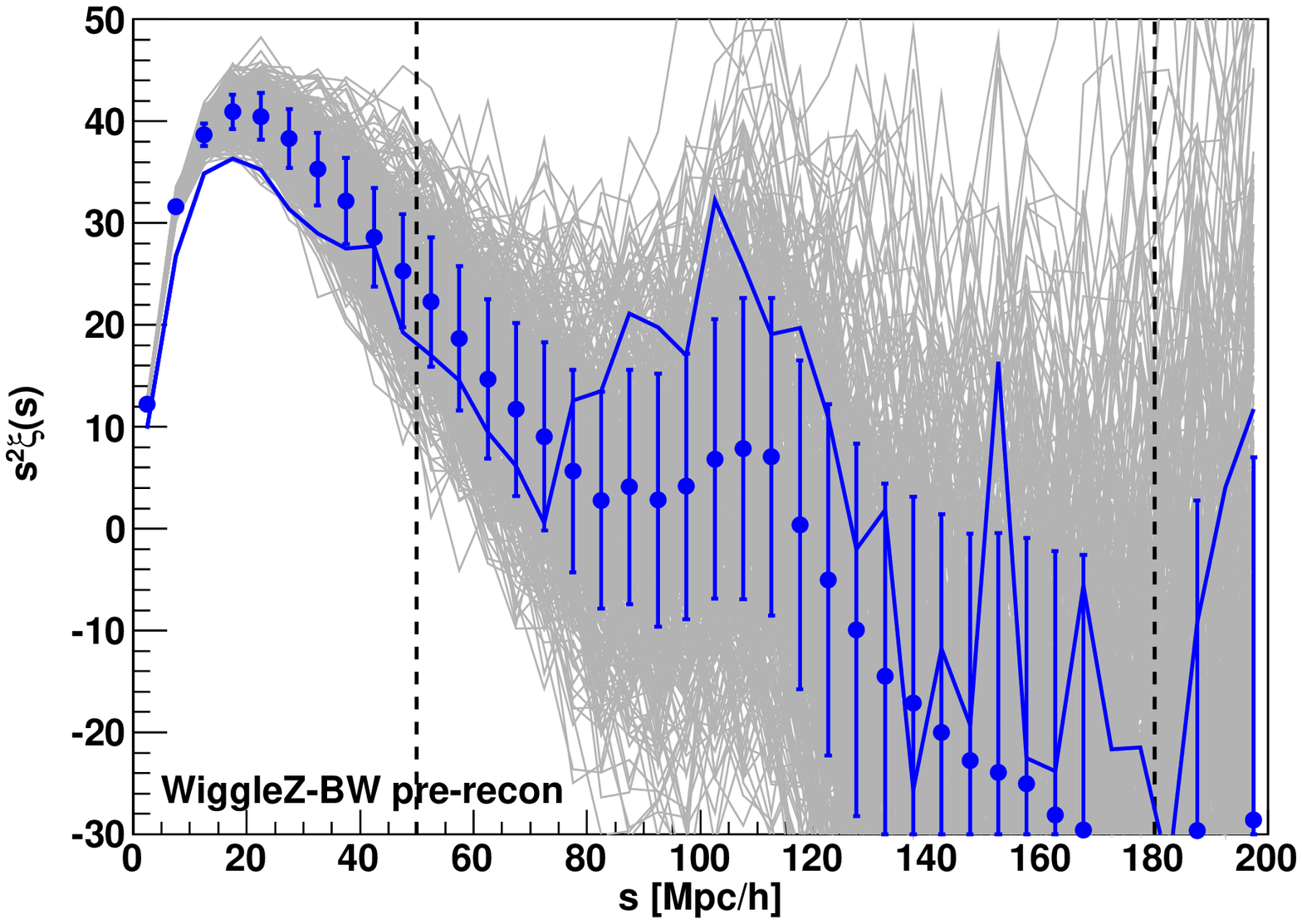,width=8cm}
\epsfig{file=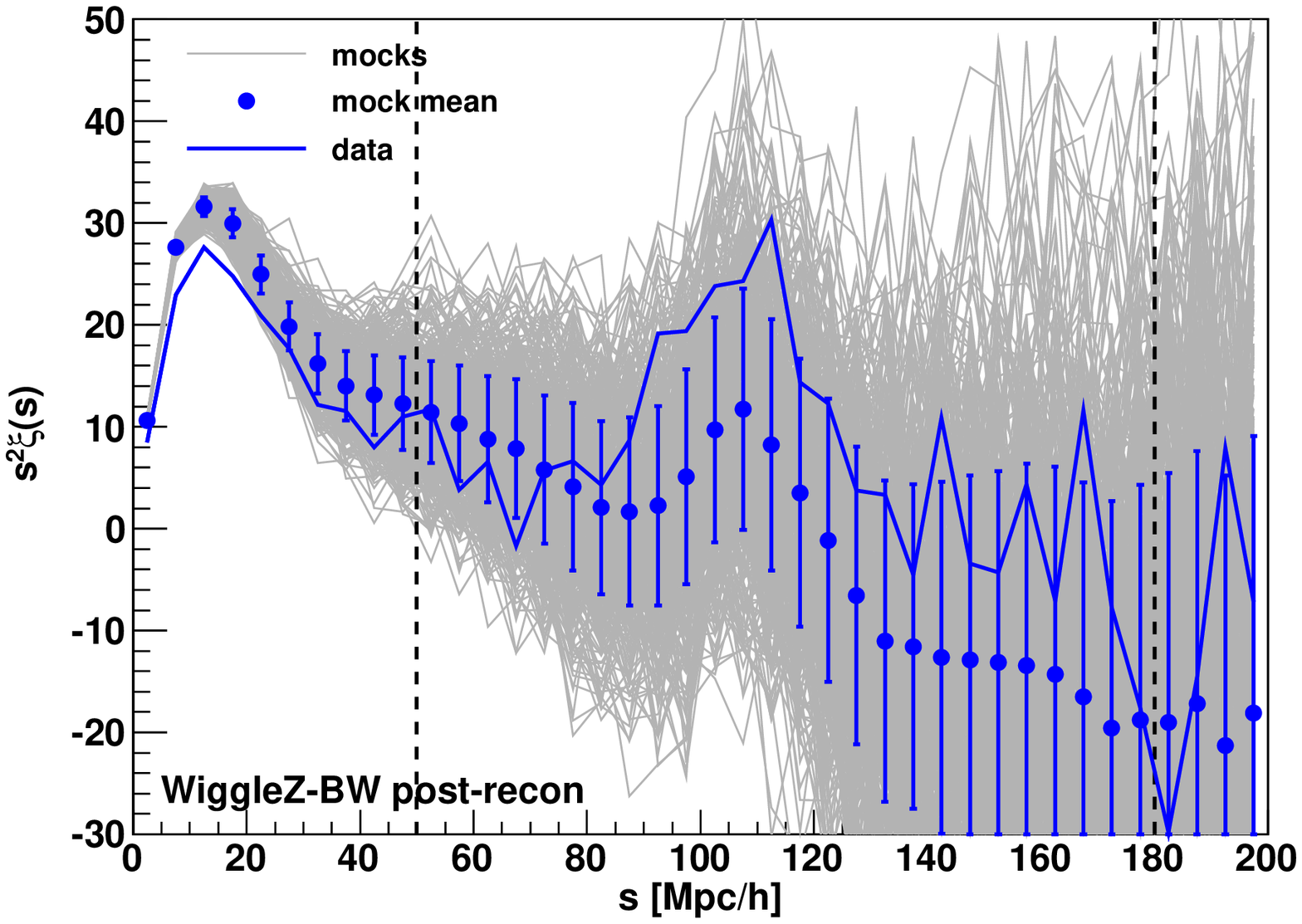,width=8cm}
\caption{Comparison between the mock realizations and the CMASS-BW (red, top) WiggleZ-BW (blue, bottom) and cross-correlations functions (black, middle). The grey lines show the $480$ mock realizations with the mean given by the colored data points and the error representing the variance. The colored lines indicate the measurement in the data. The dashed black lines mark the fitting range which goes from $50\,$Mpc$/h$ to $180\,$Mpc$/h$.}
\label{fig:mocks}
\end{center}
\end{figure*}

We produced $480$ mock catalogues for each of the five overlap regions using the COLA technique~\citep{Tassev:2013pn}. These mock catalogues will be presented in a separate paper together with the details of the COLA implementation we employed~\citep{Koda2015}. These mock catalogues have also been used in~\citet{Kazin:2014qga} and~\citet{Marin:2015}. Each simulation uses $1296^3$ particles in a $[600\,$Mpc$/h]^3$ box resulting in a particle mass of $7.5\times 10^9M_{\odot}/h$, allowing us to resolve CMASS as well as WiggleZ size halos. The halos are identified using a friend-of-friend algorithm with a linking length of $0.2$ times the mean particle separation. We use Halo Occupation distribution models to populate these halos with galaxies so that the mock realizations match the measured projected correlation functions $w_p(r_p)$, where $r_p$ is the angular separation between a galaxy pair. The fiducial cosmology of these mock catalogues is flat $\Lambda$CDM with $\Omega_m = 0.273$, $\Omega_b = 0.0456$, $H_0 = 70.5\,$Mpc/km/s, $\sigma_8 = 0.812$ and $n_s = 0.96$. The comparison of the correlation functions measured in the mock catalogues and the data correlation functions are shown in Figure~\ref{fig:mocks}. The mocks match the WiggleZ and CMASS clustering on large scales while they slightly overestimate the clustering amplitude of the cross-correlation function. The discrepancies are less significant after reconstruction (three panels on the right in Figure~\ref{fig:mocks}).

\subsection{Covariance matrix}
\label{sec:cov}

\begin{figure*}
\begin{center}
\epsfig{file=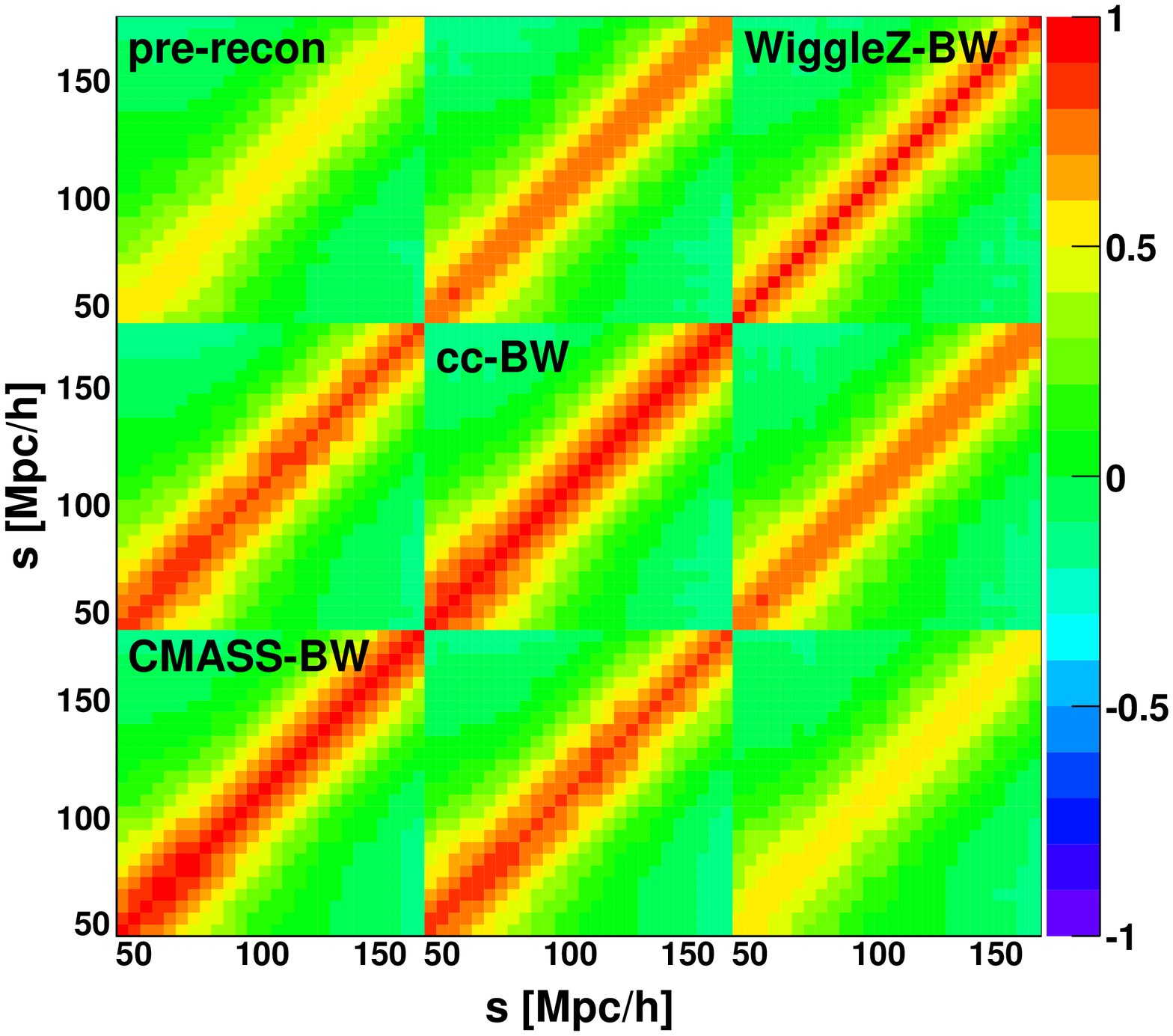,width=8cm}
\epsfig{file=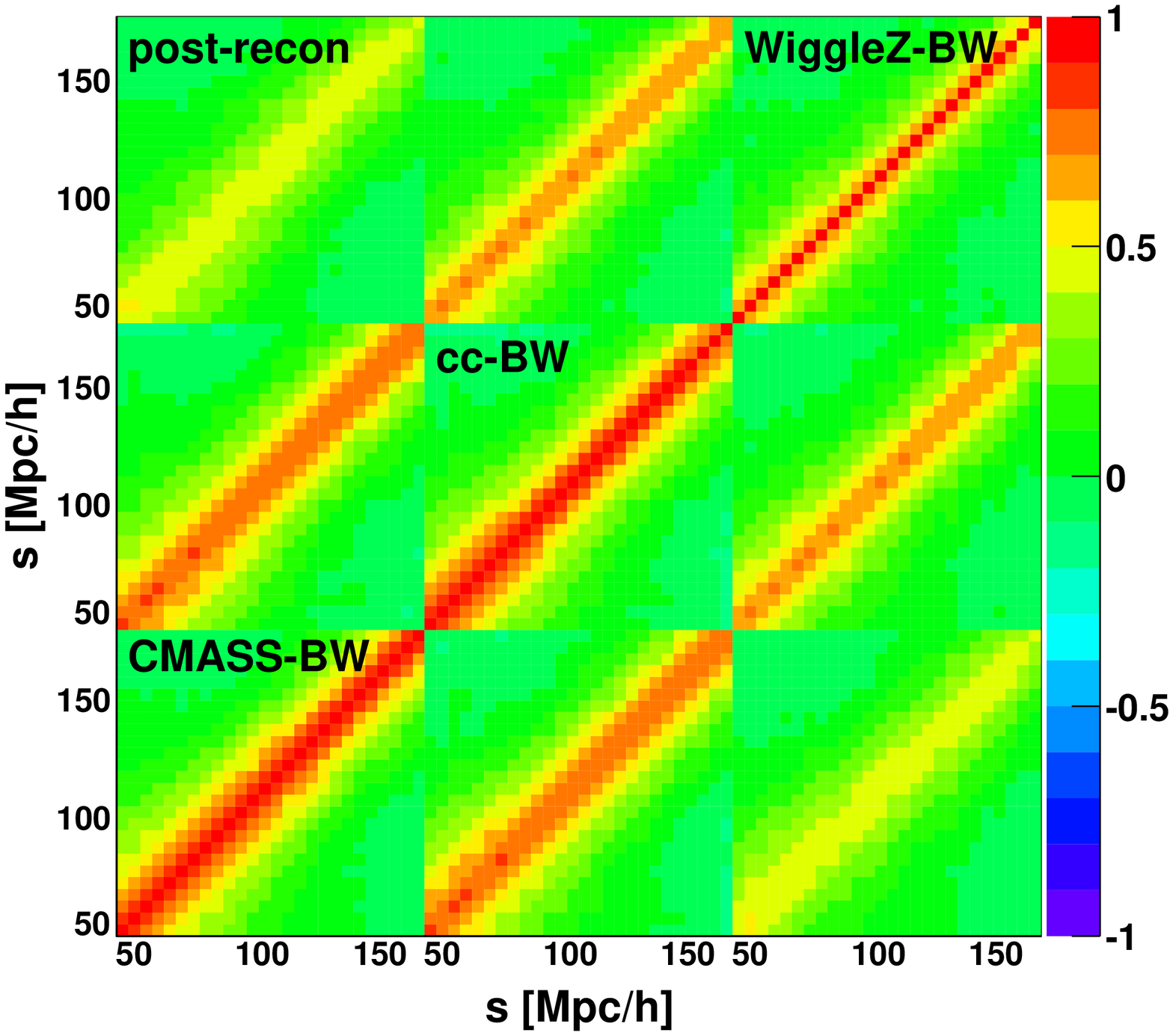,width=8cm}
\caption{The correlation matrix for the combined CMASS-BW, cross-correlation (cc-BW) and WiggleZ-BW correlation functions before (left) and after (right) density field reconstruction. These matrices are combinations of the individual matrices for the five separate regions using Eq.~\ref{eq:combine2}. For each region we use $480$ mock realizations for CMASS-BW and WiggleZ-BW. The color indicates the level of correlation, where red is high correlation, green is no correlation and blue is high anti-correlation. Since each set of CMASS-BW and WiggleZ-BW mock catalogues has been produced from the same simulation (see Section~\ref{sec:mocks}), there is a considerable amount of correlation between the three correlation functions, mimicking the situation of the real data. Given that we use a fitting range of $r = 50$ - $180\,$Mpc$/h$ with $5\,$Mpc$/h$ bins, this matrix has $26\times26$ bins for each correlation function and $78\times78$ bins in total.}
\label{fig:covs}
\end{center}
\end{figure*}

\begin{figure*}
\begin{center}
\epsfig{file=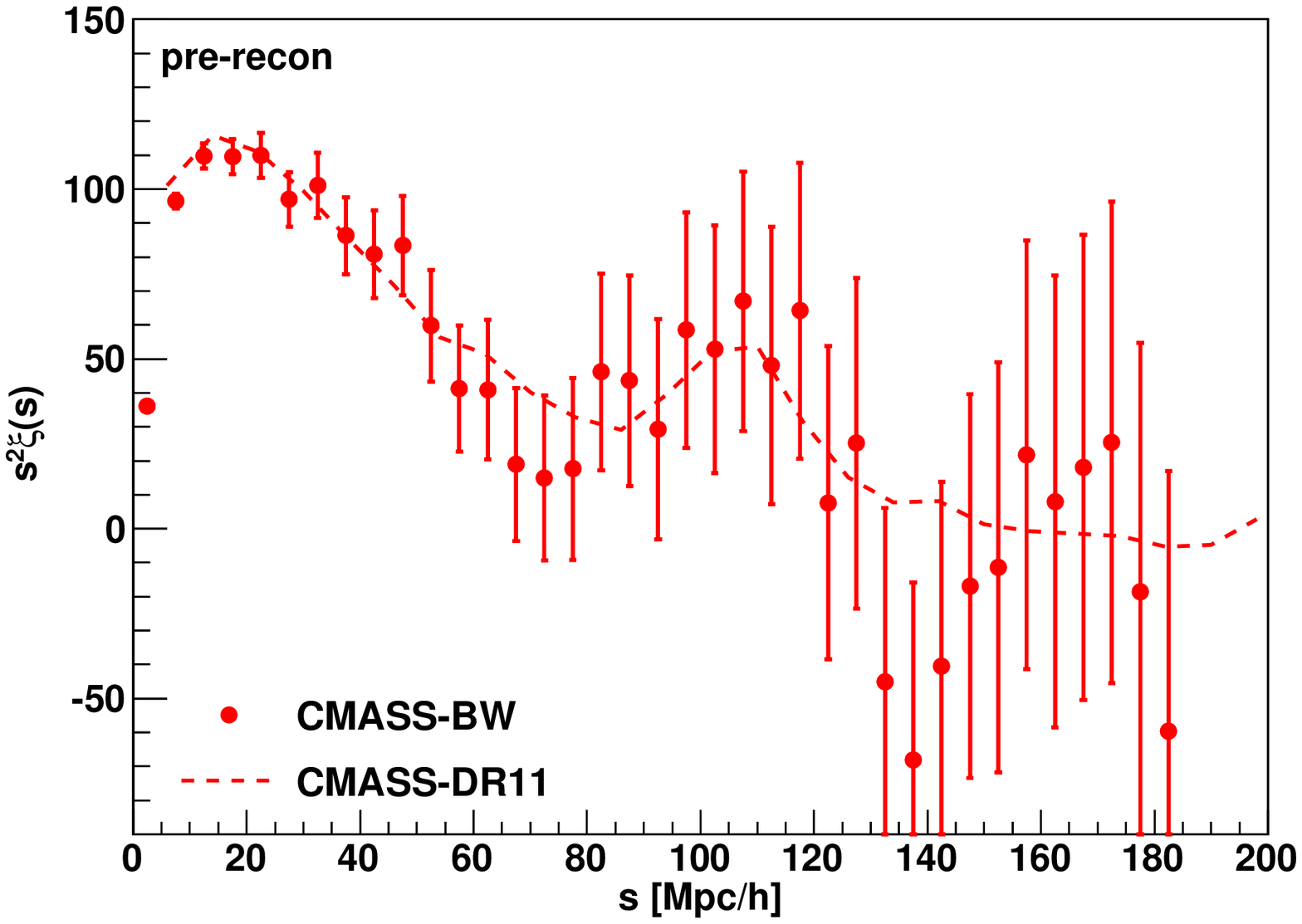,width=8cm}
\epsfig{file=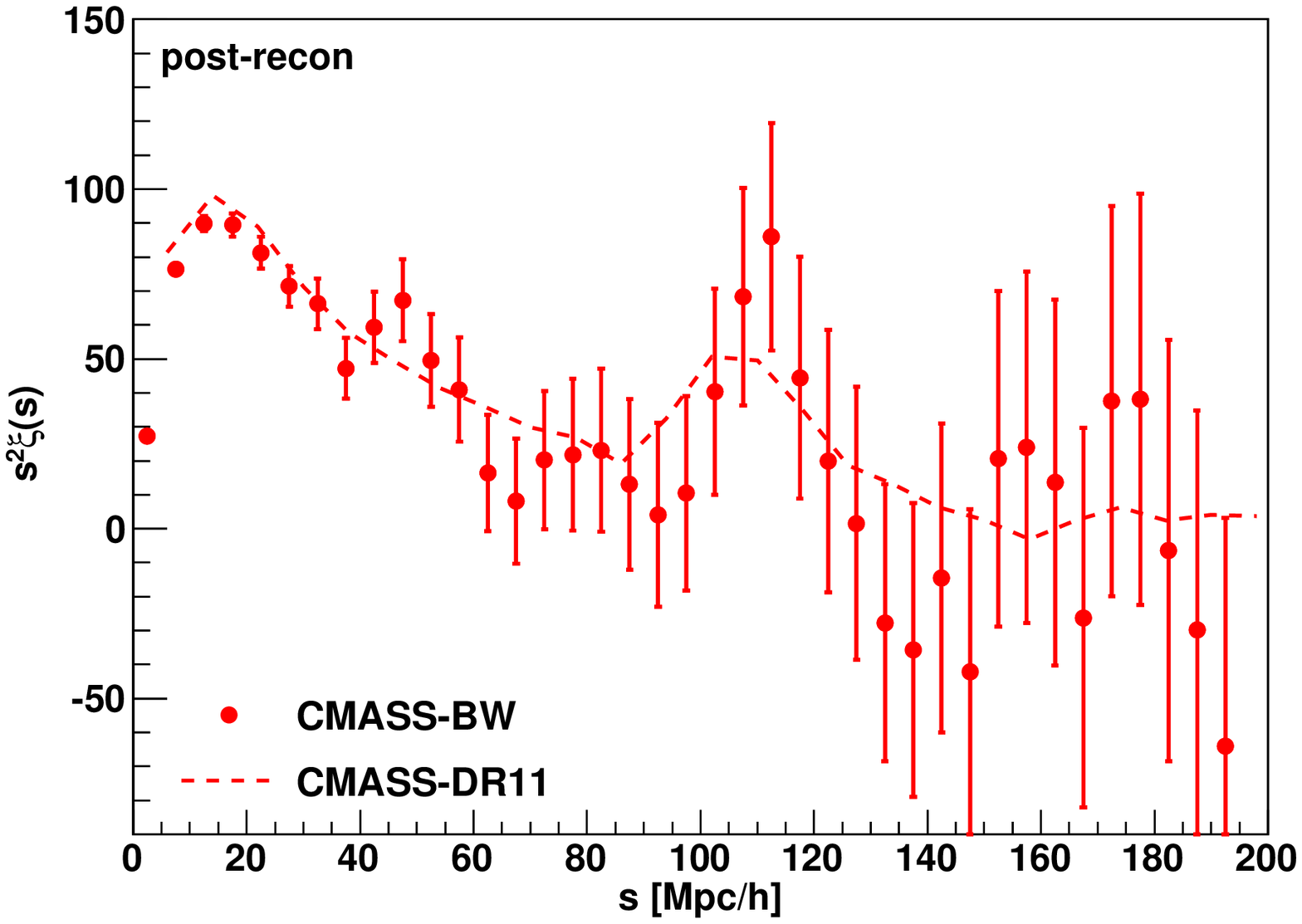,width=8cm}
\caption{Comparison of the CMASS-BW (red data points) and CMASS-DR11 (red dashed lines) correlation functions before (left) and after (right) density field reconstruction. Post reconstruction one can see that the BAO peak in CMASS-BW is at larger scales compared to CMASS-DR11, which leads to a smaller value of $D_V$ as is also visible in the resulting likelihood distribution (see Figure~\ref{fig:like}).}
\label{fig:cmp_DR11}
\end{center}
\end{figure*}

Using the mock realizations of the individual sub-regions we can produce covariance matrices for each of the auto- and cross-correlation functions. We calculate the covariance matrix using
\begin{equation}
C_{ij} = \frac{1}{479}\sum^{480}_{n=1}\left[\xi_n(s_i) - \overline{\xi}(s_i)\right]\left[\xi_n(s_j) - \overline{\xi}(s_j)\right],
\label{eq:cov}
\end{equation}
with $\xi_n(s_i)$ being the $n$-th correlation function estimate at separation $s_i$ and the sum is over all $480$ mock realizations. The mean value is defined as
\begin{equation}
\overline{\xi}(s_i) = \frac{1}{480}\sum^{480}_{n=1}\xi_n(s_i).
\end{equation}
Instead of analyzing the $10$ auto-correlation functions and five cross-correlation functions individually, we chose to combine the correlation functions to obtain two auto-correlation functions for CMASS-BW and WiggleZ-BW as well as one cross-correlation function. We combined the correlation functions of the five individual sub-regions using the covariance matrices calculated above and following the procedure outlined in~\citet{White:2010ed} and \citet{Blake:2011en}. Each sub-region is weighted by its corresponding uncertainty
\begin{equation}
C^{-1}\xi^{\rm tot}(s) = \sum^5_{\rm regions\;i}[C^i]^{-1}\xi^i(s),
\label{eq:combine}
\end{equation}
with $C^i$ being the covariance matrices of the individual sub-regions. The inverse covariance matrix for the combined correlation functions is given by
\begin{equation}
C^{-1} = \sum^5_{\rm regions\;i}[C^i]^{-1},
\label{eq:combine2}
\end{equation}
which follows from Eq.~\ref{eq:combine}. The combined covariance matrices before and after density field reconstruction are presented in Figure~\ref{fig:covs}. The combined correlation functions for CMASS-BW, WiggleZ-BW and the cross-correlation function are shown in Figure~\ref{fig:cf_cmp1} as colored data points. We also compare the CMASS-BW correlation function with the CMASS-DR11 correlation function in Figure~\ref{fig:cmp_DR11}. While the CMASS-BW correlation function before reconstruction is in excellent agreement with CMASS-DR11, we find the prominent BAO peak at slightly larger scales compared to CMASS-DR11. We will discuss this aspect further when fitting these correlation functions in Section~\ref{sec:fit}. 

It has been shown that the inverse covariance $C^{-1}$ derived from a finite number of realizations underestimates the uncertainties~\citep{Anderson:2003,Hartlap:2006kj,Percival:2013sga}. In the case of Gaussian errors and statistically independent bins, this effect can be accounted for by multiplying the variance estimated from the likelihood distribution with
\begin{equation}
m_{\sigma} = \frac{1 + B(N_{\rm bins} - N_{\rm p})}{1 + 2A + B(N_{\rm p} + 1)},
\end{equation}
where $N_{\rm mocks}$ is the number of mock realizations, $N_{\rm bins}$ is the number of bins, $N_{\rm p}$ is the number of free parameters and 
\begin{align}
A &= \frac{1}{(N_{\rm mocks} - N_{\rm bins} - 1)(N_{\rm mocks} - N_{\rm bins} - 4)},\\
B &= A(N_{\rm mocks}  - N_{\rm bins} - 2).
\end{align}
Furthermore, the sample variance needs to be multiplied by 
\begin{equation}
m_{v} = m_{\sigma}\frac{N_{\rm mocks} - 1}{N_{\rm mocks} - N_{\rm bins} - 2}.
\label{eq:scaling}
\end{equation}
Since the bins in a correlation function are not statistically independent, these correction factors are only an approximation. Given the $480$ mock realizations in our analysis, $26$ bins and $5$ free parameters (see Section~\ref{sec:fit}), we have $m_{\sigma} = 1.033$ and $m_v = 1.095$. However, when fitting all three correlation functions simultaneously ($78$ bins), this factor can rise to $m_v = 1.4$, significantly contributing to our error budget. 

\section{Density field reconstruction}
\label{sec:recon}

In linear theory, the co-moving position of the BAO peak is set after the epoch of decoupling, providing the foundation of its use as a standard ruler. There are, however, non-linear effects, which can change the BAO peak position, as well as its shape. The most significant effect is non-linear damping of the BAO peak~\citep{Eisenstein:2006nj,Seo:2008yx}. This effect is often modeled with a Gaussian damping term. Although we are interested in the BAO peak position and not in its amplitude, damping can shift the peak position, because of the non-symmetric shape of the correlation function around the BAO peak~\citep{Eisenstein:2006nj,Guzik:2006bu,Smith:2006ne,Smith:2007gi,Angulo:2007fw,Mehta:2011xf}. Additionally, \citet{Crocce:2007dt} found that mode coupling can lead to shifts in the BAO peak position. Interestingly, mode coupling as well as non-linear damping can be removed by a technique called density field reconstruction, meaning that the measured distribution of galaxies itself can be used to reduce the impact of these non-linear effects by estimating the displacements of galaxies from their initial position in the density field~\citep{Eisenstein:2006nk,Padmanabhan:2009yr,Mehta:2011xf}.
Density field reconstruction enhances the signal-to-noise ratio of the BAO signature using extra information contained in the higher order correlations of the galaxy distribution~\citep{Eisenstein:2006nk}.
We apply density field reconstruction to the observed density field following the formalism of~\citet{Padmanabhan:2012hf}. First we  smooth the observed and random fields with a Gaussian filter of the form
\begin{equation}
G(k) = \exp\left[\frac{-(k\Sigma_{\rm smooth})^2}{2}\right],
\end{equation}
where we choose $\Sigma_{\rm smooth} = 15\,$Mpc$/h$~\citep{Xu:2012hg}. The over-density field is then calculated in real-space as 
\begin{equation}
\delta(\vec{x}) = \frac{\rho_{\rm g}(\vec{x})}{\rho_{\rm r}(\vec{x})}\frac{n_{\rm r}}{n_{\rm g}} - 1,
\end{equation}
with $\rho_{g}(x)$ and $\rho_{r}(x)$ being the density of the smoothed galaxy and random distribution, respectively. The normalization is defined as
\begin{equation}
\frac{n_{\rm r}}{n_{\rm g}} = \frac{\sum^{N_{\rm r}}_i w_{i}(\vec{x})}{\sum^{N_{\rm g}}_i w_{i}(\vec{x})}.
\label{eq:norm}
\end{equation}
In linear perturbation theory, the displacement field $\vec{\Psi}(x)$ is related to the redshift-space density field by~\citep{Nusser:1993sx}
\begin{equation}
\nabla \cdot \vec{\Psi}(\vec{x}) + \beta\nabla\cdot\left[\vec{\Psi}_{\rm los}(\vec{x})\right] = -\frac{\delta(\vec{x})}{b},
\label{eq:bias}
\end{equation} 
where $\vec{\Psi}_{\rm los}$ is the line-of-sight component of the displacement field. Transforming this equation into Fourier space and using the approach $\phi(x) = \sum_k\phi(k)\exp(ikx)$ and $\delta(x)=\sum_k\delta(k)\exp(ikx)$ we get
\begin{equation}
-\phi(k)\left[k_x^2 + k_y^2 + k_z^2(1 + \beta)\right] = \frac{\delta(k)}{b},
\end{equation}
which we solve for $\phi(k)$ for every wavenumber $k$. The displacement field is than given by $\Psi(k) = -i\frac{k}{|k|}\phi(k)$, which we Fourier transfer back into configuration space. Our approach uses the plane-parallel approximation, which is valid for the small angular coverage of the five individual fields studied in this analysis~\citep{Blake:2011rj}.\\

We then apply the displacement to our galaxies by shifting their $x$, $y$ and line-of sight positions following 
\begin{align}
s_{\rm los} &= s_{\rm old} - (1+f)\vec{\Psi}_{\rm los}(\vec{x}),
\label{eq:los}\\
s_{x, y} &= s_{\rm old} - \vec{\Psi}_{x,y}(\vec{x}).
\end{align}
We do not apply the factor of $(1+f)$ in the case of the random galaxies, since the random distribution does not contain redshift space distortions. During reconstruction we use the growth rate $f = 0.7$ as well as the linear bias $b=1.9$ for CMASS-BW~\citep{Beutler:2013yhm}  and $b=1.0$ for WiggleZ-BW~\citep{Blake:2011rj}. The three plots on the right of Figure~\ref{fig:cf_cmp1} show the correlation functions for CMASS-BW (top), WiggleZ-BW (bottom) and cross-correlation (cc-BW, middle) after applying density field reconstruction.

\section{Modeling the large scale correlation function}
\label{sec:model}

Our model for the galaxy correlation function follows the procedure of~\citet{Anderson:2013zyy}. The galaxy correlation function is given by 
\begin{equation}
\xi(s) = B^2\xi_m(\alpha s) + A(s)
\label{eq:noshape}
\end{equation}
where
\begin{equation}
A(s) = \frac{a_{1}}{s^2} + \frac{a_{2}}{s} + a_{3}.
\end{equation}
The matter correlation function is obtained through~\citep{Eisenstein:2006nj}
\begin{equation}
\xi_m(s) = \int \frac{k^2dk}{2\pi^2}P(k)j_0(ks)e^{-k^2\sigma_s^2}
\label{eq:noshape2}
\end{equation}
with $\sigma_s = 2\,$Mpc$/h$ and the monopole power spectrum is given by
\begin{equation}
P(k) = P_{\rm sm,lin}(k)\left[ 1 + \left(O_{\rm lin}(k) - 1\right)e^{-\left[k^2\Sigma_{\rm nl}^2\right]/2}\right].
\end{equation}
We fix $\Sigma_{\rm nl} = 8.8\,$Mpc$/h$ before reconstruction and $\Sigma_{\rm nl} = 4.4\,$Mpc$/h$ after reconstruction~\citep{Anderson:2013zyy,Magana:2013wpa}. The function $O_{\rm lin}(k)$ represents the oscillatory part of the fiducial linear power spectrum and $P_{\rm sm}(k)$ is the smooth power spectrum monopole. To obtain $P_{\rm sm, lin}(k)$ we fit the fiducial linear power spectrum, $P_{\rm lin}(k)$, with an~\citet{Eisenstein:1997ik} no-Wiggle power spectrum, $P_{\rm nw}(k)$, together with five polynomial terms:
\begin{equation}
P_{\rm sm,lin}(k) = B^2P_{\rm nw}(k) + \frac{c_{1}}{k^3} + \frac{c_{2}}{k^2} + \frac{c_{3}}{k} + c_{4} + c_{5}k.
\end{equation}
The oscillatory part of the power spectrum is given by 
\begin{equation}
O_{\rm lin}(k) = \frac{P_{\rm lin}(k)}{P_{\rm sm,lin}(k)}.
\end{equation}
Our model in Eq.~\ref{eq:noshape} has five free parameters ($B$, $\alpha$, $a_{1-3}$). 

To turn the constraint on $\alpha$ into a physical parameter we use 
\begin{equation}
\alpha = \frac{D_V(z)r_s^{\rm fid}}{D^{\rm fid}_V(z)r_s}
\end{equation}
with 
\begin{equation}
D_V(z) = \left[(1+z)^2D_A(z)\frac{cz}{H(z)}\right]^{1/3},
\end{equation}
where $D_A(z)$ is the angular diameter distance and $H(z)$ is the Hubble parameter.

\section{Testing for BAO systematics}
\label{sec:sys}

Although the linear bias model was always believed to be sufficient for scales as large as the BAO signal, some studies using halo catalogues from N-body simulations suggest that there are scale-dependent bias effects even on BAO scales~\citep{Noh:2009bb,Desjacques:2010gz,wang2013}. This means that the BAO signal can vary, depending on the tracer chosen to map the underlying density field. In the following sections of this paper, we will fit the large scale correlation function of CMASS and WiggleZ, and compare the displacement fields derived from the two surveys. Since we restrict our analysis to a common volume, we expect the results to be correlated. Since the two surveys trace the underlying density field differently, we can test for systematic effects in the BAO analysis.

\subsection{Comparing the CMASS and WiggleZ displacement fields}
\label{sec:cmp}

In Section~\ref{sec:recon} we derived two displacement fields using the CMASS-BW and WiggleZ-BW galaxies, respectively. Here we are interested to learn (1) whether one of the displacement fields leads to better BAO constraints and (2) whether there are any systematic shifts in the BAO position depending on which displacement field is used for the reconstruction.

\begin{figure}
\begin{center}
\epsfig{file=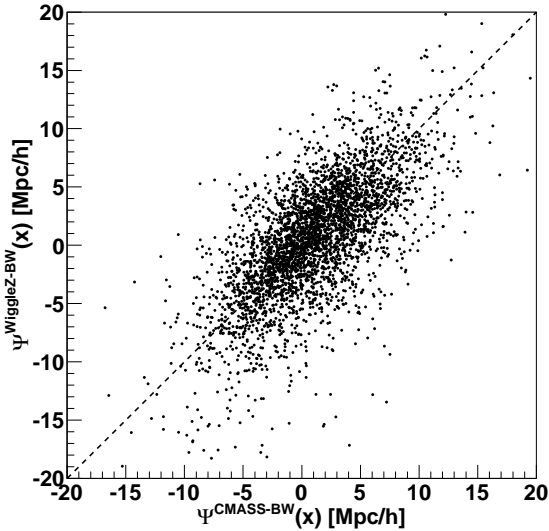,width=8cm}
\caption{Comparison of the displacements in the CMASS-BW catalogue using the CMASS-BW displacement field ($x$-axis) and the WiggleZ-BW displacement field ($y$-axis). The plot shows a random selection of $5\%$ of all galaxies in the five overlap regions.}
\label{fig:cmp_displ2}
\end{center}
\end{figure}

We apply the displacement field derived using the WiggleZ survey to the CMASS galaxies and the displacement field derived from the CMASS survey to the WiggleZ galaxies resulting in four datasets:
\begin{enumerate}[1]
\item CMASS-BW using the CMASS-BW displacement field, 
\item CMASS-BW using the WiggleZ-BW displacement field, 
\item WiggleZ-BW using the CMASS-BW displacement field, 
\item WiggleZ-BW using the WiggleZ-BW displacement field. 
\end{enumerate}
Figure~\ref{fig:cmp_displ2} compares the two displacements for each CMASS-BW galaxy. We quantify the correlation between the displacement fields using the correlation coefficient 
\begin{equation}
r({\rm A},{\rm B}) = \frac{\sum_i\left(\Psi^{\rm A}_i - \overline{\Psi}^{\rm A}\right)\left(\Psi^{\rm B}_i - \overline{\Psi}^{\rm B}\right)}{\sqrt{\sigma_{\rm A}\sigma_{\rm B}}} 
\end{equation}
where $\overline{\Psi}$ represents the mean of the displacement and $\sigma_x = \sum_i(x_i - \overline{x})^2$, summing over all galaxies $i$. The correlation coefficient for the regions (S1, S22, N9, N11, N15) = ($0.65$, $0.64$, $0.67$, $0.68$, $0.81$) for the CMASS-BW galaxies and ($0.49$, $0.74$, $0.76$, $0.73$, $0.75$) for the WiggleZ-BW galaxies. The smallest region (S1) shows the lowest correlation coefficient, indicating that volume effects do play a role in this case. 

\begin{figure}
\begin{center}
\epsfig{file=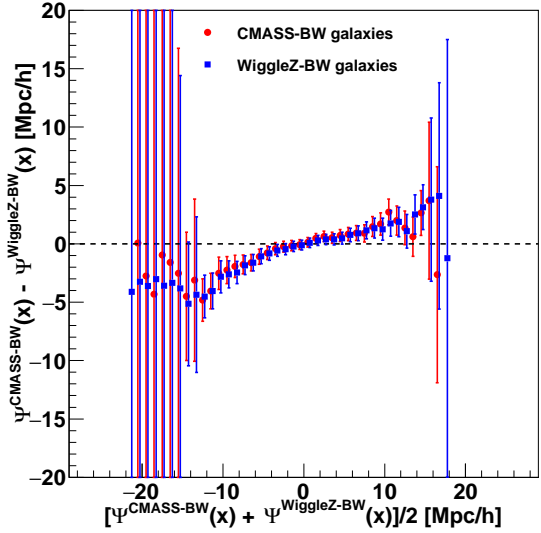,width=8cm}
\caption{Comparison of the CMASS-BW (red) and WiggleZ-BW (blue) galaxies using the CMASS-BW and WiggleZ-BW displacement fields, respectively. The $x$-axis is the mean displacement, while the $y$-axis shows the difference. The CMASS-BW data points are shifted by $0.25\,$Mpc/h to the right for clarity. The uncertainties are derived from the mock realizations.}
\label{fig:mean}
\end{center}
\end{figure}

The mean difference between the CMASS-BW and WiggleZ-BW displacement fields using CMASS galaxies is $\Delta \Psi = \Psi^{\rm CMASS-BW} - \Psi^{\rm WiggleZ-BW} = -0.047\pm0.016\,$Mpc$/h$, while for the WiggleZ catalogue $\Delta \Psi = -0.150\pm0.020\,$Mpc$/h$ (the errors are the error on the mean between all galaxies). We therefore find moderate differences between the two displacement fields. The difference between the two displacement fields does depend linearly on the amplitude of the displacement, $(\Psi_{\rm CMASS}+\Psi_{\rm WiggleZ})/2$, as shown in Figure~\ref{fig:mean} for the CMASS-BW galaxies (red) and the WiggleZ-BW galaxies (blue). Such a discrepancy could be caused by an incorrect assumption of the bias parameter when deriving the displacement field. 

\begin{figure}
\begin{center}
\epsfig{file=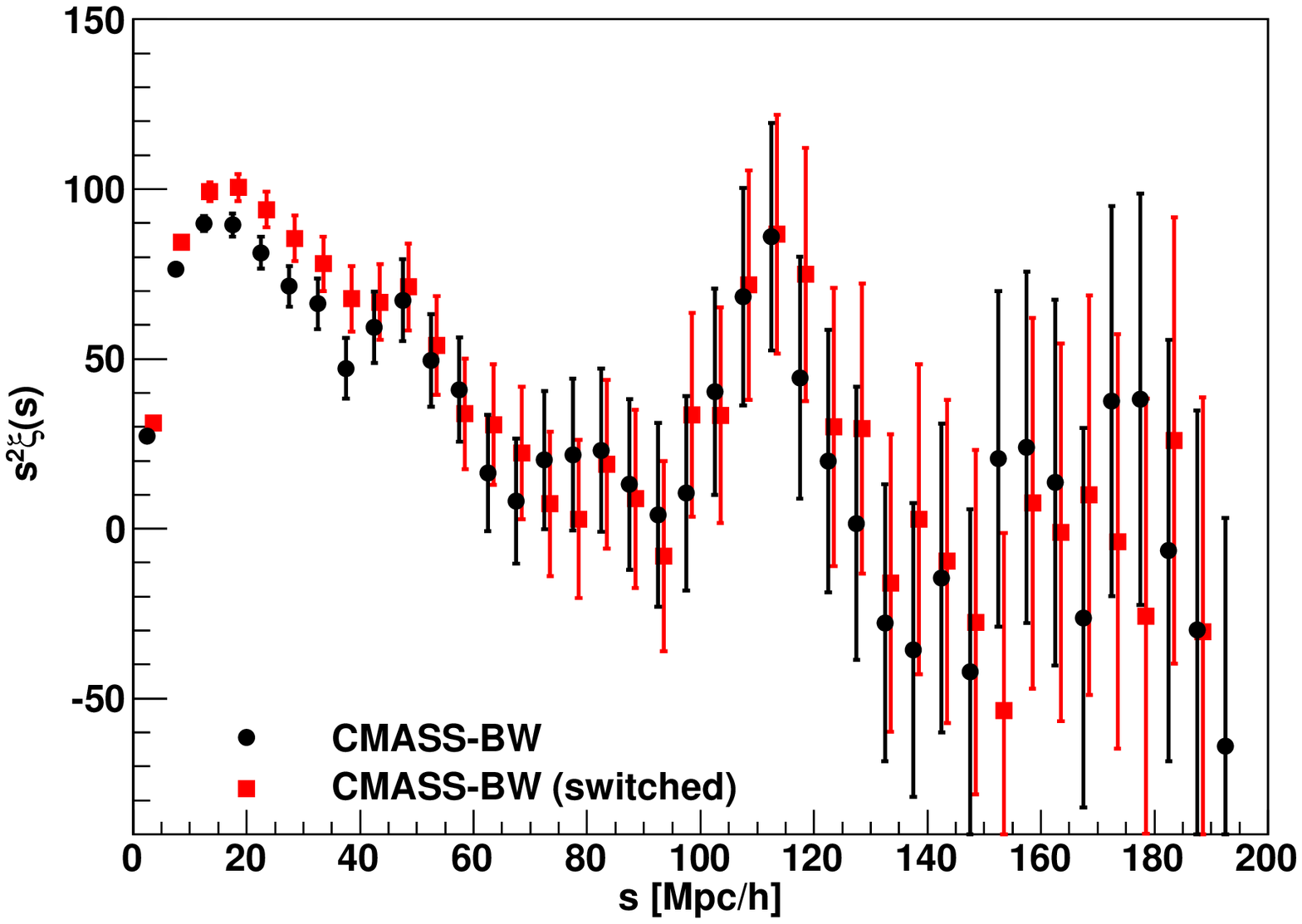,width=9cm}
\epsfig{file=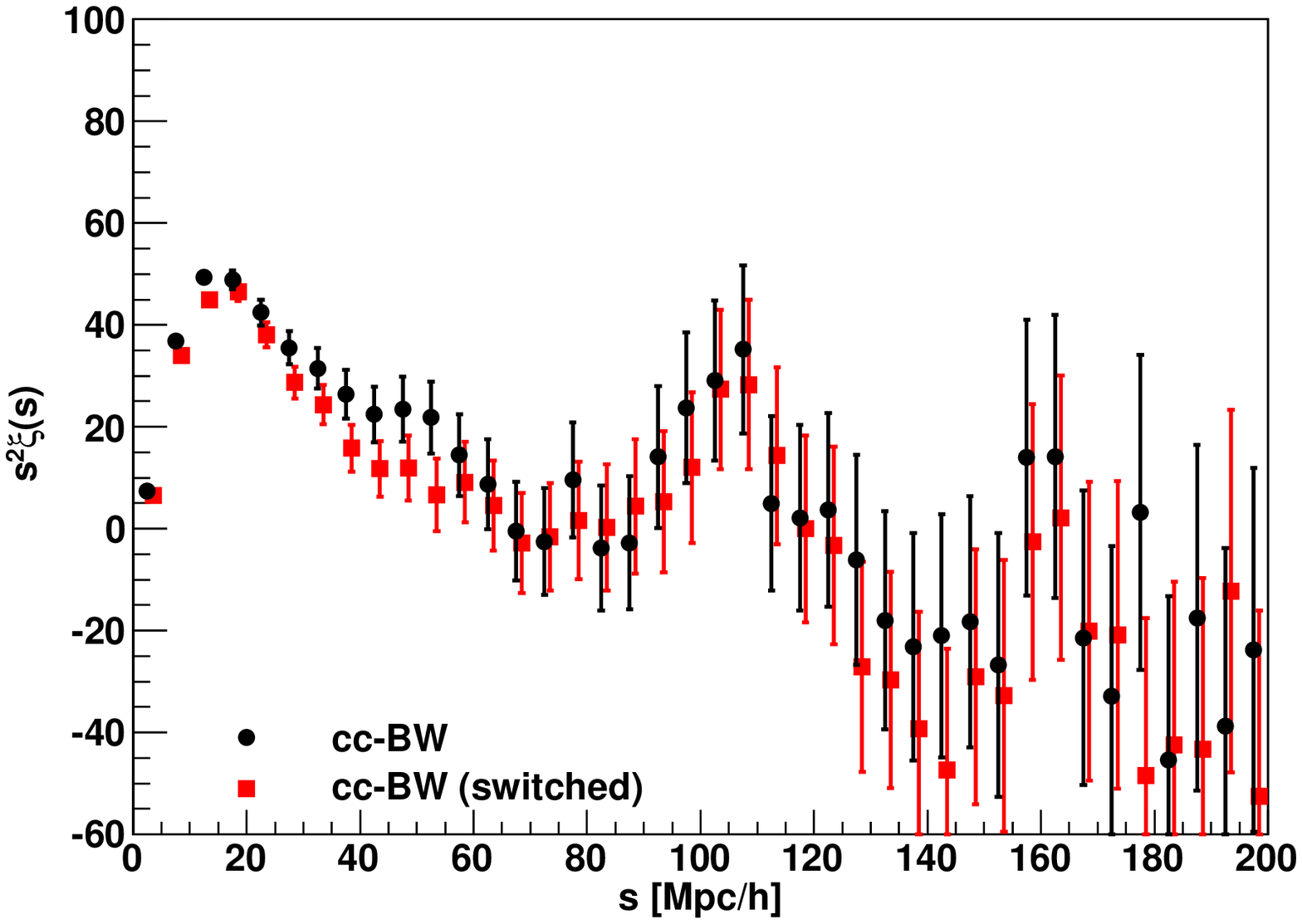,width=9cm}
\epsfig{file=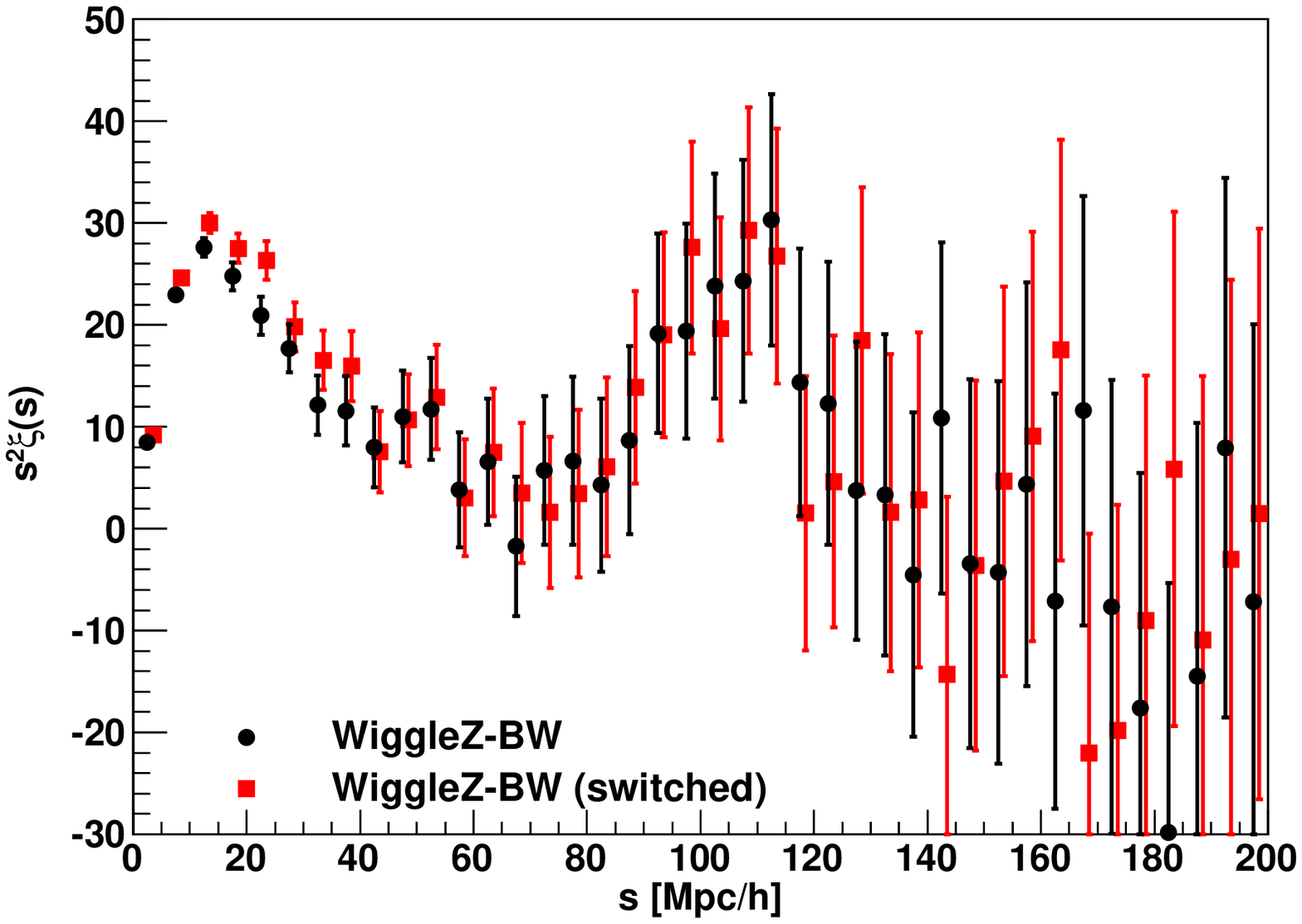,width=9cm}
\caption{Comparison of the correlation functions for CMASS-BW (top), WiggleZ-BW (bottom) and the cross-correlation function (cc-BW, middle) after density field reconstruction using the two displacement fields derived from the CMASS-BW and WiggleZ-BW galaxies, respectively. The label $``$(switched)$"$ indicates the case where the displacement field of the other survey has been used for reconstruction. The red data points are shifted to the right by $1$Mpc$/h$ for clarity.}
\label{fig:switched}
\end{center}
\end{figure}

To further investigate the impact of the two different displacement field on the BAO scale, we now calculate the correlation functions using both displacement fields. The correlation functions after combining the five different regions are presented in Figure~\ref{fig:switched}. In the next section we will fit these correlation functions and derive BAO constraints.

\subsection{Fitting the large scale correlation function}
\label{sec:fit}

\begin{figure}
\begin{center}
\epsfig{file=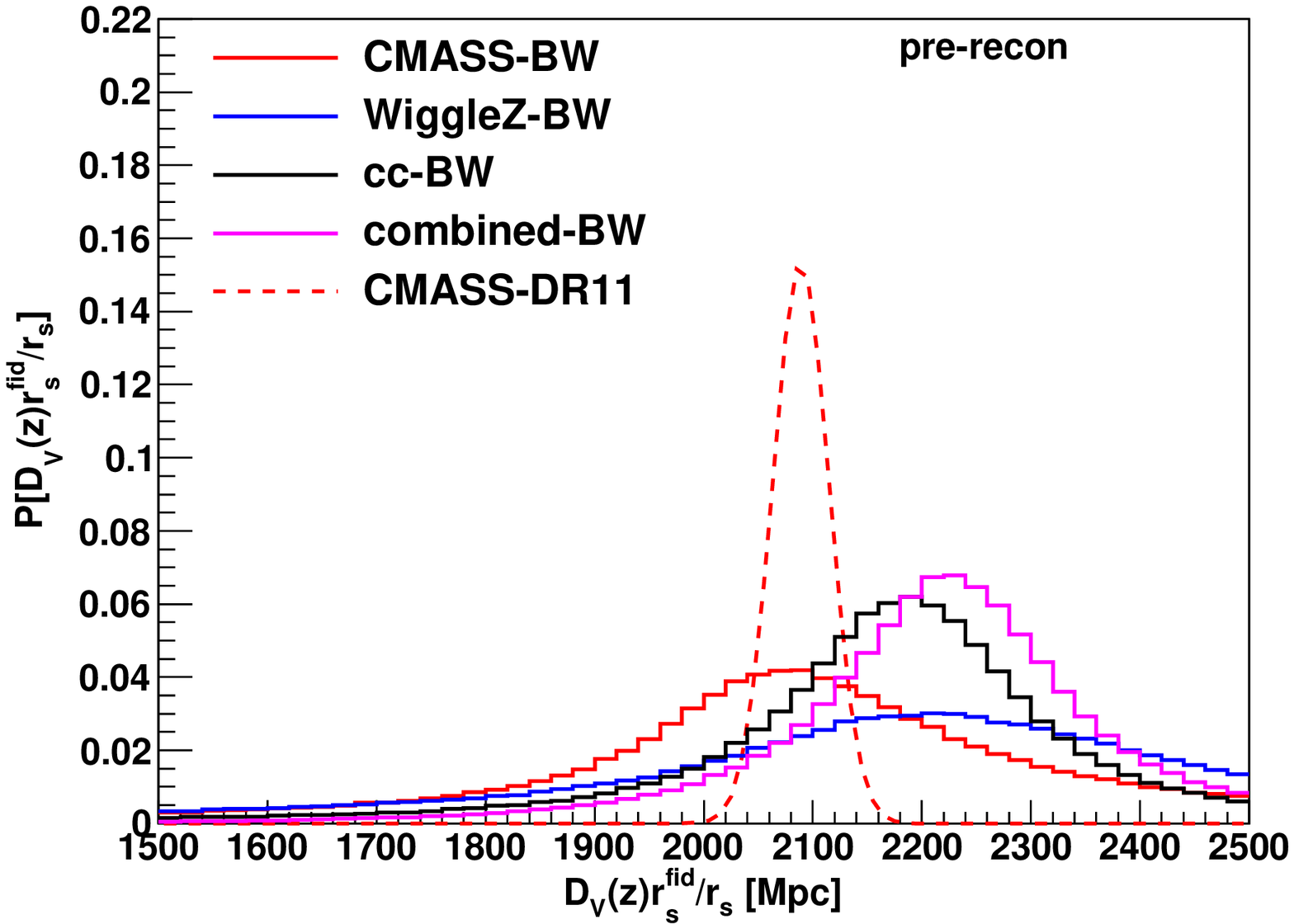,width=8.5cm}
\epsfig{file=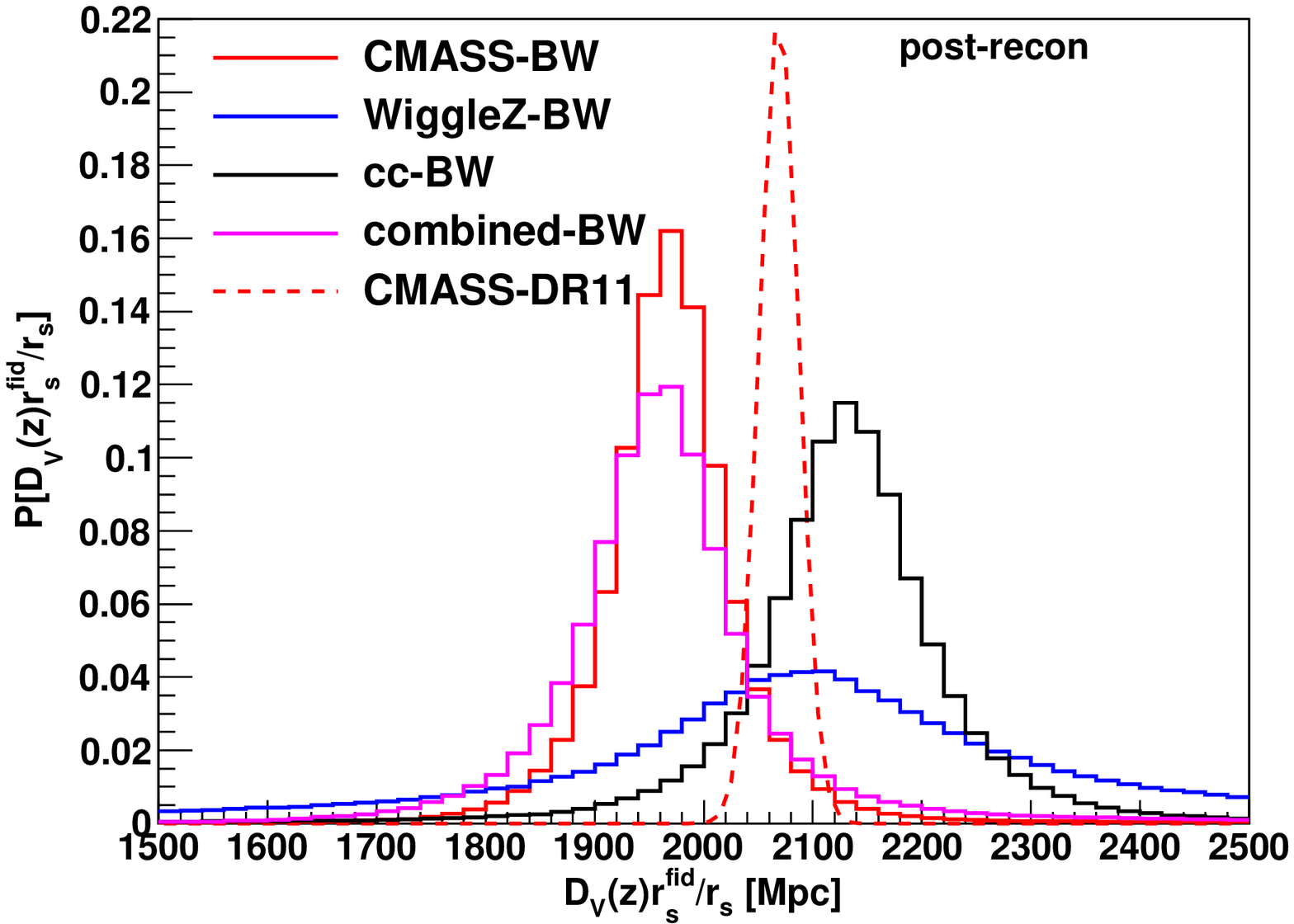,width=8.5cm}
\epsfig{file=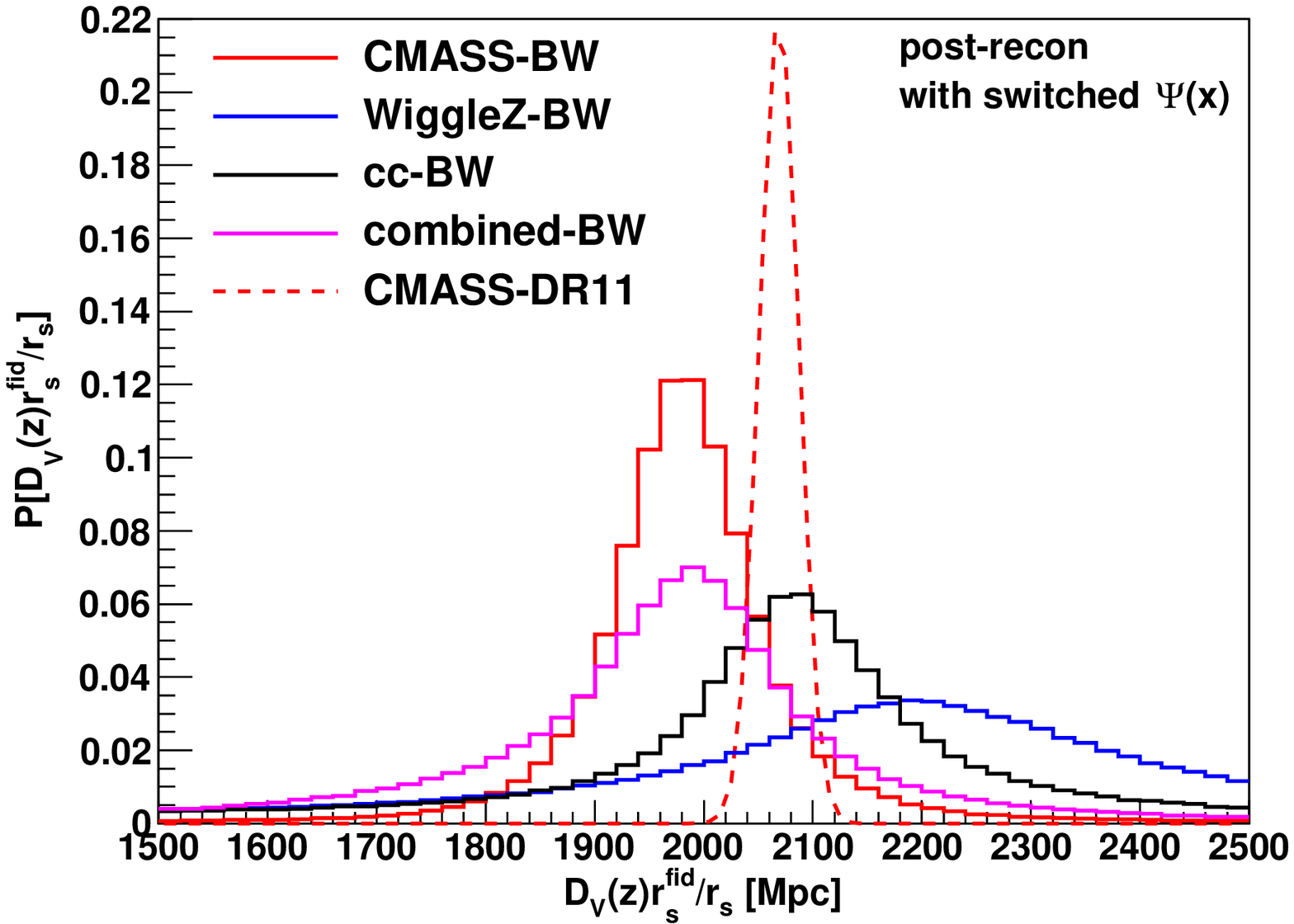,width=8.5cm}
\caption{Likelihood distribution for $D_V\frac{r^{\rm fid}_s}{r_s}$ derived from CMASS-BW (red), WiggleZ-BW (blue) and the cross-correlation function (black), before (top) and after (middle) density field reconstruction. The dashed red line shows the CMASS-DR11 constraint of~\citep{Anderson:2013zyy}. The bottom panel displays the result where the displacement fields derived from the two surveys have been switched.}
\label{fig:like}
\end{center}
\end{figure}

We start with fitting the individual correlation functions of CMASS-BW and WiggleZ-BW as well as the cross-correlation function. We search for the best fitting parameters defined by the minimum $\chi^2$, given by
\begin{equation}
\chi^2 = \sum_{ij}D^{\rm T}_{i}C^{-1}_{ij}D_{j},
\end{equation}
where $D$ is a vector containing the difference between the data and the model. Using the fitting range $50-180\,$Mpc$/h$ in $5\,$Mpc$/h$ bins results in $26$ elements for the vector $D$. We use the python-based MCMC sampler emcee~\citep{ForemanMackey:2012ig} to derive the likelihood. The results are shown in Table~\ref{tab:results}. We can clearly see that the constraints for all three correlation functions improve significantly after reconstruction. The resulting BAO constraints are worse, however, if we switch the displacement fields between the two surveys (these results are labeled as ``(switched)'' in Table~\ref{tab:results}). The same result occurs in the mock realizations, where $58\%$ of the WiggleZ-BW mock catalogues show a larger uncertainty on $\alpha$ when using the CMASS-BW displacement field for reconstruction instead of the WiggleZ-BW displacement field. Similarly $59\%$ of the CMASS-BW mock realizations show poorer constraints when using the WiggleZ-BW displacement field. The resulting likelihood distributions for all fits are presented in Figure~\ref{fig:like} including a comparison to the CMASS-DR11 result. The likelihood distributions are reasonably approximated by Gaussians.

In the limit of sample variance dominated uncertainties the three correlation functions would carry the same amount of information and only one of them would need to be analyzed. In the shot noise limit all three correlation functions would be independent and would need to be analyzed together to make maximal use of the available information. In the case of CMASS-BW and WiggleZ-BW, shot noise does contribute significantly to the error budget, so that a combined analysis is beneficial. 

Therefore we now fit all three correlation functions together using the combined covariance matrix shown in Figure~\ref{fig:covs}. The fit has $13$ free parameters, one scaling parameter, $\alpha$, the three bias parameters, $B_{\rm CMASS-BW}$, $B_{\rm cc-BW}$ and $B_{\rm WiggleZ-BW}$ as well as three polynomial terms per correlation function. The data vector for this fit is given by
\begin{equation}
D = \left(\begin{array}{cccc}\xi^{\text{CMASS-BW}}_{\rm model}(s_1) - \xi^{\text{CMASS-BW}}_{\rm data}(s_1)\cr\vdots\cr
 \xi^{\text{CMASS-BW}}_{\rm model}(s_{26}) - \xi^{\text{CMASS-BW}}_{\rm data}(s_{26})\cr
  \xi^{\text{cc-BW}}_{\rm model}(s_{1}) - \xi^{\text{cc-BW}}_{\rm data}(s_{1})\cr\vdots\cr
 \xi^{\text{cc-BW}}_{\rm model}(s_{26}) - \xi^{\text{cc-BW}}_{\rm data}(s_{26}).\cr
 \xi^{\text{WiggleZ-BW}}_{\rm model}(s_{1}) - \xi^{\text{WiggleZ-BW}}_{\rm data}(s_{1})\cr\vdots\cr
 \xi^{\text{WiggleZ-BW}}_{\rm model}(s_{26}) - \xi^{\text{WiggleZ-BW}}_{\rm data}(s_{26})\end{array}\right)
\end{equation}
containing $3\times26 = 78$ bins in total. 
The results are shown in the lower part of Table~\ref{tab:results}. In the case of pre-reconstruction, this fit is driven by the cross-correlation function, which has significantly smaller uncertainties than any of the auto-correlation functions and leads to a value of $\alpha = 1.095\pm0.068$. After reconstruction it is the CMASS-BW constraint which drives the combined fit, leading to $\alpha = 0.966\pm0.031$. The combined constraint on $\alpha$ is worse than the CMASS-BW only constraint on $\alpha$, which is mainly caused by the large scaling factor of $m_v = 1.4$ (see Eq.~\ref{eq:scaling}), needed for this fit. 

\begin{table*}
\begin{center}
\caption{Summary of the fitting results. The first sector reports the fits to CMASS-BW, WiggleZ-BW and cross-correlation (cc-BW) functions individually, while the second sector shows the combined fits to all three correlation functions. For each case we list the result pre- and post reconstruction, as well as the result where we switched the displacement fields $\Psi(x)$ used for reconstruction (switched). The errors on each parameter are obtained by marginalizing over all other parameters. The fitting range is $50$ - $180\,$Mpc$/h$ in $5\,$Mpc$/h$ bins leading to $26$ bins and $5$ free parameter (bias, three polynomials and $\alpha$) in case of the fit to the individual correlation functions. When fitting all correlation functions simultaneously (last three rows) there are $3\times 26 = 78$ degrees of freedom and $13$ free parameters. The likelihood distributions are shown in Figure~\ref{fig:like}. Our fiducial sound horizon is $r^{\rm fid}_s(z_d) = 150.18\,$Mpc.}
	\begin{tabular}{lcccc}
		\hline
		survey & $\alpha$ & $D_V(z)\frac{r^{\rm fid}_s(z)}{r_s(z)}$ [Mpc] & $\chi^2$\\
		\hline
		CMASS-BW pre-recon & $1.029^{+0.11}_{-0.085}$ & $2100^{+220}_{-170}$ & $31.2/(26-5)$\\
		CMASS-BW post-recon & $0.970\pm0.022$ & $1970\pm45$ & $22.6/(26-5)$\\
		CMASS-BW post-recon (switched) & $0.976\pm0.029$ & $1982\pm59$ & $23.1/(26-5)$\\
		cc-BW pre-recon & $1.073^{+0.056}_{-0.067}$ & $2180^{+110}_{-140}$ & $22.8/(26-5)$\\
		cc-BW post-recon & $1.050\pm0.032$ & $2132\pm65$ & $26.7/(26-5)$\\
		cc-BW post-recon (switched) & $1.023^{+0.059}_{-0.092}$ & $2080^{+120}_{-190}$ & $13.0/(26-5)$\\
		WiggleZ-BW pre-recon & $1.08^{+0.12}_{-0.15}$ & $2190^{+240}_{-300}$ & $18.1/(26-5)$\\
		WiggleZ-BW post-recon & $1.033\pm0.10$ & $2100\pm200$ & $10.4/(26-5)$\\
		WiggleZ-BW post-recon (switched) & $1.08^{+0.11}_{-0.14}$ & $2190^{+220}_{-280}$ & $15.6/(26-5)$\\
		\hline
		combined-BW pre-recon  & $1.095\pm0.068$ & $2220\pm140$ & $84.6/(78-13)$\\
		combined-BW post-recon & $0.966\pm0.031$ & $1956\pm63$ & $103.5/(78-13)$\\
		combined-BW post-recon (switched) & $0.972^{+0.047}_{-0.078}$ & $1974^{+95}_{-158}$ & $60.5/(78-13)$\\
		\hline
	  \end{tabular}
\label{tab:results}
\end{center}
\end{table*}

\subsection{Comparison to mock realizations}
\label{sec:cmpmocks}

The question now is whether the measured $\alpha$ values for CMASS-BW and WiggleZ-BW are consistent. We can test this, by using the $480$ correlated mock realizations, which we used to calculate the covariance matrix in Section~\ref{sec:cov}. We calculate the correlation function for each mock catalogue and repeat the fitting procedure described in the last section. Figure~\ref{fig:mocks2} shows the distribution of the different constraints on $\alpha$ for the mock realizations, together with the results found for the data (red data points). The ellipses in these plots are the $1\sigma$ standard deviation including the correlation between the different measurements. We only plot results which have a value of $\alpha$ between $0.6 < \alpha < 1.4$ as well as an error on $\alpha$ less than $25\%$. For CMASS-BW we have $318$ out of $480$ mock catalogues which fulfill these criteria, while for WiggleZ-BW there are $242$ and $302$ for the cross-correlation.

\begin{figure}
\begin{center}
\epsfig{file=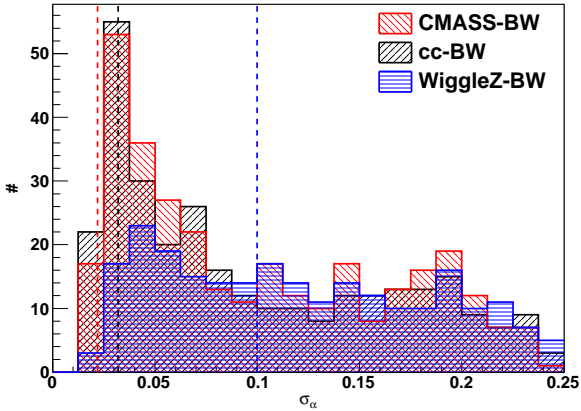,width=8.5cm}
\caption{The error on $\alpha$ measured in the $480$ mock realizations of CMASS-BW (red), WiggleZ-BW (blue) and the cross-correlation function (black). The uncertainty $\sigma_{\alpha}$ is taken to be the mean of the upper and lower $68\%$ confidence levels. The values measured in the data are shown by the correspondingly colored vertical dashed lines.}
\label{fig:sigalpha}
\end{center}
\end{figure}

Using the mock realizations the standard deviations for $\alpha$ are $\sigma_{\alpha} = (0.039, 0.036, 0.055)$ for CMASS-BW, cc-BW and WiggleZ-BW, respectively. While the mock realizations predict the best BAO constraint to be in the cross-correlation function, in the data the most accurate distance scale measurement (post-reconstruction) is in CMASS-BW. This result is, however, consistent with sample variance, and we have many mock realizations which show a similar behavior. Figure~\ref{fig:sigalpha} compares the distribution of errors for the mock realizations with the data. The signal-to-noise ratio of the auto-correlation functions is given by
\begin{equation}
\frac{n_{\rm A}P_{\rm AA}}{n_{\rm A}P_{\rm AA} + 1},
\end{equation}
while the signal-to-noise ratio of the cross-correlation function scales with~\citep{Smith:2008ut}
\begin{equation}
2\frac{\sqrt{n_{\rm A}n_{\rm B}}P_{\rm AB}}{\sqrt{(n_{\rm A}P_{\rm AA} + 1)(n_{\rm B}P_{\rm BB} + 1) + n_{\rm A}n_{\rm B}P^2_{\rm AB}}}.
\end{equation}
Assuming $nP = 3$ for CMASS-BW and $nP = 1$ for WiggleZ-BW as well as $P_{\rm CMASS-BW} = 4P_{\rm WiggleZ-BW}$ ($P_{\rm cc-BW} = 2P_{\rm WiggleZ-BW}$), we find $0.75$, $1.04$ and $0.5$ for CMASS-BW, cc-BW and WiggleZ-BW, respectively. This rough estimate of the expected signal-to-noise ratio agrees well with the result of the mock realizations. We can therefore conclude that in our actual data measurements the WiggleZ-BW constraint has an error slightly larger than expected, the cross-correlation function error is close to the expected error and the CMASS-BW error is smaller than expected.

To determine whether the three different BAO constraints agree, we construct the vector $T = (\alpha_{\rm CMASS-BW} - \alpha_{\rm cc-BW}, \alpha_{\rm CMASS-BW} - \alpha_{\rm WiggleZ-BW})$. Using the mock catalogues we obtain a correlation coefficient for the two components of the vector $T$ of $60.6\%$, which allows us to construct a covariance matrix and calculate $\chi^2 = T^TC^{-1}T$ with two degrees of freedom. Post reconstruction we get $\chi^2/\text{d.o.f.} = 1.25/2$ when using the mean uncertainty predicted from the mocks, and $4.1/2$ when using (averaged) $1\sigma$ errors from the actual measurements. The three different BAO measurements are therefore statistically consistent.

The BAO peak in CMASS-BW leads to a distance constraint of $D_Vr_s^{\rm fid}/r_s = 1970\pm45\,$Mpc (post reconstruction), which is low compared to the CMASS-DR11 result of $D_Vr_s^{\rm fid}/r_s = 2068\pm 20\,$Mpc~\citep{Anderson:2013zyy}\footnote{This value has been corrected by the ratio of the fiducial sound horizons to allow direct comparison with our constraint.}. Since CMASS-BW uses about $7\%$ of the CMASS-DR11 galaxies, we can approximate the correlation coefficient, $r^2$, to be about $0.07$, implying that these constraints are fairly independent. We can quantify the significance of the deviation between two correlated measurements with
\begin{equation}
\frac{\Delta X}{\sqrt{\sigma^2_A + \sigma^2_B - 2\times r(\sigma_A,\sigma_B) \sigma_A\sigma_B}},
\label{eq:cor}
\end{equation}
where $r(\sigma_A,\sigma_B)$ is the correlation coefficient. This leads to a significance of $2.1\sigma$ for the deviation between CMASS-DR11 and CMASS-BW.

The WiggleZ survey has a distance constraint of $D_Vr_s^{\rm fid}/r_s = 2100\pm200\,$Mpc, which we can compare to $D_Vr_s^{\rm fid}/r_s = 2221^{+97}_{-104}\,$Mpc measured in~\citet{Kazin:2014qga}. The WiggleZ constraint has a slightly different redshift range compared to our WiggleZ-BW constraint ($0.4 < z < 0.8$) and does include the additional sky region S3 (see Figure~\ref{fig:sky}), therefore a direct comparison is not possible.

\citet{Ross:2013vla} split the CMASS sample based on $k + e$ corrected $i$-band absolute magnitudes and $[r - i]_{0.55}$ colors, yielding two sub-samples with bias $b = 1.65$ (blue) and $b= 2.3$ (red). Studies of the BAO scale in these sub-samples revealed no statistically significant deviations, in agreement with our findings. Different to~\citet{Ross:2013vla} our study is based on two entirely different surveys and therefore also includes possible systematics due to instrumentation, telescope site conditions or reduction pipeline. 

\begin{figure}
\begin{center}
\epsfig{file=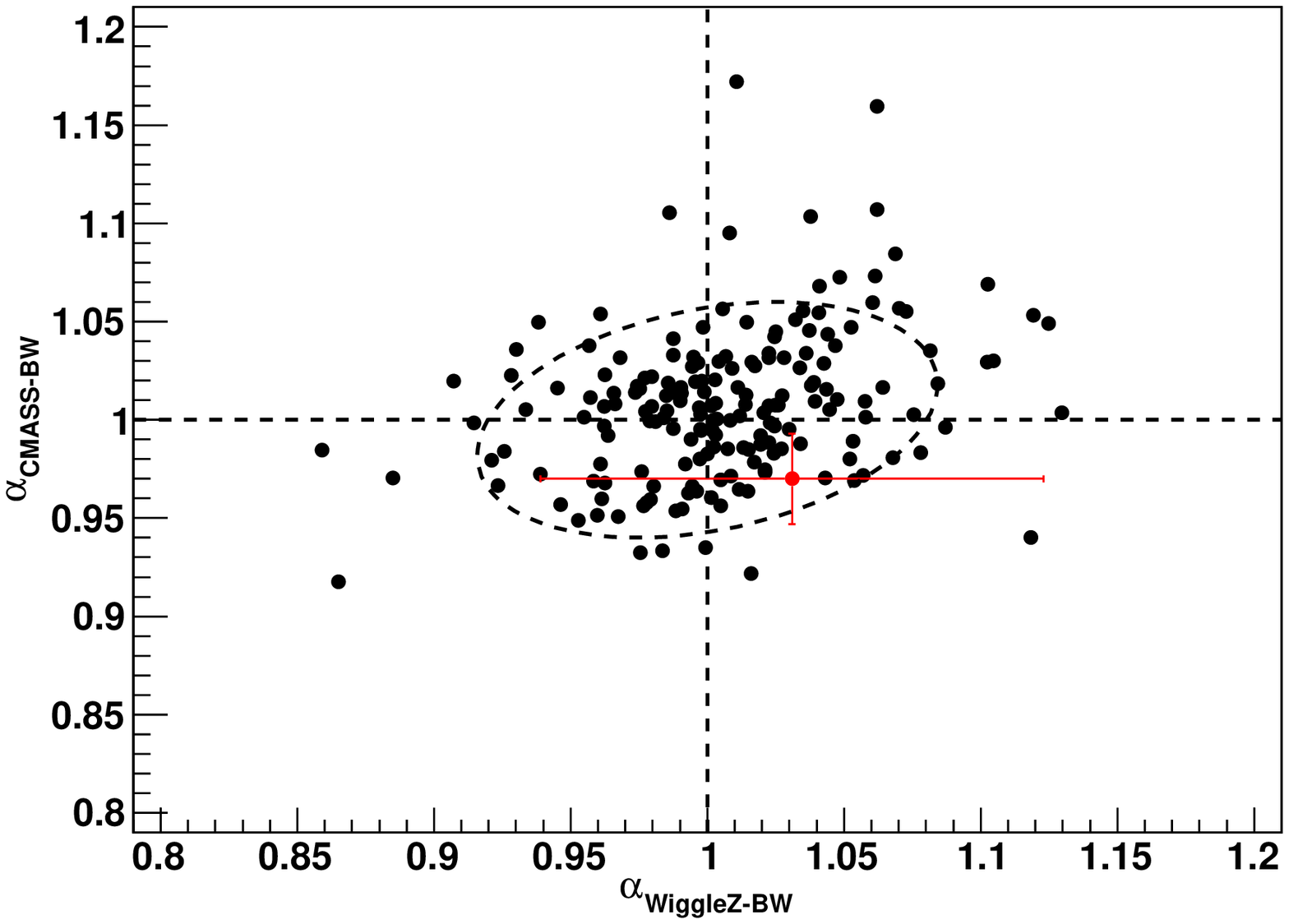,width=8.5cm}
\epsfig{file=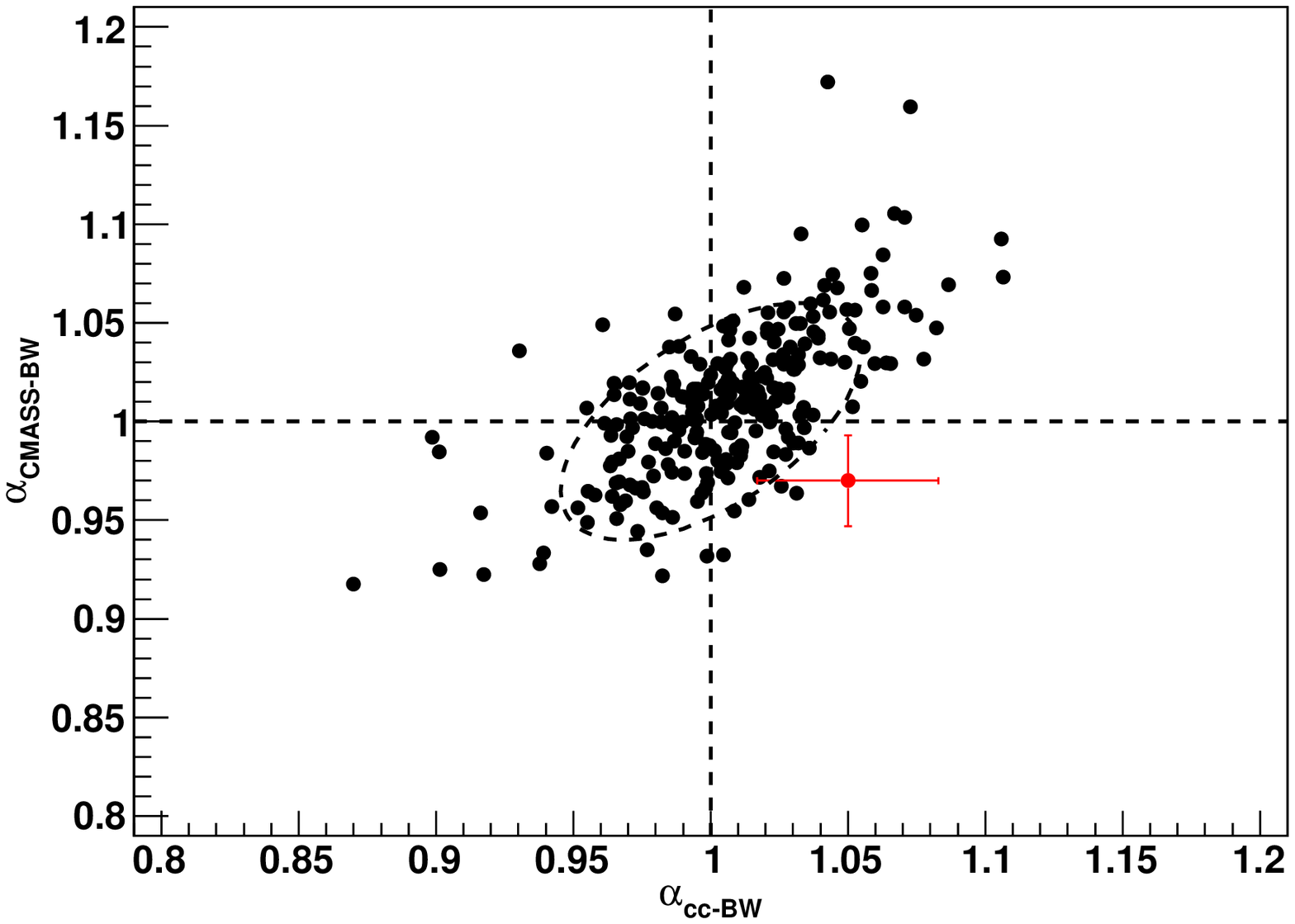,width=8.5cm}
\epsfig{file=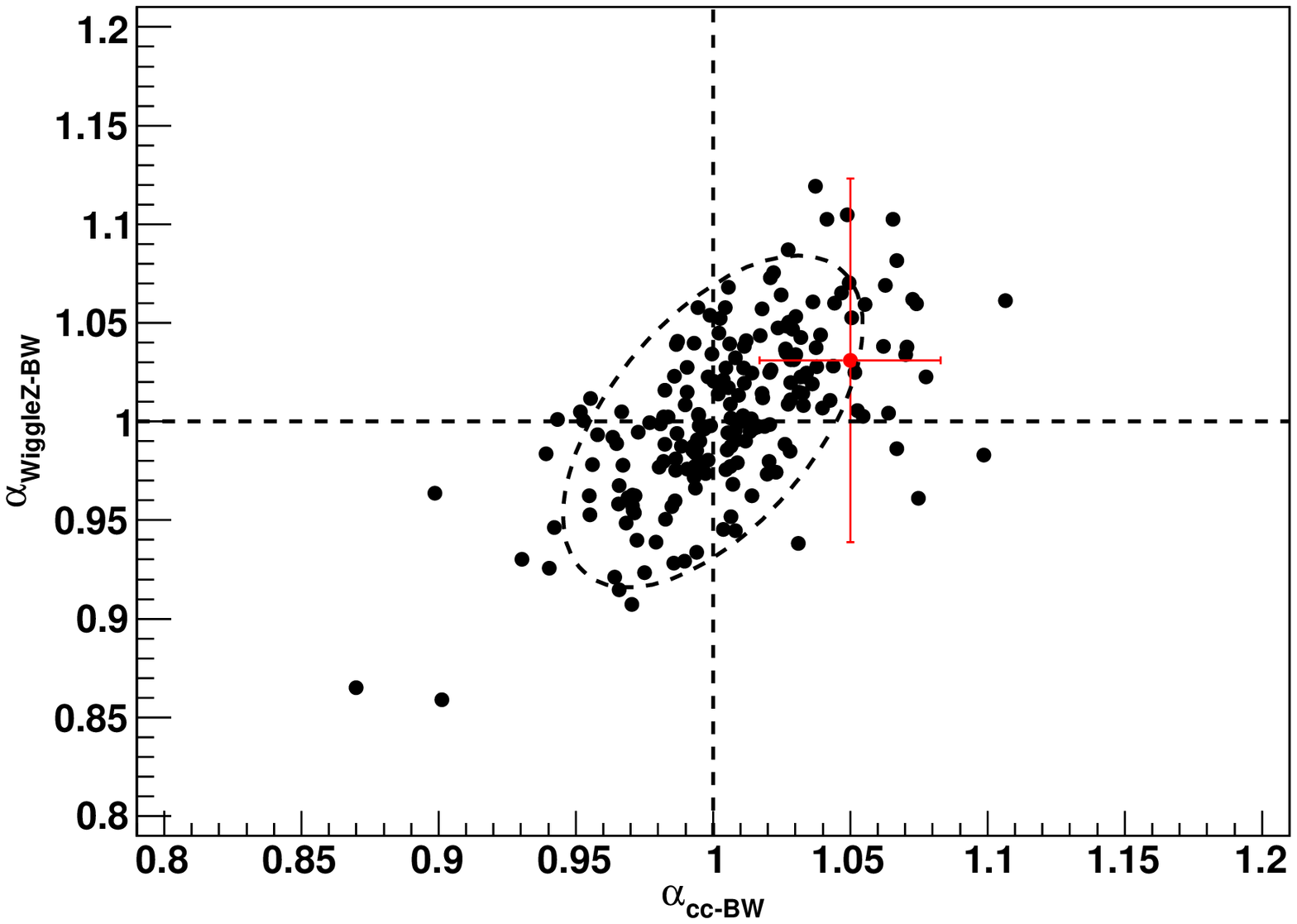,width=8.5cm}
\caption{These plots show the distribution of $\alpha$ for the mock realizations of CMASS-BW, WiggleZ-BW and cc-BW. We only plot results which have a value of $\alpha$ between $0.6 < \alpha < 1.4$ as well as an error on $\alpha$ less than $25\%$. The ellipse represents the $1\sigma$ distribution drawn from the variance and correlation coefficient. Note that the ellipse has not been derived from the black points drawn in these plots but instead from jack-knife samples to avoid outliers. The red datapoint shows our measurement post-reconstruction from Table~\ref{tab:results}. The agreement between the red data points and the black ellipse is only required if the cosmology of the mocks is the true cosmology.}
\label{fig:mocks2}
\end{center}
\end{figure}

\section{Determining the correlation between BAO constraints in CMASS and WiggleZ}
\label{sec:covariance}

In the last section we set constraints on the quantity $D_V\frac{r^{\rm fid}_s}{r_s}$ using the CMASS-BW and WiggleZ-BW auto-correlation functions as well as their cross-correlation function. The constraint from the WiggleZ-BW auto-correlation function uses almost all WiggleZ information within the redshift range $0.43 <  z < 0.7$, excluding only region S3, which is small in comparison. The entire CMASS sample, however, covers a sky area much larger than the overlap region. In this section we will determine the correlation of WiggleZ-BW and cross-correlation constraints found in the last section with the CMASS-DR11 constraint of~\citet{Anderson:2013zyy}. We will then construct a covariance matrix which allows us to use our results together with the result of CMASS-DR11 for cosmological parameter constraints. We will also provide an estimate of the correlation for the future CMASS-DR12 constraint.

First we divide the two surveys into the following sub-regions:
\begin{enumerate}[1]
\item BOSS-CMASS, excluding overlap region.
\item BOSS-CMASS, in overlap region (CMASS-BW).
\item WiggleZ, excluding overlap region.
\item WiggleZ, in overlap region] (WiggleZ-BW).
\end{enumerate}
For each region, the parameter of interest is the constraint on the scaling parameter $\alpha$. Assuming no correlation between the different regions, the final constraints for each survey are given by
\begin{align}
\alpha_{\rm B} &= 1 + \frac{V_1 (\alpha_1 - 1) + V_2 (\alpha_2 - 1)}{V_1 + V_2},\\
\alpha_{\rm W} &= 1 + \frac{V_3 (\alpha_3 - 1) + V_4 (\alpha_4 - 1)}{V_3 + V_4},
\end{align}
where the subscripted numbers refer to the four survey sub-regions described above, and $V$ is the volume given in Table~\ref{tab:volume}. The subscripted $B$ stands for BOSS-CMASS and $W$ stands for WiggleZ. The correlation between $\alpha_{\rm B}$ and $\alpha_{\rm W}$ is given by
\begin{equation}
r^2(\alpha_{\rm B}, \alpha_{\rm W}) \approx \frac{r^2(\alpha_2,\alpha_4)V_2 V_4}{V_{\rm B}V_{\rm W}},
\label{eq:cor}
\end{equation}
where the correlation coefficient between the constraints in the overlap regions, $r^2(\alpha_2,\alpha_4)$, can be calculated from the $480$ mock realizations as
\begin{equation} 
r^2(\alpha_2,\alpha_4) = \frac{\sum^{480} (\alpha_2 - \overline{\alpha}_2)(\alpha_4 - \overline{\alpha}_4)}{\sqrt{\sum^{480}(\alpha_2 - \overline{\alpha}_2)^2\sum^{480} (\alpha_4 - \overline{\alpha}_4)^2}}.
\label{eq:corco}
\end{equation}
For practical reasons we use a jack-knife approach, in which we determine $\alpha$ for the mean of $N-1$ realizations, excluding each of the $480$ mock realizations in turn. The correlation coefficients between CMASS-BW and WiggleZ-BW in the overlap region is $r^2(\alpha_2,\alpha_4) = 0.301$. Similarly, we can define the correlation coefficient between the auto-correlation functions and the cross-correlation function giving $r^2(\alpha_C,\alpha_4) = 0.570$ and $r^2(\alpha_2,\alpha_C) = 0.584$, where the subscripted $C$ stands for the cross-correlation function. To obtain the correlation coefficient between our WiggleZ-BW constraint and the CMASS DR11 constraint of~\citet{Anderson:2013zyy}, we use the volume of CMASS-DR11 (as given in Table~\ref{tab:volume}) in Eq.~\ref{eq:cor} and set $V_4 = V_W$. 
Therefore the data vector $D = (\text{CMASS-DR11}, \text{cc-BW}, \text{WiggleZ-BW}) = (2056, 2132, 2100)\,$Mpc has the following correlation matrix:
\begin{align} 
R_{\rm DR11} &= \left( \begin{array}{ccc}
1 &   &  \\
r^2(\alpha_B,\alpha_C) & 1 &  \\
r^2(\alpha_B,\alpha_4)  & r^2(\alpha_C,\alpha_4) & 1\end{array} \right),\notag\\
&= \left( \begin{array}{ccc}
1 &   &  \\
0.043 & 1 &  \\
0.022  & 0.570 & 1\end{array} \right).
\label{eq:cormatrix}
\end{align}
The covariance matrix is given by $C = V^{\rm T}RV$, where the vector $V$ contains the variance of the individual constraints. In our case we have $V_{\rm DR11} = (20, 65, 200)\,$Mpc, where we adopted the CMASS-DR11 uncertainty (left) from~\citet{Anderson:2013zyy} together with the WiggleZ-BW (right) and cross-correlation function (middle) uncertainties from Table~\ref{tab:results}. This approach leads to the following covariance matrix
\begin{equation} 
C_{\rm DR11} = \left( \begin{array}{ccc}
400 &  & \\
56 &  4225 & \\
88 & 7410 & 40000 \end{array} \right)
\end{equation}
and the inverse is given by
\begin{equation} 
C^{-1}_{\rm DR11} = \left( \begin{array}{ccc}
250.47 &  & \\
-3.48 & 35.11 & \\
0.09 & -6.50 & 3.70  \end{array} \right)\times 10^{-5}.
 \label{eq:invcov}
\end{equation}
Since the overlap volume between CMASS and WiggleZ will not change with DR12, this formalism can be rescaled to obtain the correlation between our results and future CMASS data releases. For example, using a cosmic volume of $2.322\,h^{-3}$Gpc$^3$ for DR12 we find the following correlation matrix
\begin{equation} 
R_{\rm DR12} = \left( \begin{array}{ccc}
1 &   &  \\
0.038 & 1 &  \\
0.020 & 0.570 & 1\end{array} \right).
\label{eq:cormatrix}
\end{equation}
This covariance matrix is only correct assuming that any correlation between these surveys can be scaled with volume. 

So far we have only used WiggleZ galaxies in the redshift range $0.43 < z < 0.7$, ignoring a significant fraction of WiggleZ galaxies at higher and lower redshifts. We can combine our results with the high redshift ($0.6 < z < 1.0$) WiggleZ measurement reported in~\citet{Kazin:2014qga} and given by $D_Vr_s^{\rm fid}/r_s = 2516\pm86\,$Mpc. This measurement has an effective redshift of $z_{\rm eff} = 0.73$. The effective volume of WiggleZ in the redshift range $0.6 < z < 0.7$ (overlap between the high redshift WiggleZ measurement and the CMASS redshift range) is $6.1\times 10^7h^{-3}$Mpc$^3$. The correlation matrix including the high redshift WiggleZ data point (labeled by ext.) would be 
\begin{equation} 
R^{\rm ext}_{\rm DR11} = \left( \begin{array}{cccc}
1 &   &  & \\
0.043 & 1 & & \\
0.022 & 0.57 & 1 &\\
0.017 & 0.42 & 0.51 & 1\end{array} \right)\\
\end{equation}
with the data vector 
\begin{equation}
D^{\rm ext} =  \begin{pmatrix}\text{CMASS-DR11} \\ \text{cc-BW} \\ \text{WiggleZ-BW} \\ \text{WiggleZ-highz}\end{pmatrix} = \begin{pmatrix}2056\\ 2132 \\ 2100\\ 2516\end{pmatrix}\text{Mpc}
\end{equation}
and the variance vector is $V^{\rm ext}_{\rm DR11} = (20, 65, 200, 86)\,$Mpc. This result makes the additional assumption that the correlation coefficient we found for the CMASS redshift range can be scaled to the overlap redshift range $0.6 < z < 0.7$. We do not combine our measurements with the low and medium redshift bins reported in~\citet{Kazin:2014qga} since both overlap with the BOSS-LOWZ redshift range. Therefore the results reported in this section can be combined with the BOSS-LOWZ~\citep{Anderson:2013zyy,Tojeiro:2014eea} measurement straightforwardly.

The assumed fiducial cosmologies used in the different measurements above are not the same, resulting in different fiducial sound horizons. The sound horizon used in the CMASS-DR11 analysis is $r_s^{\rm fid} = 149.28\,$Mpc, \citet{Kazin:2014qga} have $r_s^{\rm fid}=148.6\,$Mpc and our analysis uses $r_s^{\rm fid} = 150.18\,$Mpc. When comparing the measurements of $D_V$ above with a cosmological model one has to include the ratio of the fiducial sound horizon and the sound horizon of that model.

\section{The relative velocity effect}
\label{sec:relvel}

In this section we discuss one possible source of systematic uncertainty for BAO constraints, the relative velocity effect~\citep{Tseliakhovich:2010bj}. We first introduce the idea of the relative velocity effect and discuss our model, before fitting the model to the data.

While dark matter perturbations start to grow directly after the end of inflation, baryon perturbations cannot grow until they decouple from the photons, about $380\,000$ years later. 
The different velocities of dark matter and baryons after decoupling means that there is a relative velocity between the two components~\citep{Tseliakhovich:2010bj,Fialkov:2014rba}. The relative velocity can allow baryons to escape the dark matter potentials and prevent the formation of the first stars in regions with high relative velocity~\citep{Fialkov:2011iw,McQuinn:2012rt,Naoz:2011if}. This modulation would select regions with small relative velocity to first undergo reionisation. Since the relative velocity effect decays with $1+ z$, it mainly affects the high-redshift Universe. However, it has been speculated that galaxies which form at high redshift carry this selection process down to low redshift, perhaps through processes such as altering the metal abundances or supernovae feedback~\citep{Yoo:2011tq}. In Fourier space the relative velocity effect has an oscillatory pattern on large scales which is out of phase with the Baryon Acoustic Oscillations~\citep{Yoo:2013qla}. In configuration space the relative velocity effect modifies the clustering amplitude primarily below the sound horizon, leading to a shift of the BAO peak~\citep{Slepian:2014dda}. 

The hypothesis is that old galaxies still carry the selection of the relative velocity effect, while young galaxies do not. Under this hypothesis we can measure the relative velocity bias by comparing clustering statistics of BOSS and WiggleZ, since BOSS mainly targeted (old) luminous red galaxies, while WiggleZ selected young star forming galaxies (ELGs). Our analysis method therefore assumes that BOSS galaxies are affected by the relative velocity effect, while WiggleZ galaxies are not.

\subsection{Modeling}

\begin{figure*}
\begin{center}
\epsfig{file=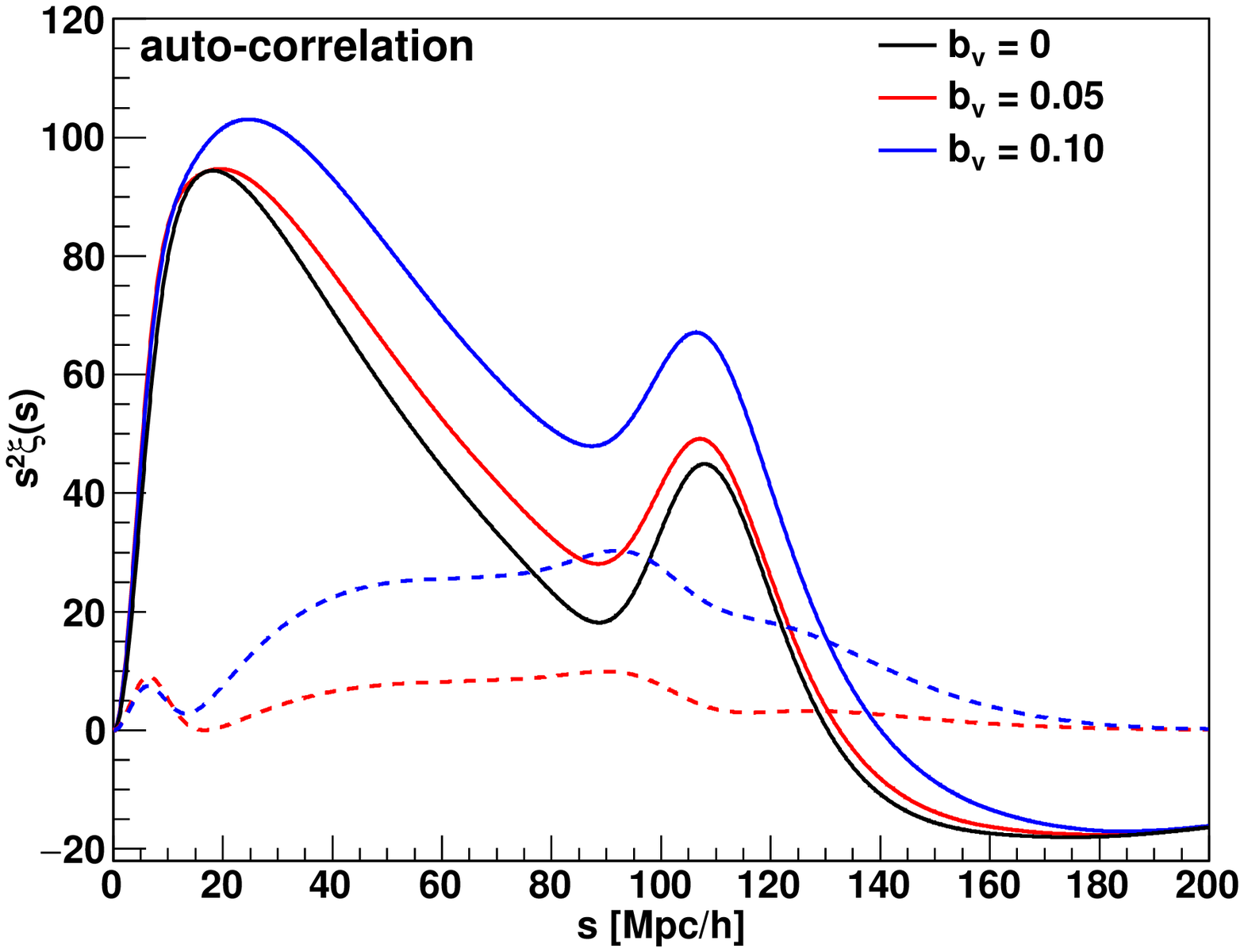,width=8.5cm}
\epsfig{file=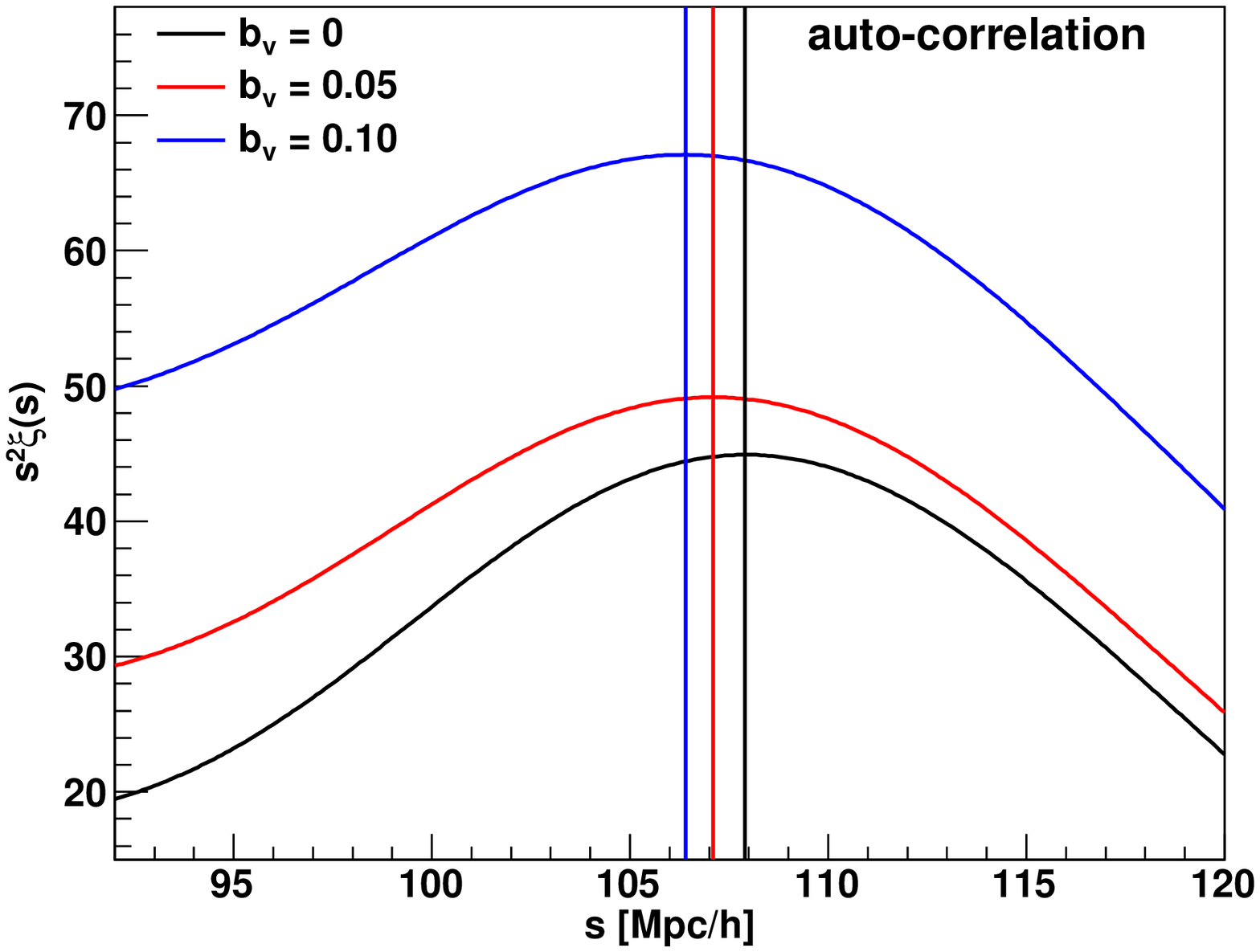,width=8.5cm}
\caption{Comparison of the auto-correlation functions for different values of the velocity bias parameter. The plot on the left shows the entire correlation function, while the plot on the right focuses on the BAO peak. The relative velocity effect causes an increase in the correlation function amplitude as well as a shift in the BAO peak position towards smaller scales. For these figures we assumed $b_1 = 2$ and $b_2 = -0.4$. The dashed lines represent the correction terms for different values of the velocity bias; the dashed lines added to the black solid line results in the correspondingly colored solid lines.}
\label{fig:peak1}
\end{center}
\end{figure*}

\begin{figure*}
\begin{center}
\epsfig{file=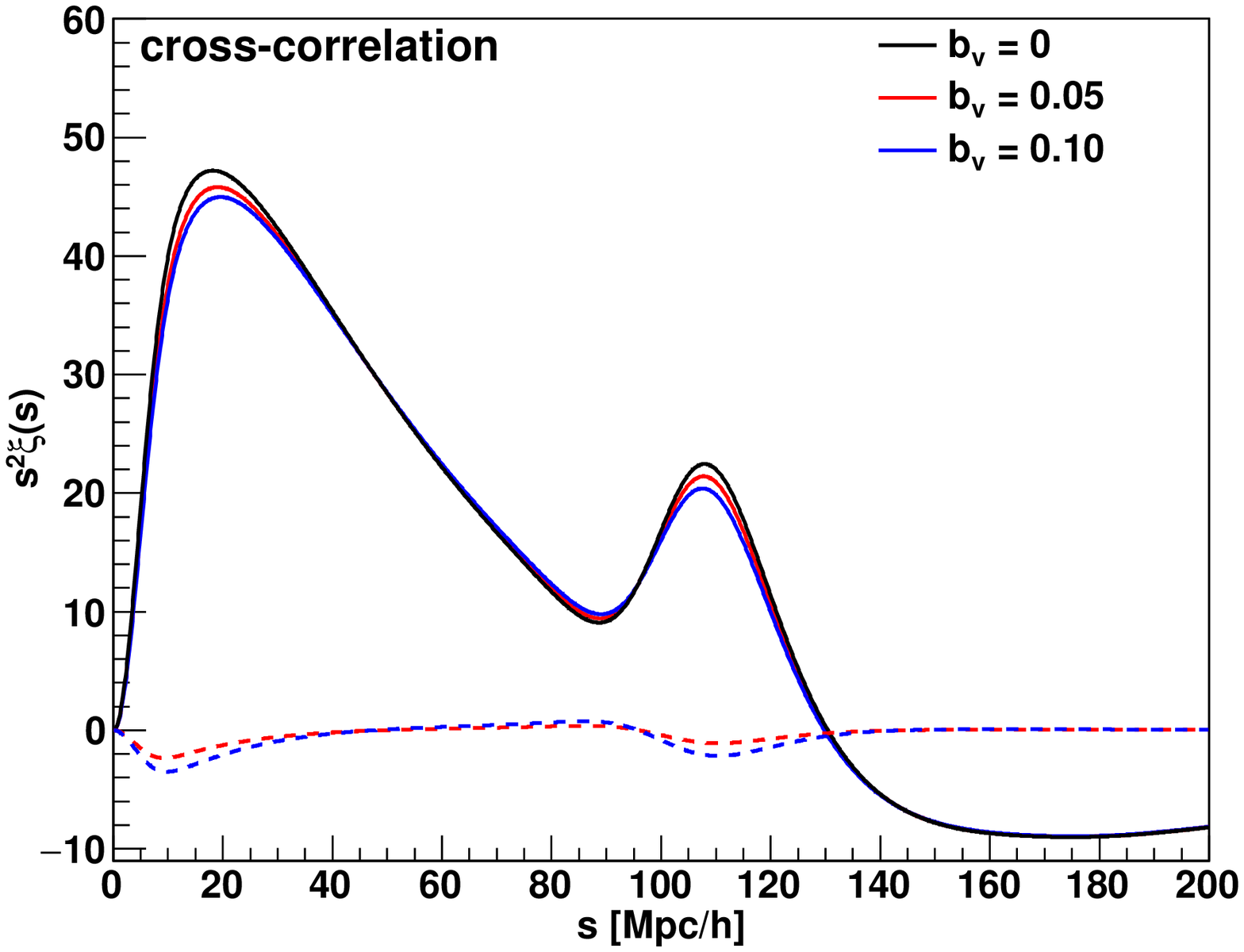,width=8.5cm}
\epsfig{file=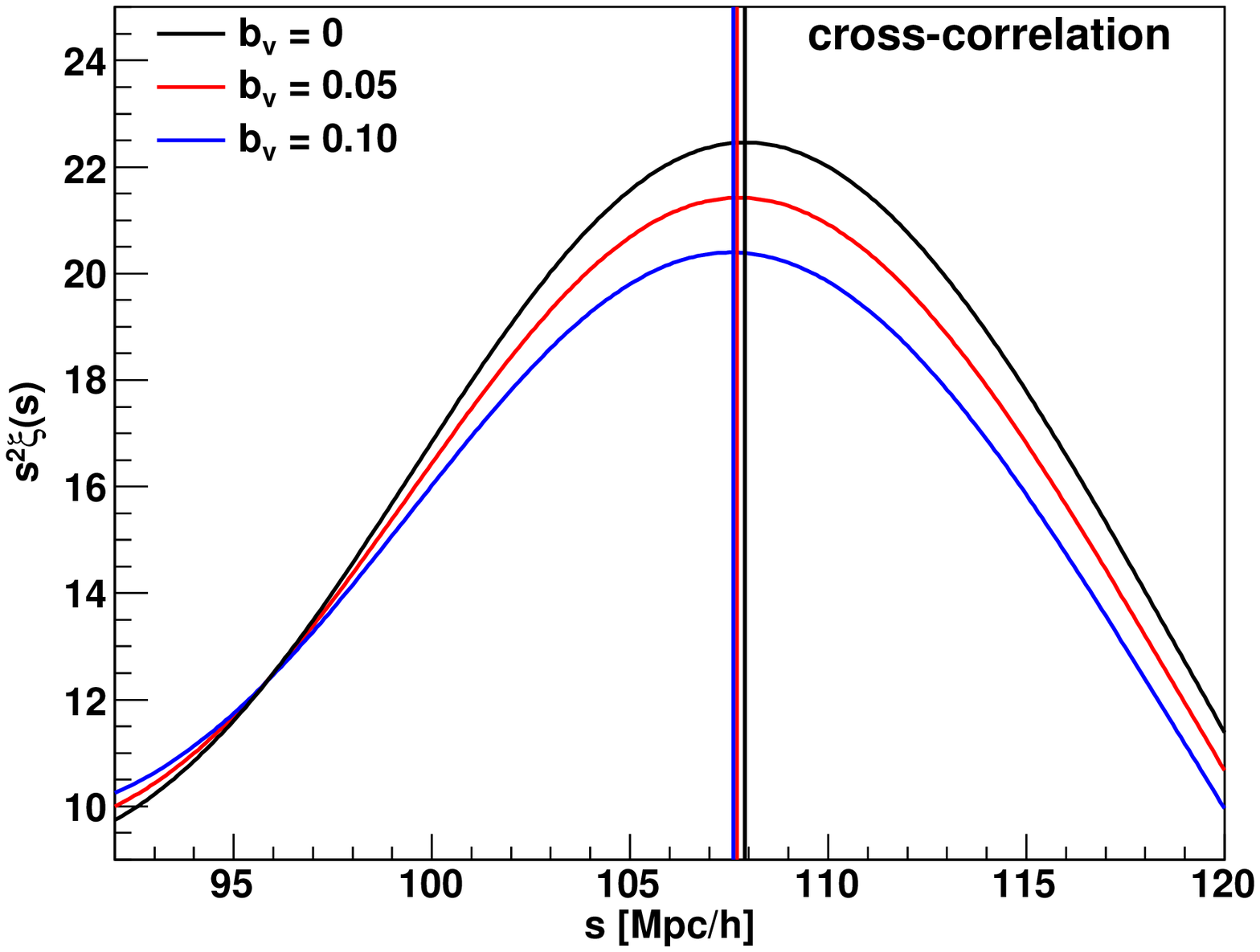,width=8.5cm}
\caption{Comparison of the cross-correlation functions for different values of the velocity bias parameter. The plot on the left shows the entire correlation function, while the plot on the right focuses on the BAO peak. The relative velocity effect causes a small shift in the BAO peak position towards smaller scales. Unlike for the auto-correlation function, the relative velocity effect does not change the amplitude of the cross-correlation function. For these plots we assumed $b_1^A = 2$, $b_2^A = -0.4$, $b_1^B = 1$ and $b_2^B = 1$. The dashed lines represent the correction terms for different values of the velocity bias; the dashed lines added to the black solid line results in the correspondingly colored solid lines.}
\label{fig:peak2}
\end{center}
\end{figure*}

\begin{figure}
\begin{center}
\epsfig{file=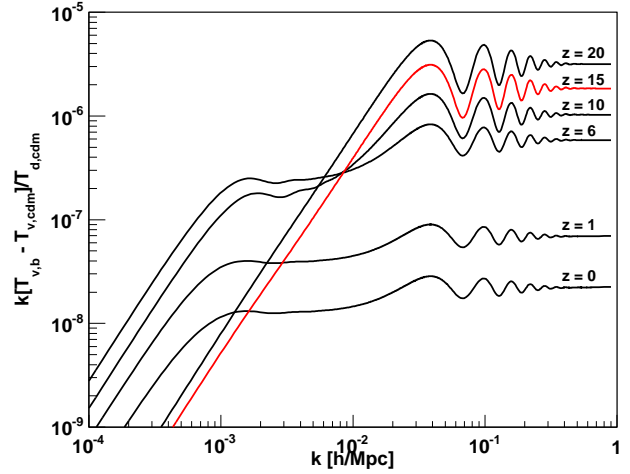,width=9cm}
\caption{The velocity transfer function $T_v = T_{v_{\rm b}} - T_{v_{\rm cdm}}$ relative to the cold dark matter transfer function $T_{\rm cdm}$. While at high redshift this ratio only changes in amplitude, at low redshift it has additional contributions from reionisation. Assuming that the relative velocity effect has been imprinted before reionisation, the velocity transfer function can be evaluated at any redshift (above reionisation), since the amplitude is rescaled through Eq.~\ref{eq:relnorm}.}
\label{fig:z_dep}
\end{center}
\end{figure}

To model the correlation function including the relative velocity effect we follow the implementation of~\citet{Yoo:2013qla}. In this model, the galaxy density field is given by 
\begin{equation}
\delta_g(\vec{x}) = b_1\delta_m(\vec{x}) + \frac{b_2}{2}\left[\delta_m^2(\vec{x}) - \sigma_m^2\right] + b_v \left[u_r^2 - \sigma_{u_r}\right],
\label{eq:deltarel}
\end{equation}
where the relative velocity $u_r$ is computed at the linear order and the matter density is computed to the second order. The auto-power spectrum from such a density field can be written as 
\begin{equation}
\begin{split}
P_{\rm g}(k) &= b_1^2P_{\rm NL}(k) + \int \frac{d^3q}{(2\pi)^3}P_{m}(q)P_{m}(|k-q|)\\
&\Big[\frac{1}{2}b_2^2 + 2b_1b_2F_2(q,k-q)+\\
&+4b_1b_vF_2(q,k-q)G_u(q,k-q)\\
&+ 2b_2b_vG_{\rm u}(q,k-q) + 2b_v^2G^2_{\rm u}(q,k-q)\Big]
\end{split}
\label{eq:relmodel}
\end{equation}
with the kernels
\begin{align}
G_{\rm u}(k_1,k_2) &= - \frac{T_{ur}(k_1)}{T_m(k_1)}\frac{T_{ur}(k_2)}{T_m(k_2)}\frac{\vec{k}_1\cdot\vec{k}_2}{k_1k_2}\\
F_{2}(k_1,k_2) &= \frac{5}{7}+\frac{2}{7}\left(\frac{\vec{k}_1\cdot\vec{k}_2}{k_1k_2}\right)^2 + \frac{\vec{k}_1\cdot\vec{k}_2}{2}\left(\frac{1}{k_1^2} + \frac{1}{k_2^2}\right).
\end{align}
The dimensionless relative velocity transfer function $T_{u_r}$ is defined as 
\begin{equation}
T_{u_r} = \frac{T_{v_b}-T_{v_{\rm cdm}}}{\sigma_{v_r}},
\end{equation}
where $T_{v_b}$ and $T_{v_{\rm cdm}}$ are the velocity transfer functions of baryons and cold dark matter, respectively and the normalization is given by
\begin{equation}
\sigma^2_{v_r}(z) = \frac{1}{3}\int \frac{dk}{k} T^2_{v_r}(k,z)A_s\left(\frac{k}{0.002\,\text{Mpc}}\right)^{n_s-1}.
\label{eq:relnorm}
\end{equation}
The transfer functions describe the evolution of each mode with redshift as 
\begin{equation}
\delta(\vec{k},z) = T(k,z)\delta_{\rm pri}(\vec{k}),
\end{equation}
where $\delta_{\rm pri}$ is the primordial density perturbation. Before reionisation the linear relative velocity transfer function does not change its shape but only its amplitude (see Figure~\ref{fig:z_dep}). Assuming the relative velocity effect has been imprinted before reionisation, any redshift above reionisation can be chosen to calculate the relative velocity transfer function\footnote{CMBfast and early versions of CAMB do not include the effect of reionisation to the baryon transfer function (see Section 2.3 in~\citealt{Lesgourgues:2011rg}).}. We calculate the velocity transfer function $T_{u_r}$ at $z = 15$, while the matter transfer function in the denominator of the velocity kernel is calculated at $z = 0.57$.

We can turn this model into configuration space using a Fourier transform as given in Eq.~\ref{eq:noshape2}. Figure~\ref{fig:peak1} shows the correlation function for different values of the relative velocity bias $b_v$. The relative velocity bias causes an increase in the correlation function amplitude as well as a shift of the BAO peak position towards smaller scales.

Using the definition of Eq.~\ref{eq:deltarel} we can write a model for the cross-correlation function of a tracer $A$, which carries the relative velocity effect, with another tracer $B$, which does not carry this effect:
\begin{equation}
\begin{split}
P^{\rm AB}_{\rm g}(k) &= b^{\rm A}_1b^{\rm B}_1P_{\rm NL}(k) + \int \frac{d^3q}{(2\pi)^3}P_{m}(q)P_{m}(|k-q|)\\
&\Big[\frac{1}{2}b^{\rm A}_2b^{\rm B}_2 + (b^A_1b^{\rm B}_2 + b_2^{\rm A}b_1^{\rm B})F_2(q,k-q)+\\
&+2b^{\rm B}_1b^{\rm A}_rF_2(q,k-q)G_u(q,k-q)\\
&+ b^{\rm B}_2b^{\rm A}_rG_{\rm u}(q,k-q)\Big].
\end{split}
\label{eq:relmodel2}
\end{equation}
Figure~\ref{fig:peak2} shows the cross-correlation function for different values of the relative velocity bias $b_v$. As was the case for the auto-correlation function, the relative velocity bias causes a shift of the BAO peak position towards smaller scales. In the cross-correlation function the shift is about $1/3$ of the shift present in the auto-correlation function. Unlike the auto-correlation, the relative velocity effect does not change the amplitude of the cross-correlation function. This means that the cross-correlation function between two tracers does not have the bias $b_1^{\rm A}b_1^{\rm B}$ and the correlation coefficient $r^2 = \xi^2_{\rm AB}/(\xi_{\rm A}\xi_{\rm B})$ is predicted to be smaller than unity. We therefore have two effects which can be used to constrain the relative velocity effect: (1) the BAO peak position and (2) the amplitude of the cross-correlation function relative to the auto-correlation functions.

\subsection{Constraining the relative velocity effect - BAO fits}
\label{sec:relvelfitting1}

\begin{table*}
\begin{center}
\caption{Summary of the fitting results for the relative velocity effect. The first two rows show the results when using only the BAO peak position to constrain $b_v$ before density field reconstruction. The last two rows present the same fit after reconstruction. The third and fourth rows list the fit parameters including the shape (and amplitudes) of the correlation functions. The fifth row provides the fit parameters to the correlation coefficient $r^2$. All uncertainties are defined by the $68\%$ confidence levels. The fitting ranges are shown in the second column.}
	\begin{tabular}{cccccc}
		\hline
		 fit condition & fitting range & $b_v$& $b^{\rm CMASS}_2$ & $b^{\rm WiggleZ}_2$ & $\chi^2$\\
		\hline
		\multicolumn{6}{l}{pre-recon}\\
		BAO only & $50$ - $180\,$Mpc$/h$ & $-0.067 < b_v < 0.010$ & $0.0^{+6.1}_{-1.4}$ & $14.5^{+4.7}_{-11.5}$ & $82.4/(78-16)$\\
		BAO only & $50$ - $180\,$Mpc$/h$ & $-0.31 < b_v < 0.060$ & $1.0$ & $-0.4$ & $84.0/(78-14)$\\
		shape & $50$ - $180\,$Mpc$/h$ & $-0.059 < b_v < 0.096$ & $-2.0^{+15}_{-14}$ & $-0.7^{+7.2}_{-6.7}$ & $89.4/(78-6)$\\
		shape & $50$ - $180\,$Mpc$/h$ & $-0.12 < b_v < 0.037$ & $1.0$ & $-0.4$ & $94.7/(78-4)$\\
		$r^2$ & $20$ - $60\,$Mpc$/h$ & $-0.086 < b_v < 0.062$ & $1.0$ & $-0.4$ & $7.4/(9-4)$\\
		\multicolumn{6}{l}{post-recon}\\
		BAO only & $50$ - $180\,$Mpc$/h$ & $-0.21<b_v<0.02$ & $7.6^{+7.8}_{-8.3}$ & $-0.5^{+2.8}_{-3.1}$ & $98.9/(78-16)$\\
		BAO only & $50$ - $180\,$Mpc$/h$ & $-0.22<b_v<0.10$ & $1.0$ & $-0.4$ & $103.5/(78-14)$\\
		\hline
	  \end{tabular}
\label{tab:results2}
\end{center}
\end{table*}

\begin{figure}
\begin{center}
\epsfig{file=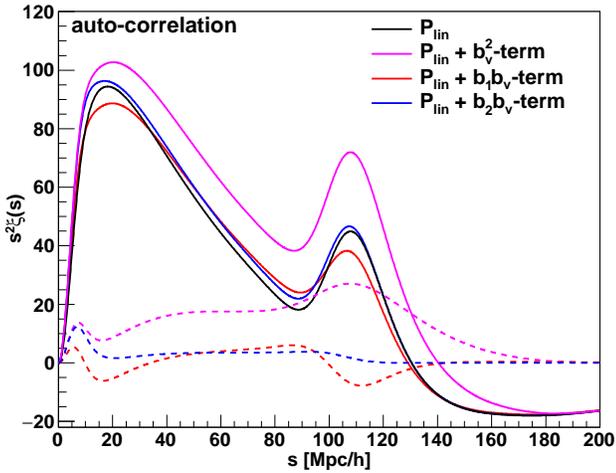,width=9cm}
\caption{The effect of the three relative velocity terms of Eq.~\ref{eq:relmodel} to the large scale correlation function. The black line shows the linear input correlation functions, while the magenta, red and blue lines include the $b_v^2$, $b_1b_v$ and $b_2b_v$ terms, respectively. The dashed lines represent the effect of the relative velocity terms (solid colored lines minus the black solid line). We fix $b_v=0.1$, $b_1 = 2$ and $b_2 = 1$.}
\label{fig:terms}
\end{center}
\end{figure}

\begin{figure}
\begin{center}
\epsfig{file=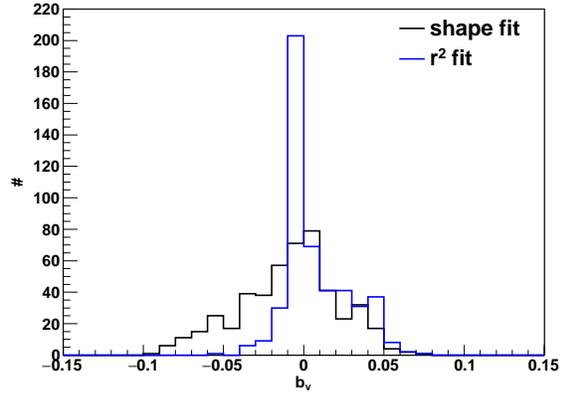,width=8cm}
\caption{Distribution of the relative velocity bias $b_v$ obtained from the $480$ mock catalogues using the ``shape'' and $r^2$ fitting methods (see Table~\ref{tab:results2} and text in Section~\ref{sec:relvelfitting1} and~\ref{sec:relvelfitting2}).}
\label{fig:bvhist}
\end{center}
\end{figure}

As mentioned in the last section, we have two effects which can be exploited to constrain the relative velocity effect, the shift in the BAO peak and the relative amplitudes of the individual correlation functions. While the shift in the BAO peak position can be considered as robust, there are effects other than the relative velocity bias which could change the amplitude. Any stochasticity, $\delta_s$, in the galaxy density field, which is not correlated with the matter density, $\delta_g = b_g\delta_m + \delta_s$, and which does not correlate with the density field of the other survey, would lead to a reduction in the amplitude of the cross-correlation function. We will therefore perform multiple fits. First we will show the constraint on the relative velocity effect just using the BAO peak position, and  then include the relative amplitudes of the correlation functions.

We start with the BAO peak position. For this fit we convert the models of Eq.~\ref{eq:relmodel} and~\ref{eq:relmodel2} into configuration space and introduce additional polynomial terms to marginalize over the shape of the correlation functions, similar to the discussion in Section~\ref{sec:model}:
\begin{equation}
\xi'_{\rm relvel}(s) = B^2\xi_{\rm relvel}(\alpha s) + A(s).
\end{equation}
We also marginalize over the amplitude of the three correlation functions by giving each correlation function a separate bias parameter. Higher order terms for the cross-correlation function are always set by the bias of the auto-correlation functions. In total we have $14$ free parameters ($B_B$, $B_C$, $B_W$, $A_B$, $A_C$, $A_W$, $\alpha$, $b_v$), where the polynomial terms $A$ have three parameters each. Since we assume that the relative velocity is only present in CMASS, the  relative velocity parameter, $b_v$, only affects the CMASS and cross-correlation function model. We perform fits where we additionally vary the parameter $b_2$, but since this parameter is not well constrained we often fix it to $1.0$ for CMASS-BW and $-0.4$ for WiggleZ-BW~\citep{Yoo:2013qla,Marin:2013bbb}. The result of the fits are presented in the first two rows of Table~\ref{tab:results2} before reconstruction, and the last two rows after reconstruction. Regardless of how we treat the parameter $b_2$, we obtain constraints on $b_v$ which are consistent with zero. This result is not surprising since the shift of the BAO peak due to the relative velocity effect is a shift to smaller scales. Our data, however, show a BAO peak at larger scales for CMASS-BW compared to WiggleZ-BW.

\subsection{Constraining the relative velocity effect - shape fits}
\label{sec:relvelfitting2}

Next we fit the correlation functions without marginalizing the relative amplitudes. In this case we include a bias parameter for CMASS-BW and WiggleZ-BW, but not for the cross-correlation function. The amplitude of the cross-correlation function is given by the product of the CMASS-BW and WiggleZ-BW bias parameters. Since this fit does not marginalize over the relative amplitudes and shape of the individual correlation functions, we also include the velocity dispersion parameter $\sigma_v$ as
\begin{align}
P_{\rm g, final}(k) &= P_{\rm g}(k)\exp(-k^2\sigma^2_v/2),\cr
P^{\rm AB}_{\rm g, final}(k) &= P^{\rm AB}_{\rm g}(k)\exp(-k^2\sigma^2_v/2).
\end{align}
Given the simplicity of our model, the parameter $\sigma_v$ absorbs small-scale effects like non-linear structure formation which are bias dependent. We therefore include three different $\sigma_v$ parameters, one for each correlation function. The parameter $\sigma_v$ introduces stochasticity on small scales which could mimic the relative velocity effect. We verify that this model yields a relative velocity bias consistent with zero when applied to our mock catalogues. Figure~\ref{fig:bvhist} shows the distribution of $b_v$ obtained from the 480 mock catalogues. We only perform these fits pre-reconstruction, since we do not have a model for the post-reconstruction correlation function. The results are included in Table~\ref{tab:results2} with the label ``shape''. While the constraints on $b_v$ become tighter compared to the ``BAO only'' fits, they are still consistent with zero. 

For the ``shape'' fit the correction factor of Eq.~\ref{eq:scaling} is $1.4$ and contributes significantly to our error budget. To avoid this additional source of error we can fit the correlation coefficient $r^2$ instead of the individual correlation functions. This approach reduces the number of bins, which reduces the correction factor of Eq.~\ref{eq:scaling}. The correlation coefficient should also be fairly independent of the underlying cosmological model, since any effect common to the correlation functions cancels. However, the parameters $b_1$, $b_2$ and $b_v$ are degenerate when using $r^2$. From Figure~\ref{fig:terms} we can see that it might be possible to separately constrain $b_1$ and $b_2$ given that $b_2$ does behave very differently on small scales and around the BAO peak. However, while on small scales there is concern about the applicability of our model, on large scales the uncertainties are too large to exploit these effects. Thus we cannot vary all three parameters simultaneously. We therefore fix the value of $b_1 = 1.9$ and $b_2 = 1.0$ for CMASS-BW and $b_1 = 1$ and $b_2 = -0.4$ for WiggleZ-BW. The term proportional to $b_v^2$ is usually significantly larger than the $b_1b_v$ and $b_2b_v$ terms, justifying to some extent our choice of fixing $b_1$ and $b_2$ (see Figure~\ref{fig:terms}). 

The mock realizations are in good agreement with the expected value of the correlation coefficient above $20\,$Mpc$/h$. The scales above $60\,$Mpc$/h$ have large uncertainties and can be neglected for this fit, leading to the fitting range $20$ - $60\,$Mpc/h. We again verify that our model can reproduce a relative velocity bias of zero when applied to our mock catalogues. The distribution of maximum likelihood $b_v$ for the $480$ mock realizations is included in Figure~\ref{fig:bvhist} (blue line). The best fitting parameters are included in Table~\ref{tab:results2}. The relative velocity bias is again consistent with zero.

So far the only constraint on the relative velocity bias, $b_v$, in the literature has been reported by~\citet{Yoo:2013qla} using the CMASS-DR9 power spectrum. They found $b_v < 0.033$ at the $95\%$ confidence level, consistent with our result. However, their constraint was found by fixing all cosmological parameters to the Planck values, while our BAO-only constraint can be considered model independent. 

\section{Conclusion}
\label{sec:conclusion}

We have investigated the galaxy clustering in the overlap region between the BOSS-CMASS and WiggleZ galaxy surveys. Having two galaxy samples in the same volume with different galaxy properties as well as survey selection effects presents a valuable opportunity to test for possible systematic uncertainties in our analysis of the Baryon Acoustic Oscillation scale. We can summarize our results as follows:
\begin{itemize}
\item We detect a BAO signal in both auto-correlation functions as well as the cross-correlation function of CMASS and WiggleZ using only the overlap region between the two surveys. The BAO detection in the cross-correlation function represents the first BAO detection in the cross-correlation function of two completely different galaxy surveys. After applying density field reconstruction we find distance constraints of $D_V\frac{r_s^{\rm fid}}{r_s} = (1970\pm45, 2132\pm65,2100\pm200)$Mpc for CMASS, the cross-correlation and WiggleZ, respectively. The three constraints are consistent with each other and with the distribution found in the mock realizations. The results are also robust against switching the displacement field of the two surveys during density field reconstruction. We therefore cannot see signs of systematic uncertainties.
\item We use our correlated mock realizations to determine the correlation between CMASS and WiggleZ. Using these correlations we derived a covariance matrix for the CMASS-DR11 and our WiggleZ and cross-correlation constraints. While in the past the WiggleZ constraints have often been ignored when constraining cosmological models given the overlap (and hence correlation) with the CMASS results, our covariance matrix now allows one to make use of the WiggleZ information for cosmological constraints. Since the overlap region between the two surveys will not grow with future CMASS data releases, the covariance derived in this paper can easily be rescaled to obtain the covariance between our WiggleZ constraints and future CMASS data releases. We already provide a correlation matrix for the expected CMASS-DR12 results.
\item Using the measured correlation functions we test for the relative velocity effect, which is a possible source of systematic uncertainty for BAO measurements. We perform various fits using the effect of the relative velocity bias on the BAO peak position as well as the relative amplitudes of the auto- and cross-correlation functions. We cannot detect any signs of a relative velocity bias.
\end{itemize}

\section*{Acknowledgments}

FB would like to thanks Jaiyul Yoo for fruitful discussions and Julien Lesgourgues, Antony Lewis and Thomas Tram for help with the effect of reionisation to the relative velocity transfer function. CB acknowledges the support of the Australian Research Council through the award of a Future Fellowship.
FM and JK were supported by the Australian Research Council Centre of Excellence for All-Sky Astrophysics (CAASTRO) through project number CE110001020. 

Funding for SDSS-III has been provided by the Alfred P. Sloan Foundation, the Participating Institutions, the National Science Foundation, and the U.S. Department of Energy. SDSS-III is managed by the Astrophysical Research Consortium for the Participating Institutions of the SDSSIII Collaboration including the University of Arizona, the Brazilian Participation Group, Brookhaven National Laboratory, University of Cambridge, Carnegie Mellon University, University of Florida, the French Participation Group, the German Participation Group, Harvard University, the Instituto de Astrofisica de Canarias, the Michigan State/Notre Dame/JINA Participation Group, Johns Hopkins University, Lawrence Berkeley National Laboratory, Max Planck Institute for Astrophysics, Max Planck Institute for Extraterrestrial Physics, New Mexico State University, New York University, Ohio State University, Pennsylvania State University, University of Portsmouth, Princeton University, the Spanish Participation Group, University of Tokyo, University of Utah, Vanderbilt University, University of Virginia, University of Washington, and Yale University.

\setlength{\bibhang}{2em}
\setlength{\labelwidth}{0pt}

\appendix

\label{lastpage}

\end{document}